\documentclass[]{aa}  

\usepackage{graphicx}
\usepackage{txfonts}
\usepackage{multirow}
\usepackage{pdflscape}
\usepackage{hyperref}

\graphicspath{ {figures/} }

\begin{document} 

   \title{A machine-learning photometric classifier for massive stars in nearby galaxies\footnote{Table A.1 ia only available in electronic form at the CDS via anonymous ftp to cdsarc.u-strasbg.fr (130.79.128.5) or via \url{http://cdsweb.u-strasbg.fr/cgi-bin/qcat?J/A+A/}}}
   
   \titlerunning{A machine-learning classifier for massive stars in nearby galaxies I}

   \subtitle{I. The method}

   \author{Grigoris Maravelias\inst{1,2},
   Alceste Z. Bonanos\inst{1}, 
   Frank Tramper\inst{1,3}, 
   Stephan de Wit\inst{1}, 
   Ming Yang\inst{1}, 
   \and Paolo Bonfini \inst{2,4,5}
          }

    \authorrunning{Maravelias et al.}

   \institute{IAASARS, National Observatory of Athens, GR-15236, Penteli, Greece    \\
   \email{maravelias@noa.gr}
    \and Institute of Astrophysics, FORTH, GR-71110, Heraklion, Greece
    \and Institute of Astronomy, KU Leuven, Celestijnenlaan 200D, 3001, Leuven, Belgium
    \and Computer Science Department, University of Crete, GR-71003, Heraklion, Greece
    \and Carrera Group, Jacksonville Beach, Florida, USA
    }

  \date{Received 26 May, 2021; accepted 08 March, 2022} 

 
  \abstract
    {Mass loss is a key parameter in the evolution of massive stars. Despite the recent progress in the theoretical understanding of how stars lose mass, discrepancies between theory and observations still hold. Moreover, episodic mass loss in evolved massive stars is not included in models, and the importance of its role in the evolution of massive stars is  currently undetermined.  }
   {A major hindrance to determining the role of episodic mass loss is the lack of large samples of classified stars. Given the recent availability of extensive photometric catalogs from various surveys spanning a range of metallicity environments, we aim to remedy the situation by applying machine-learning techniques to these catalogs.}
   {We compiled a large catalog of known massive stars in M31 and M33 using  IR (\textit{Spitzer}) and optical (Pan-STARRS) photometry, as well as \textit{Gaia} astrometric information, which helps with foreground source detection. We grouped them into seven classes (Blue, Red, Yellow, B[e] supergiants, Luminous Blue Variables, Wolf-Rayet stars, and outliers, e.g., quasi-stellar objects and background galaxies). As this training set is highly imbalanced, we implemented synthetic data generation to populate the underrepresented classes and improve separation by undersampling the majority class. We built an ensemble classifier utilizing color indices as features. The probabilities from three machine-learning algorithms (Support Vector Classification, Random Forest, and Multilayer Perceptron) were combined to obtain the final classification. 
   }
   {The overall weighted balanced accuracy of the classifier is $\sim83\%$. Red supergiants are always recovered at $\sim94\%$. Blue and Yellow supergiants, B[e] supergiants, and background galaxies achieve $\sim50 - 80\%$. Wolf-Rayet sources are detected at $\sim45\%$, while Luminous Blue Variables are recovered at $\sim30\%$ from one method mainly. This is primarily due to the small sample sizes of these classes.  In addition, the mixing of spectral types, as there are no strict boundaries in the features space (color indices) between those classes, complicates the classification. In an independent application of the classifier to other galaxies (IC 1613, WLM, and Sextans A), we obtained an overall accuracy of $\sim70\%$. This discrepancy is attributed to the different metallicity and extinction effects of the host galaxies. Motivated by the presence of missing values, we investigated the impact of missing data imputation using a simple replacement with mean values and an iterative imputer, which proved to be more capable. We also investigated the feature importance to find that $r-i$ and $y-[3.6]$ are the most important, although different classes are sensitive to different features (with potential improvement with additional features). 
   }
   {The prediction capability of the classifier is limited by the available number of sources per class (which corresponds to the sampling of their feature space), reflecting the rarity of these objects and the possible physical links between these massive star phases. Our methodology is also efficient in correctly classifying sources with missing data as well as at lower metallicities (with some accuracy loss), making it an excellent tool for accentuating interesting objects and prioritizing targets for observations.  
   }


   \maketitle
%

\section{Introduction}

Although rare, massive stars ($M_*>8-10M_\odot$) play a crucial role in multiple astrophysical domains in the Universe. Throughout their life they continuously lose mass via strong stellar winds that transfer energy and momentum to the interstellar medium. As the main engines of nucleosynthesis, they produce a series of elements and shed chemically processed material as they evolve through various phases of intense mass loss. And they do not simply die: they explode as spectacular supernovae, significantly enhancing the galactic environment of their host galaxies. Their end products, neutron stars and black holes, offer the opportunity to study extreme physics (in terms of gravity and temperature) as well as gamma-ray bursts and gravitational wave sources. As they are very luminous, they can be observed in more distant galaxies, which makes them the ideal tool for understanding stellar evolution across cosmological time, especially for interpreting observations from the first galaxies (such as those to be obtained from \textit{James Webb Space Telescope}). 

While the role of different stellar populations on galaxy evolution has been thoroughly investigated in the literature \citep{Bruzual2003, Maraston2005}, a key ingredient of models, the evolution of massive stars beyond the main sequence, is still uncertain \citep{Martins2013, Peters2013}. Apart from the initial mass, the main factors that determine the evolution and final stages of a single massive star are metallicity, stellar rotation, and mass loss \citep{Ekstrom2012, Georgy2013, Smith2014}. Additionally, the presence of a companion, which is common among massive stars with binary fractions of $\sim50-70\%$ (\citealt{Sana2012, Sana2013, Dunstall2015}), can significantly alter the evolution of a star through strong interactions \citep{deMink2014,Eldridge2017}. Although all these factors critically determine the future evolution and the final outcome of the star, they are, in many cases, not well constrained.

In particular, mass loss is of paramount importance as it determines not only the stellar evolution but the enrichment and the formation of the immediate circumstellar environment (for a review, see \citealt{Smith2014} and references therein). Especially in the case of single stars, their strong radiation-driven winds during the main-sequence phase remove material continuously but not necessarily in a homogeneous way, due to clumping \citep{Owocki1999}. On top of that, there are various transition phases in the stellar evolution of massive stars during which they experience episodic activity and outbursts, such as Wolf-Rayet stars (WRs), Luminous Blue Variables (LBVs), Blue supergiants (BSGs), B[e] supergiants (B[e]SGs), Red supergiants (RSGs), and Yellow supergiants (YSGs). This contributes to the formation of complex structures, such as shells and bipolar nebulae in WRs and LBVs (\citealt{Gvaramadze2010, Wachter2010}) and disks in B[e]SGs (\citealt{Maravelias2018}). But how important the episodic mass loss is, how it depends on the metallicity (in different galaxies), and what links exist between the different evolutionary phases are still open questions. 

To address these questions, the  European Research Council-funded project ASSESS\footnote{\url{https://assess.astro.noa.gr/}} (\textit{"Episodic Mass Loss in Evolved Massive stars: Key to Understanding the Explosive Early Universe"}) aims to determine the role of episodic mass by: (i) assembling a large sample of evolved massive stars in a selected number of nearby galaxies at a range of metallicities through multiwavelength photometry, (ii) performing follow-up spectroscopic observations on candidates to validate their nature and extract stellar parameters, and (iii) testing the observations against the assumptions and predictions of the stellar evolution models. In this paper we present our approach for the first step, which is to develop an automated classifier based on multiwavelength photometry. 

One major complication for this work is the lack of a sufficiently large number of massive stars with known spectral types. Some of these types are rare, which makes the identification of new sources  in nearby galaxies even more difficult. Moreover, spectroscopic observations at these distances are challenging due to the time and large telescopes required. On the other hand, photometric observations can provide information for thousands of stars but at the cost of a much lower (spectral) resolution, leading to a coarser spectral-type classification (e.g., \citealt{Massey2006, Bonanos2009, Bonanos2010, Yang2019}). Using the Hertzsprung–Russell diagram (HRD) and color-color diagrams, one needs a detailed and careful approach to properly determine the boundaries between the different populations and identify new objects, a process that is not free from contaminants (e.g., \citealt{Yang2019}).

To circumvent this problem, we can use a data-driven approach. In this case, data can be fed to more sophisticated algorithms that are capable of "learning" from the data and finding the mathematical relations that best separate the different classes. These machine-learning methods have been extremely successful in various problems in astronomy (see Sect. \ref{s:algorithms}). Still, though, applications of these techniques tailored for the classification of massive stars with photometry are, to the best of our knowledge, scarce if not almost nonexistent. \cite{Morello2018} studied the \textit{k}-nearest neighbors on IR colors to select WRs from other classes, while \cite{Dorn-Wallenstein2021} explored other techniques to obtain a wider classification based on \textit{Gaia} and IR colors for a large number of Galactic objects. This work provides an additional tool, focusing on massive stars in nearby galaxies. It presents the development of a photometric classifier, which will be used in a future work to provide the classification for thousands of previously unclassified  sources\footnote{Code and other notes available at: \url{https://github.com/gmaravel/pc4masstars}}.

In Sect. \ref{s:data} we present the construction of our training sample (spectral types, foreground removal,  and photometric data). In Sect. \ref{s:ml} we provide a quick summary of the methods used and describe the class and feature selection, as well as the implementation and the optimization of the algorithms. In Sect. \ref{s:results} we show the performance of our classifier for the M31 and M33 galaxies (on which it was trained) and its application to an independent set of galaxies (IC 1613, WLM, and Sextans A). In Sect. \ref{s:discussion} we discuss the necessity of a good training sample and labels, as well as the feature sensitivity. Finally, in Sect. \ref{s:summary} we summarize and conclude our work.

\section{Building the training sample}
\label{s:data}

In the following section we describe the steps we followed to create our training sample, starting for the available IR photometric catalogs, removing foreground sources using \textit{Gaia} astrometric information, and collecting spectral types from the literature.

\begin{figure*}[hbt!]
    \centering
        \includegraphics[width=\textwidth]{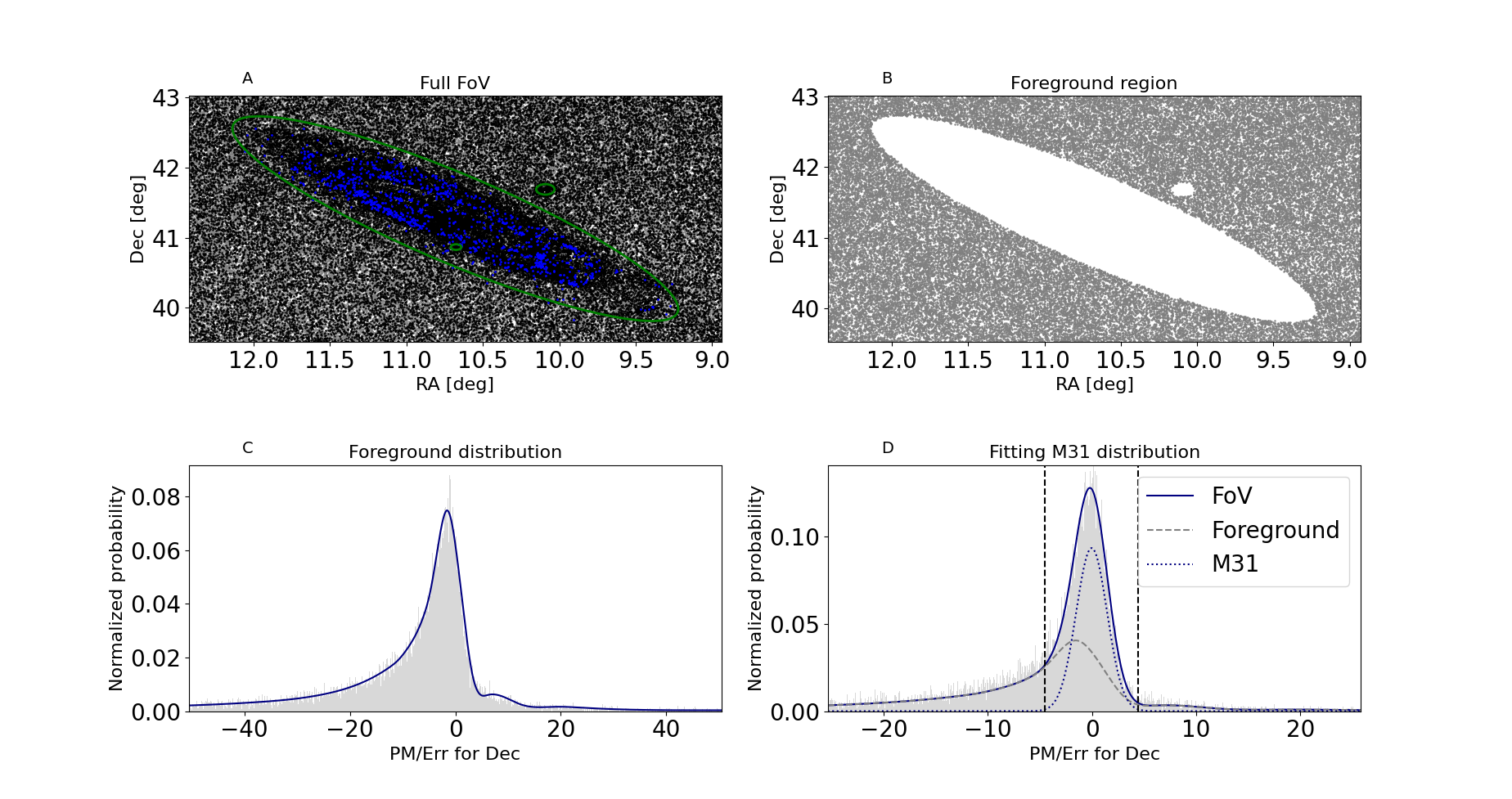}\\ 
        \includegraphics[width=\textwidth]{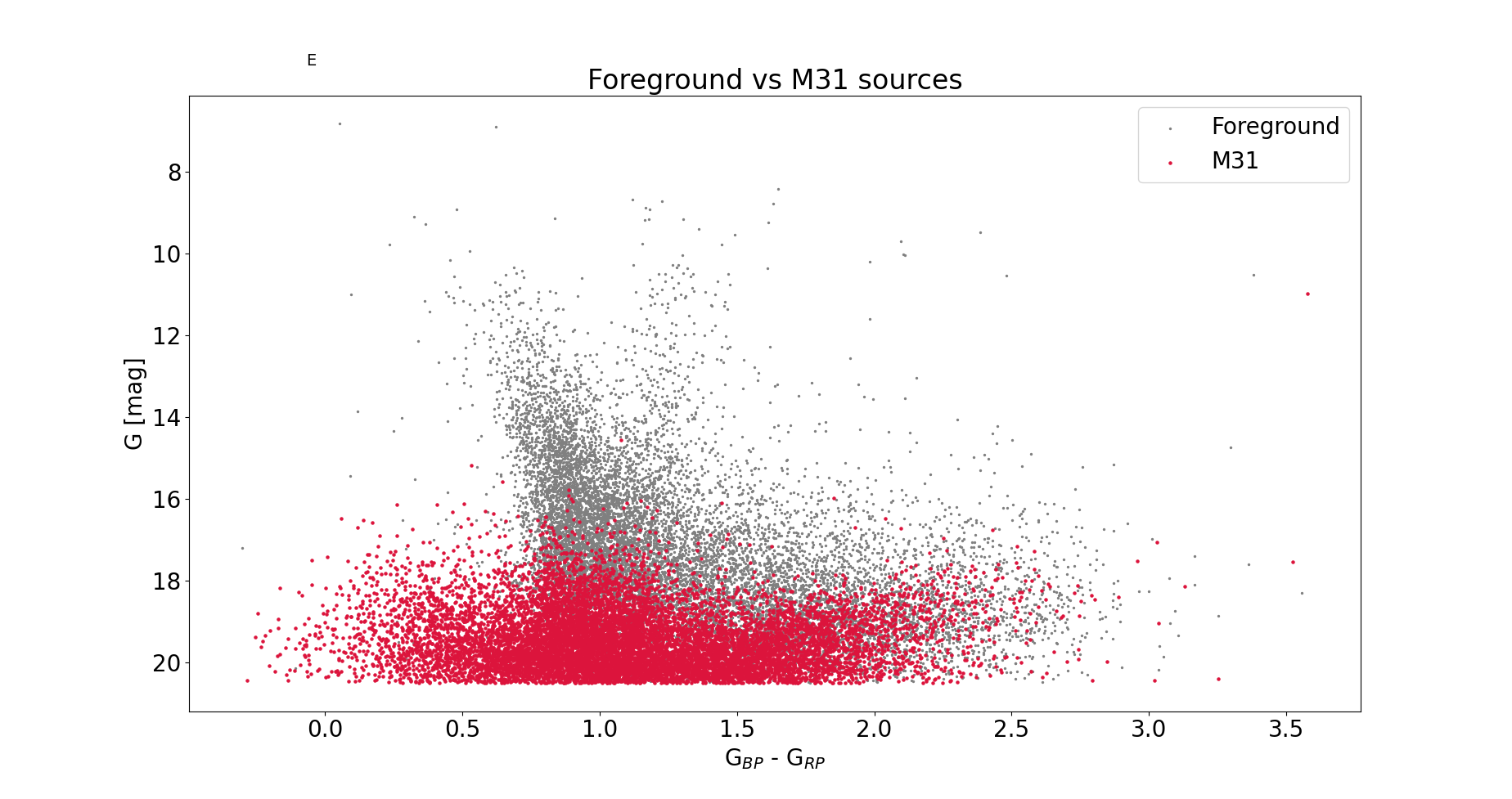} 
        \caption{ Using \textit{Gaia} to identify and remove foreground sources. (A) Field-of-view of \textit{Gaia} sources (black dots) for M31. The big green ellipse marks the boundary we defined for the M31 galaxy, and the smaller green ellipses define M110 and M32 (inside the M31 ellipse) and are excluded. The blue dots highlight  the sources in M31 with known spectral classification. (B) Foreground region, excluding the sources inside M110.  (C) Distribution of the proper motion over its error for Dec, for all \textit{Gaia} sources in the foreground region, fitted with a spline function. (D) Distribution of the proper motion over its error for Dec (solid line), for all sources along the line-of-sight of M31, which includes both foreground and galactic (M31) sources. We fitted this with a scaled spline, to account for the number of foreground sources expected inside M31 (dashed line), and a Gaussian function (dotted line). The vertical dashed lines correspond to the $3\sigma$ threshold of the Gaussian. Any source with values outside this region is flagged as a potential foreground source.  (E) \textit{Gaia} CMD of all sources identified as galactic (red points) and foreground (gray). The majority of the foreground sources lie on the yellow branch of the CMD, which is exactly the position at which we expect the largest fraction of the contamination.}
    \label{f:gaia_process}
\end{figure*}

\subsection{Surveys used}
\label{s:surveys}

Infrared bands are ideal probes for distinguishing massive stars and particular those with dusty environments \citep{Bonanos2009, Bonanos2010}. The use of IR colors is a successful method for target selection, as demonstrated by \citealt{Britavskiy2014,  Britavskiy2015}. We based our catalog composition on mid-IR photometry ($3.6\, \mu m$, $4.5\, \mu m$, $5.8\, \mu m$, $8.0\, \mu m$, $24\, \mu m$), using pre-compiled point-source catalogs from the \textit{Spitzer} Space Telescope \citep{Khan2015, Khan2017, Williams2016}, which have only recently become publicly available. This allows us to use positions derived from a single instrument, a necessity for cross-matching since spectral typing comes from various works and instruments. The cross-match radius applied in all cases was 1", since this corresponds to a significant physical separation that grows gradually as a function of distance. Additionally, we only kept sources with single matches, as it is impossible to choose the correct match to the \textit{Spitzer} source when two or more candidates exist within the search radius (accounting for about 2-3\% of all sources in M31 and M33).  

Although the inclusion of near-IR data would help better sampling of the spectral energy distribution of our sources, this is currently impossible given the shallowness of Two Micron All-Sky Survey (2MASS) for our target galaxies, and the -- unfortunate -- lack of any other public all sky near-IR survey. Some data (for a particular band only; $J_{\rm{UK}}$) were collected from the United Kingdom Infra-Red Telescope (UKIRT) Hemisphere Survey\footnote{\url{http://wsa.roe.ac.uk/uhsDR1.html}} \citep{Dye2018}. 

The data set was supplemented with optical photometry ($g,r,i,z,y$) obtained from the Panoramic Survey Telescope and Rapid Response System (Pan-STARRS; \citealt{Chambers2016}), using sources with ${\rm \texttt{nDetections}}\geq2$ to exclude spurious detections\footnote{Although DR2 became available after the compilation of our catalogs, it contains information from the individual epochs. DR1 provides the object detection and their corresponding photometry from the stacked images, which we opted to use.}. We also collected photometry from \textit{Gaia} Data Release 2 (DR2) (G, G$_{\rm{BP}}$, G$_{\rm{RP}}$; \citealt{Gaia2016, Gaia2018b}).  

We investigated all other available surveys in the optical and IR but the aforementioned catalogs provided the most numerous and consistent sample with good astrometry for our target galaxies (M31, M33, IC 1613, WLM, and Sextans A). Significant populations of massive stars are well known for the Magellanic Clouds and the Milky Way but there are issues that prohibited us from using them. The Clouds are not covered by the Pan-STARRS survey, which means that photometry from other surveys should be used that would make the whole sample inhomogeneous (with all possible systematics introduced by the different instrumentation, data reductions, etc). Although Milky Way is covered by both Pan-STARRS and \textit{Spitzer} surveys, there are hardly any data available for the most interesting sources, such as B[e]SGs, WRs, and LBVs, through the \textit{Spitzer} Enhanced Imaging Products (which focus on the quality of the products and not completeness). Therefore, we limited ourselves to M31 and M33 galaxies when building our training sample.

\subsection{Removing foreground stars}

The source lists compiled from the photometric surveys described in the previous section, contain mostly genuine members of the corresponding galaxies. It is possible though that foreground sources may still contaminate these lists.  To optimize our selection, we queried the \textit{Gaia} DR2 catalog \citep{Gaia2016, Gaia2018b}. With the statistical handling of the astrometric data we were able to identify and remove most probable foreground sources in the line-of-sight of our galaxies. 

We first defined a sufficiently large box around each galaxy: $3.5\,{\rm deg} \times 3.5\,{\rm deg} $ for M31 and $ 1.5\, {\rm deg} \times 1.5\, {\rm deg} $ for M33, which yielded 145837 and 34662 sources, respectively. From these we first excluded all sources with nonexistent or poorly defined proper motions (${\rm pmra\_error} \geq 3.0\, {\rm mas}$, $ {\rm pmdec\_error} \geq 3.0\, {\rm mas}$) or parallax (${\rm parallax\_error} \geq 1.5 {\rm mas}$), sources with large  astrometric excess noise ($ {\rm astrometric\_excess\_noise} \geq 1.0$; following the cleaning suggestions by \citealt{Lindegren2018}), or sources that were fainter than our limit set in the optical (${\rm phot\_g\_mean\_mag} \geq  20.5$). These quality cuts left us with 78375 and 26553 sources in M31 and M33, respectively. 

The boundary for each galaxy was determined as the ellipse at which the star density dropped significantly at approximately the density of the background. This boundary was visually inspected also so that it masks the main body (disk) of each galaxy where our targets are expected to be located (and to exclude contaminating regions inside and outside the galaxy, namely M32 and M110 galaxies for M31, see Fig. \ref{f:gaia_process}, Panel A; for M33 see Fig. \ref{f:gaia_process-M33}). Therefore, we could securely assign the remainder of stars as foreground objects (see Fig. \ref{f:gaia_process}, Panel B). From these we obtained the distributions on the proper motions in RA and Dec (over their corresponding errors) and the parallax (over its error). We fitted these distributions with a spline to allow more flexibility
(see Dec for example in Fig. \ref{f:gaia_process}, Panel C).

Similarly, we plotted the distributions for all sources within the ellipse, which contained both galactic and foreground sources. To fit these we used a combination of a Gaussian and a spline function (see Fig. \ref{f:gaia_process}, Panel D). The spline was derived from the sources outside the galaxy (Fig. \ref{f:gaia_process}, Panel C), but when used for the sources within the ellipse it was scaled down according to the ratio of the area outside and inside the galaxy (assuming that the foreground distribution does not change). From the estimated widths of the Gaussian distributions (M31: $ {\rm pmRA/error} = 0.04 \pm 1.28$, $ {\rm pmDEC/error} = -0.03 \pm 1.48$, $ {\rm parallax/error} = 0.21 \pm 1.39$; M33: $ {\rm pmRA/error} = 0.12 \pm 1.18$, $ {\rm pmDEC/error} = 0.05 \pm 1.31$, $ {\rm parallax/error} = -0.03 \pm 1.16$) we defined as foreground sources those with values larger than $3\sigma$ in any of the above quantities. For the parallax we took into account the systematic 0.03 mas offset induced by the global zero point found by \cite{Lindegren2018}. This particular cut was applied only to sources with actual positive values, as zero or negative values are not decisive for exclusion. In the \textit{Gaia} color-magnitude diagram (CMD) of Fig. \ref{f:gaia_process}, Panel E, we show all sources identified as members of the host galaxy (red points) and foreground (gray). The majority of the foreground sources lie on the yellow branch of the CMD, which is exactly the position at which we expect the largest fraction of the contamination.

This process was successful in the cases of M31 and M33 due to the numerous sources that allow their statistical handling. In the other galaxies, where the field-of-view is substantially smaller, the low numbers of sources led to a poorer (if any) estimation of these criteria. Consequently, for those galaxies we considered as foreground sources those with any of their \textit{Gaia} properties (${\rm pmRA/error}$, $ {\rm pmDEC/error}$, or $ {\rm parallax/error} $) larger than  $3\sigma$ of the largest measured errors, following the most conservative approach. In practice, this means that we used the same criteria to characterize foreground sources as with M31. 


\subsection{Collecting spectral types}
\label{s:data-spec_types}

The use of any supervised machine-learning application requires a training sample. It is of paramount importance that the sample be well defined such that it covers the parameter space spanned by the objects under consideration. For this reason, we performed a meticulous search of the literature to obtain a sample as complete as possible to our knowledge with known spectral types (that were used as labels). The vast majority of collected data are  found in  M31 and M33. The source catalogs were retrieved primarily from \cite{Massey2016} as part of their Local Group Galaxy Survey (LGGS) survey, complemented by other works (see Table \ref{t:spectypes_refs} for the full list of numbers and references used). 

In all cases we carefully checked for and removed duplicates, while in a few cases we updated the classification of some sources based on newer works (e.g., candidate LBVs to B[e]SGs based on \citealt{Kraus2019a}). The initial catalogs for M31 and M33 contain 1142 and 1388 sources with spectral classification, respectively (see Fig. \ref{f:gaia_process}, Panel A, blue dots). Within these sources we purposely included some outliers (such as background galaxies and quasi-stelar objects; e.g., \citealt{Massey2019}). 

A significant fraction of these sources ($\sim64\%$) have \textit{Gaia} astrometric information. Applying the criteria of the previous section, we obtained 58 (M31) and 76 (M33) sources marked as foreground\footnote{ There are 696 (M31) and 926 (M33) sources with \textit{Gaia} information. The identification of 58 (M31) and 76 (M33) sources as foreground corresponds to a $\sim8\%$ contamination. Given that there are 446 (M31) and 462 (M33) additional sources but without \textit{Gaia} values we expect another $\sim72$ sources to be foreground (according to our criteria) that remained in our catalog.}. After removing those, we were left with 1084 M31 and 1312 M33 sources, which are cross-matched with the photometric catalogs, considering single matches only at 1" (see Sect. \ref{s:surveys}). After this screening process our final sample consists of 527 (M31) and 562 (M33) sources.

\begin{table}
    \centering
    \caption{List of references with their corresponding number of sources that contribute to our collected sample.}
    \label{t:spectypes_refs}
    \begin{tabular}{llr}
    Galaxy (total) & Reference & \# sources \\
    \hline
    \hline
    \multirow{3}{*}{WLM (36)} & \cite{Bresolin2006} & 20 \\
                               & \cite{Britavskiy2015} & 9 \\
                               & \cite{Levesque2012} & 7 \\
    \hline    
    \multirow{9}{*}{M31 (1142)} & \cite{Massey2016}  & 966  \\
                          & \cite{Gordon2016} & 82 \\
                          & \cite{Neugent2019} & 37  \\
                          & \cite{Drout2009} & 18 \\
                          & \cite{Massey2019} & 17 \\
                          & \cite{Kraus2019a} & 11 \\
                          & \cite{Humphreys2017} & 6 \\
                          & \cite{Neugent2012} & 3 \\
                          & \cite{Massey2009} & 2 \\
    \hline
    \multirow{4}{*}{IC 1613 (20)} & \cite{Garcia2013}  & 9 \\
                                   & \cite{Bresolin2007} & 9 \\
                                   & \cite{Herrero2010} & 1 \\
                                   & \cite{Britavskiy2014} & 1 \\
    \hline
    \multirow{16}{*}{M33 (1388)} & \cite{Massey2016}  & 1193 \\
                          & \cite{Massey1998} & 49 \\
                          & \cite{Neugent2019} & 46  \\    
                          & \cite{Humphreys2017} & 24 \\
                          & \cite{Massey2007} & 13 \\
                          & \cite{Gordon2016} & 12 \\
                          & \cite{Drout2012} & 11 \\
                          & \cite{Massey2019} & 10 \\
                          & \cite{Kraus2019a} & 7 \\
                          & \cite{Massey1998a} & 6 \\
                          & \cite{Kourniotis2018} & 4 \\
                          & \cite{Humphreys2014} & 4 \\
                          & \cite{Massey1996} & 3 \\
                          & \cite{Martin2017} & 2 \\
                          & \cite{Neugent2011} & 2 \\
                          & \cite{Bruhweiler2003} & 2 \\

    \hline  
    \multirow{4}{*}{Sextans A (16)} & \cite{Camacho2016}  & 9  \\
                                     & \cite{Britavskiy2015} & 5 \\
                                     & \cite{Britavskiy2014} & 1 \\
                                     & \cite{Kaufer2004} & 1 \\
    \hline
    \hline
    \end{tabular}
\end{table}

We compiled spectral types for three more galaxies, WLM, IC 1613, and Sextans A (see Table \ref{t:spectypes_refs}), to use as test cases. Among a larger collection of galaxies, these three offered the most numerous (albeit small) populations of classified massive stars: 36 sources in WLM, 20 in IC 1613, and 16 in Sextans A. Although a handful more sources could potentially be retrieved for other galaxies the effort to collect the data (individually from different works) would not match the very small increase in the sample. 

We present the first few lines of the compiled list of objects for guidance regarding its form and content in Table \ref{t:catalog_sptypes}.

\section{Application of machine learning}
\label{s:ml}

In this section we provide a short description of the algorithms chosen for this work (for more details, see, e.g., \citealt{Baron2019, Ball2010}). The development of a classifier for massive stars requires the inclusion of "difficult" cases, such as those that are short-lived (e.g., YSGs with a duration of a few thousand years; \citealt{Neugent2010, Drout2009}) or very rare (e.g., LBVs, \citealt{Weis2020}; and B[e]SGs, \citealt{Kraus2019a}). To secure the training of the algorithms with specific targets, we preferred the use of supervised algorithms. However, any algorithm needs the proper input, which is determined by the class and feature selection. Finally, we show the implementation and the optimization of the methods. 

\subsection{Selected algorithms}
\label{s:algorithms}

Support Vector Machines \citep{Cortes1995} is one of the most well-established methods used in a wide range of topics. Some indicative examples include classification problems for variable stars \citep{Pashchenko2018}, black hole spin \citep{Gonzalez2019}, molecular outflows \citep{Zhang2020}, and supernova remnants \citep{Kopsacheili2020}. The method searches for the line or the hyperplane (in two or multiple dimensions, respectively) that separates the input data (features) into distinct classes. The optimal line (hyperplane) is defined as the one that maximizes the support vectors (i.e., the distance of each point with the boundary), which leads to the optimal distinction between the classes. One manifestation of the method, designed better for  classification purposes, such as our problem, is the Support Vector Classification (SVC; \citealt{Ben-Hur2002}), which uses a kernel to better map the decision boundaries between the different classes.

Astronomers are great machines when it comes to classification processes. A well-trained individual can easily identify the most important features for a particular problem (e.g., spectroscopic lines) and, according to specific (tree-like) criteria, can make fast and accurate decisions to classify sources. 
However, their strongest drawback is low efficiency as they can only process one object at a time. Although automated decision trees can be much more efficient than humans, they tend to overfit, that is to say, they learn the data they are trained  on too well and can fail when applied to unseen data. A solution to overfitting is Random Forest (RF; \citealt{Breiman2001}), an ensemble of decision trees, each one trained on a random subset of the initial features and sample of sources. Some example works include \cite{Jayasinghe2018} and \cite{Pashchenko2018} for variable stars, \cite{Arnason2020} to identify new X-ray sources in M31, \cite{Moller2016} on supernovae Type Ia classification, \cite{Plewa2018} and \cite{Kyritsis2022} for stellar classification. When RF is called to action, the input features of an unlabeled object propagate through each decision tree and provide a predicted label. The final classification is the result of a majority vote among all labels predicted by independent trees. Therefore, RF overcomes the problems of single decision trees as they generalize very well and can handle large numbers of features and data efficiently.
 
Neural networks originate from the idea of simulating the biological neural networks in animal brains \citep{McCulloch1943}. The nodes (that are located in layers) are connected and process an input signal according to their weight, which was assigned to them during the training process. Initial applications in astronomy were first performed in the 1990s (e.g., \citealt{Odewahn1992} on star and galaxy discrimination, \citealt{Storrie-Lombardi1992} on galactic morphology classification) but recent advance in computational power as well as in software development, allowing easy implementation, have revolutionized the field. Deeper and more complex neural network architectures have been developed, such as using deep convolutional networks to classify stellar spectra \citep{Sharma2020} and supernovae along with their host galaxies \citep{Muthukrishna2019}, generative adversarial networks to separate stars from quasars \citep{Makhija2019}, and recurrent neural networks for variable star classification \citep{Naul2018}. For the current project, a relatively simple shallow network with a few fully connected layers -- Multilayer Perceptron (MLP) -- proved sufficient.

 
In summary, the aforementioned techniques are based on different concepts; for example, SVC tries to find the best hyperplane that separates the classes, RF decides the classification result based on the thresholds set at each node (for multiple trees), while neural networks attempt to highlight the differences in the features that best separate the classes. We implemented an ensemble meta-algorithm that combines the results from all three, different, approaches. Initially each  method provides a classification result with a probability distribution across all selected classes (see Sect. \ref{s:class_selection}). Then these are further combined to obtain the final classification (described in detail in Sect. \ref{s:combining_models}).

\subsection{Class selection} 
\label{s:class_selection}

\begin{table}
    \centering
    \caption{Groups of spectral types of our initial sample (Col. 1) and their corresponding number of sources (Col. 2). Combining them into classes (Col.s 3) leads to the total number of sources combined per class (Col. 4; see Sect. \ref{s:class_selection}) and the final numbers (Col. 5) after removing 44 objects without full photometry in all bands.}
    \label{t:spectral_classes}
    \begin{tabular}{lc|ccc }
    \hline
    \hline 
    Group & initial \# & Class & class \# & final \# w/phot\\
     $[1]$ & $[2]$ & $[3]$ & $[4]$ & $[5]$\\    
    \hline 
    O   & 17    & BSG   &  \multirow{12}{*}{261} & \multirow{12}{*}{250} \\ 
    Oc  & 1         & -     &    &  \\ 
    Oe  & 2         & BSG   &    &  \\ 
    On  & 6         & BSG   &    &  \\ 
    B   & 156   & BSG   &    &  \\ 
    Bc  & 11    & -         &    &  \\ 
    Be  & 7         & BSG       &    &  \\ 
    Bn  & 18    & BSG   &    &  \\
    A   & 51    & BSG   &    &  \\
    Ac  & 3         & -     &    &  \\
    Ae  & 2         & BSG       &    &  \\ 
    An  & 2         & BSG       &    &  \\
    \hline
    WR  & 50    & WR &  \multirow{3}{*}{53} & \multirow{3}{*}{42} \\ 
    WRc & 3         & -  &    &  \\
    WRn & 3         & WR &    &  \\
    \hline
    LBV  & 6    & LBV   &  \multirow{2}{*}{6} & \multirow{2}{*}{6} \\ 
    LBVc & 18   & -         &    &  \\ 
    \hline
    BeBR        & 6         & BeBR       &  \multirow{2}{*}{17} & \multirow{2}{*}{16} \\ 
    BeBRc       & 11    & BeBR   &    &  \\ 
    \hline
    F   & 21     & YSG   &  \multirow{5}{*}{103} & \multirow{5}{*}{99} \\
    Fc  & 4          & -         &    &  \\ 
    G   & 15     & YSG   &    &  \\ 
    YSG & 67     & YSG   &    &  \\ 
    YSGc & 16    & -     &    &  \\ 
    \hline
    K   & 67     & RSG   & \multirow{7}{*}{512} & \multirow{7}{*}{496} \\
    Kc  & 3          & -         &    &  \\ 
    M   & 142    & RSG   &    &  \\ 
    Mc  & 5          & -         &    &  \\ 
    RSG & 250    & RSG   &    &  \\ 
    RSGb & 53    & RSG   &    &  \\ 
    RSGc & 36    & -     &    &  \\ 
    \hline
    AGN & 2         & GAL        & \multirow{4}{*}{24}  & \multirow{4}{*}{23} \\ 
    QSO & 17    & GAL    &    &  \\ 
    QSOc & 1    & -          &    &  \\ 
    GAL  & 5    & GAL    &    &  \\ 
    \hline
    Total  & 1077 &      & 976 &  932 \\
    
    \hline
    \hline    
    \end{tabular}
\end{table}
 
 When using supervised machine-learning algorithms it is necessary to properly select the output classes. In our case we are particularly interested in evolved massive stars, because the magnitude-limited observations of our target galaxies mainly probe the upper part of the HRD. In our compiled catalog we had a large range of spectral types, from detailed ones (such as M2.5I, F5Ia, and B1.5I) up to more generic terms (such as RSGs and YSGs). Given the small numbers per individual spectral type, as well as the continuous nature of spectral classification, which makes the separation of neighboring types difficult, we lack the ability to build a classifier sensitive to each individual spectral type. To address that we combined spectral types in broader classes, without taking into account luminosity classes (i.e., main-sequence stars and supergiants for the same spectral type were assigned to the same group). This is a two-step process as we first assigned all types to certain groups, and then, during the application of the classifier, we experimented with which classes are best detectable with our approach (given the lack of strict boundaries between these massive stars, which is a physical limitation and not a selection bias). For the first step, we grouped the 1089 sources (both in M31 and M33) as follows.

First, sources of detailed subtypes were grouped by their parent type (e.g., B2 I and B1.5 Ia to the B group; A5 I and A7 I to the A group; M2.5 I and M2-2.5 I to the M group, etc.). Some individual cases with uncertain spectral type were assigned as follows: three K5-M0 I sources to the K group, one mid-late O to the O group, one F8-G0 I to the F group, and one A9I/F0I to the A group.

Second, all sources with emission or nebular lines were assigned to the parent type group with an "e" or "n" indicator (e.g., B8 Ie to the Be group, G4 Ie to the Ge group, B1 I+Neb to the Bn group, and O3-6.5 V+Neb to the On group).

Third, sources with an initial classification as RSGs or YSGs were assigned directly to their corresponding group.

Fourth, RSG binaries with a B companion \citep{Neugent2019} were assigned to the RSGb group.

Fifth, secure LBVs and B[e]s were kept as separate groups (as LBVs and BeBRs, respectively). A source classified as HotLBV was assigned to the LBV group.  
   
Sixth, all sources classified as WRs (of all subtypes), including some individual cases (WC6+B0 I, WN4.5+O6-9, WN3+abs, WNE+B3 I, WN4.5+O, and five Ofpe/WN9), were grouped under one group (WR), except three sources that are characterized by nebular lines and were assigned to the WRn group. 

Seventh, galaxies (GALs), active galactic nuclei (AGNs) and quasi-stellar objects (QSOs) were grouped under their corresponding groups.

Eighth, all sources with an uncertainty flag (":" or "c") were assigned to their broader group followed by a "c" flag to indicate that these are candidates (i.e., not secure) classifications, such as Ac, Bc, YSGc, WRc, and QSOc. One source classified as B8Ipec/cLBV was assigned to the LBVc group. 

Finally, complex or very vague cases were disregarded. This entailed eight "HotSupergiant" sources and one source from each of the following types: "WarmSG," "LBV/Ofpe/WN9," "Non-WR(AI)," and "FeIIEm.Line(sgB[e])." 

Thus, after removing the 12 sources from the last step we are left with 1077, split into 35  groups (see Table \ref{t:spectral_classes}, Col. 1 and their corresponding numbers in Col. 2). However, these groups may contain similar objects, or in many cases a limited number of sources that may not be securely classified. To optimize our approach we experimented extensively by combining (similar) groups to broader classes to obtain the best results. 

All hot stars (i.e., O,B,A groups, including sources with emission "e" and nebular "n" lines) were combined under the BSG class after removing the uncertain sources (indicated as candidates). For the YSG class we considered all sources from the F, G, and YSG groups, again excluding only the candidates (i.e., members of the Fc and YSGc groups, especially as many of the YSGc are highly uncertain; \citealt{Massey2016}). For the RSG class we combined the K, M, RSG, and RSGb groups, excluding the candidates (i.e., Kc, Mc, and RSGc). The BeBR class includes both the secure and the candidate sources, because they show the same behavior (see Sect. \ref{s:feature_selection}) and there are more constraints to characterize a source as B[e] (see \citealt{Kraus2019a}).
More specifically the BeBRc sources were actually the result of constraining further the classification of candidate LBVs \citep{Kraus2019a}. Therefore, we kept only the secure LBVs (LBV group) to form their corresponding class. For the WR class we used all available sources, although they are of different types, as a further division would not be efficient. The last class, GAL, includes all nonstellar background objects (galaxies, AGNs, QSOs, except for the one candidate QSO) that were used as potential outliers. We do not expect any other type of outlier (but for an $\sim8\%$ foreground contamination) since at the distances of our target galaxies we are actually probing the brighter parts of the HRD where the supergiant stars are located. The number of sources finally selected for  each class is shown in Table \ref{t:spectral_classes} (Col. 4), where we used the class name to indicate which groups contribute to the class (Col. 3) while a "-" shows that a particular group is ignored. The total number of selected sources is 976.

\subsection{Imbalance treatment}
\label{s:imbalance_treatment}

What is evident from Table \ref{t:spectral_classes} is that we have an imbalanced sample of classes, which is very typical in astronomical applications (see also \citealt{Dorn-Wallenstein2021} for similar problems). In particular, the RSG class is the most populated one (with $\sim500$ spurces), followed by the BSG class (with $\sim250$ sources), accounting for almost 80\% of the total sample. The YSG class includes about a hundred sources, but the WR, GAL, BeBR, and, most importantly, LBV classes include a few tens at most. To tackle this we can either use penalizing metrics of performance (i.e., evaluations in which the algorithm provides different weights to specific classes) or train the model using adjusted sample numbers (by over-sampling the least populated classes, and simultaneously  under-sampling the most populated one). We experimented with both approaches and found a small gain when using the resampling approach. 

A typical approach to oversampling is duplicating objects. Although this may be a solution in many cases, it does not help with sampling the feature space better (i.e., it does not provide more information). An alternative approach is to create synthetic data. To this purpose, in this work we used a commonly adopted algorithm, the Synthetic Minority Oversampling TEchnique (SMOTE; \citealt{SMOTE}), which generates more data objects by following these steps: (i) it randomly selects a point (A) that corresponds to a minority class, (ii) it finds k-nearest neighbors (of the same class), (iii) it randomly chooses one of them (B), and (iv) it creates a synthetic point randomly along the line that connects A and B in the feature space. The benefits of this approach are  that the feature space is better sampled and all features are taken into account to synthesize the new data points. On the other hand, it is limited in how representative the initial sample per class is of each class's feature space. In any case, the number of points to be added is arbitrary and can very well match the majority class. 

At the same time this procedure can create noise, especially when trying to oversample classes with very few sources (e.g., LBVs with only six sources available in total). Better results are obtained when the oversampling of the minority classes is combined with undersampling the majority class. For the latter we experimented with two similar approaches: the Tomek links \cite{Tomek} and the Edited Nearest Neighbors (ENN; \citealt{ENN}). In the first one, the method identifies the pairs of points that are closest to each other (in the feature space) and belong to different classes (the Tomek links). These are noisy instances or pairs located on the boundary between the two classes (in a multi-class problem it is the one-versus-rest scheme that is used, i.e., the minority compared to all other classes collectively referred to as the majority class). By removing the point corresponding to the majority class the class separation increases, and the number of majority class points are reduced. In the ENN approach the three-nearest neighbors to a minority point are found and removed when belonging to the majority class. Thus, the ENN approach is a bit more aggressive than Tomek links, as it removes more points. 

In conclusion, the combination of SMOTE, which creates synthetic points from the minority class to balance the majority class, and the undersampling technique (either Tomek links or ENN), which cleans irrelevant points in the boundary of the classes, help to increase the separation. For the implementation we used the \texttt{imbalanced-learn} package\footnote{\url{https://github.com/scikit-learn-contrib/imbalanced-learn}} \citep{Lematre2017} and more specifically the ENN approach \texttt{imblearn.combine.SMOTEENN()}, which provided slightly better results from Tomek links. We used \texttt{k\_neighbors=3} for SMOTE (due to the small number of LBVs). We opted to use the default values for \texttt{sampling\_strategy}, which corresponds to ``not majority'' for SMOTE (which means that all classes are resampled except for RSGs) and ``all'' for ENN function, which cleans the majority points (considering one-versus-rest classes). In Table \ref{t:resampling} we provide an example of the numbers and fractions of sources per class available before and after resampling (the whole sample).

\begin{table}
    \caption{Number and fraction of sources per class  before and after resampling to treat for imbalance (using the SMOTE ENN approach). The fractions correspond to the total number of sources used in the original and resampled sets, respectively.}
    \label{t:resampling}
    \begin{tabular}{c|cccc}
    \hline
    \hline
    Class &  \multicolumn{2}{c}{Original sources} & \multicolumn{2}{c}{Resampled sources} \\
          &  (\#)  & (\%)  &  (\#) & (\%) \\
    \hline
BSG     &  250  &  26.8 &   496  &  14.9    \\
YSG     &  99   &  10.6 &   488  &  14.6    \\     
RSG     &  496  &  53.2 &   493  &  14.8    \\
BeBR    &  16   &  1.7  &   495  &  14.9    \\
LBV     &  6    &  0.6  &   444  &  13.3    \\
WR      &  42   &  4.5  &   453  &  13.6    \\
GAL     &  23   &  2.4  &   452  &  13.6    \\
    \hline
    \hline
    \end{tabular}
\end{table}

\subsection{Feature selection}
\label{s:feature_selection}

\begin{figure*}
    \centering
        \includegraphics[width=0.8\textwidth]{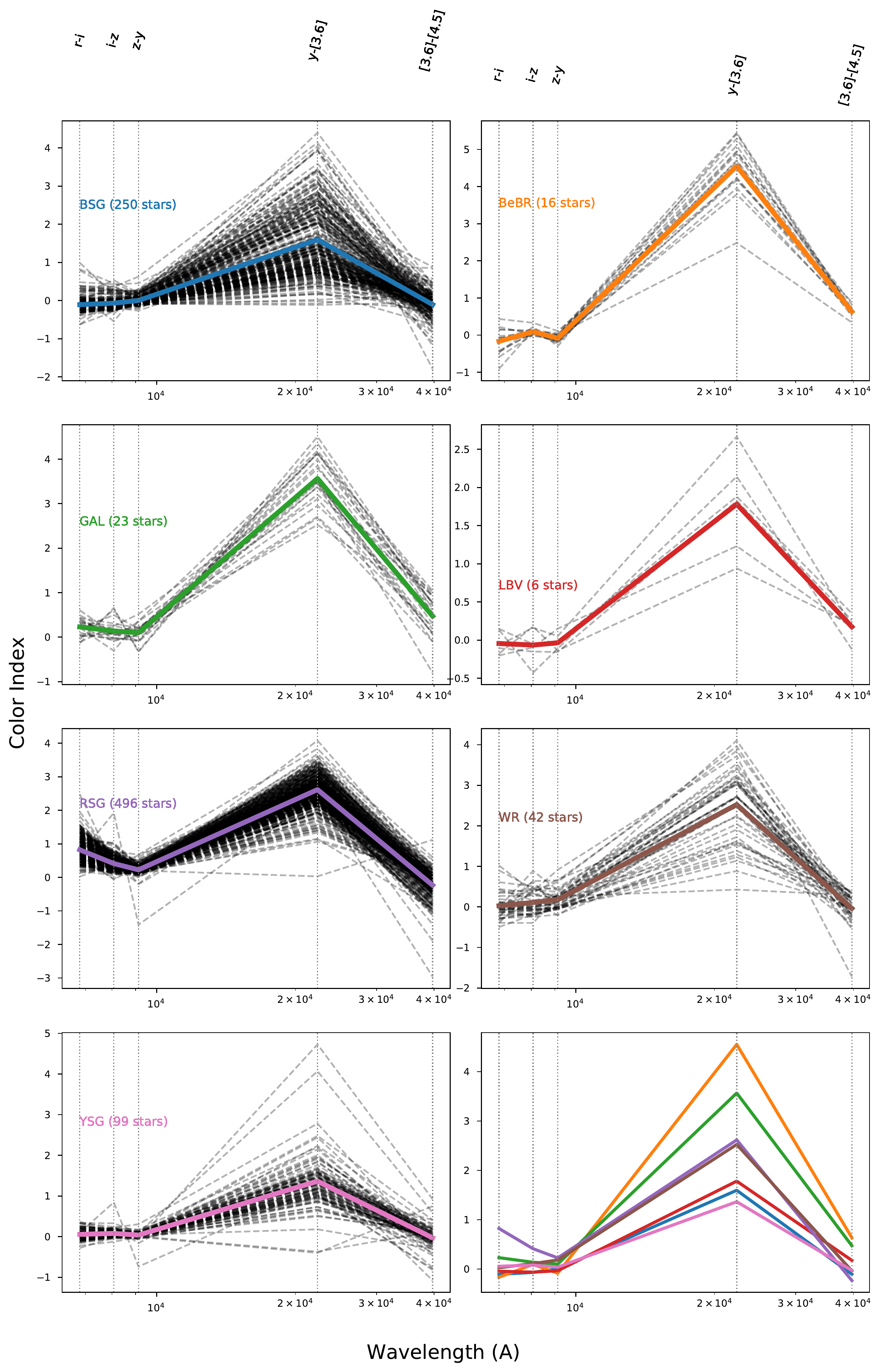}
        
        \caption{Color indices (features) vs. wavelength per class. The dashed black lines correspond to the individual sources, and the solid colored lines corresponds to their average. The last panel contains only the averaged lines to highlight the differences between the classes, with the most pronounced differences in the $y-[3.6]$ index (as BeBRs are the brightest IR sources, on average, followed by the GAL, RSG, and WR classes; see text for more). The number of sources in each panel corresponds to the total number of selected sources (see Table \ref{t:spectral_classes}, Col. 5). The vertical dashed lines correspond to the average wavelength per color index, as shown at the top of the figure.} 
    \label{f:SEDs-all}
\end{figure*}

\begin{table*}
     \caption{Data availability per class and photometric band. The first column lists the classes used and the second one the corresponding number of sources in the sample. For each class, the subsequent columns provide the fractions of sources with secure measurements in the corresponding photometric bands and their errors (which do not include objects with problematic measurements and upper limits).}
     \label{t:sample_phot_completness}

\begin{tabular}{lccccccccccc}
\hline
Class &  Sources        & [3.6] & $\sigma_\textrm{[3.6]}$ & [4.5] & $\sigma_\textrm{[4.5]}$ & [5.8] & $\sigma_\textrm{[5.8]}$ & [8.0] & $\sigma_\textrm{[8.0]}$ & [24] & $\sigma_\textrm{[24]}$  \\
  & (\#) & (\%)  & (\%) & (\%) & (\%)  & (\%) & (\%) & (\%) & (\%) & (\%) & (\%)  \\
\hline
BSG     & 261  & 100    & 100   & 100   & 100   & 100   & 80    & 100   & 70      & 100   & 41    \\
YSG     & 103  & 100    & 100   & 100   & 100   & 100   & 92    & 100   & 78      & 99    & 30    \\
RSG     & 512  & 100    & 100   & 100   & 100   & 99    & 99    & 100   & 93      & 99    & 41    \\
BeBR & 17  & 100    & 100       & 100   & 100   & 100   & 100   & 100   & 100     & 100   & 94    \\
LBV     & 6        & 100        & 100   & 100   & 100   & 100   & 100   & 100     & 100   & 100   & 66    \\
WR      & 53   & 100    & 100   & 100   & 100   & 100   & 94    & 100   & 86      & 100   & 43    \\
GAL     & 24   & 100    & 100   & 100   & 100   & 100   & 100   & 100   & 100     & 100   & 100   \\
\hline
 \end{tabular}

\vspace{2mm}

\begin{tabular}{lccccccccccc}
 \hline
 Class & Sources & $g$ & $\sigma_g$ & $r$ & $\sigma_r$ & $i$ & $\sigma_i$ & $z$ & $\sigma_z$ & $y$  & $\sigma_y$ \\
  & (\#) & (\%)  & (\%) & (\%) & (\%)  & (\%) & (\%) & (\%) & (\%) & (\%) & (\%)  \\
\hline
BSG  &  261  &  96  &  96  &  96  &  96  &  96  &  96  &  96  &  96  &  96  &  96   \\
YSG     & 103    &  96  &  96  &  96  &  96  &  96  &  96  &  96  &  96  &  96  &  96   \\
RSG     & 512    &  96  &  93  &  97  &  97  &  97  &  97  &  97  &  97  &  97  &  97   \\ 
BeBR & 17    &  100  &  100  &  100  &  100  &  100  &  100  &  100  &  100  &  94  &  94  \\ 
LBV     & 6          &  100  &  100  &  100  &  100  &  100  &  100  &  100  &  100  &  100  &  100  \\
WR      & 53     &  83  &  83  &  84  &  83  &  88  &  88  &  90  &  88  &  86  &  84  \\
GAL     & 24     &  100  &  100  &  100  &  100  &  100  &  100  &  100  &  95  &  100  &  100  \\
\hline
\end{tabular}

\vspace{2mm}

\begin{tabular}{lccccc}
     \hline
Class & Sources &  $J_{\rm{UK}}$ & G & G$_{\rm{BP}}$ & G$_{\rm{RP}}$ \\
  & (\#) & (\%)  & (\%) & (\%) & (\%)   \\
\hline
BSG     & 261  & 81     & 90    & 87    & 87 \\
YSG     & 103  & 82     & 96    & 95    & 95 \\
RSG     & 512  & 84     & 96    & 94    & 94 \\
BeBR & 17  & 70 & 100   & 100   & 100 \\
LBV     & 6        & 66 & 83    & 66    & 66 \\
WR      & 53   & 75     & 71    & 50    & 50 \\
GAL     & 24   & 83     & 95    & 83    & 83 \\
\hline
\hline
 \end{tabular}
 \end{table*}

Feature selection is a key step in any machine-learning problem. To properly select the optimal features in our case we first examined data availability. In Table \ref{t:sample_phot_completness} we list the different classes (Col. 1) and the number of available sources per class (Col. 2). In the following columns we provide the fractions of objects with photometry in the corresponding bands and with proper errors (i.e., excluding problematic sources and upper limits), per survey queried (\textit{Spitzer}, Pan-STARRS, UKIRT Hemisphere Survey, and \textit{Gaia}). To build our training sample, we required the sources to have well-measured values across all bands. To avoid significantly decreasing the size of the training sample (by almost half in the case of the LBV and BeBR classes), we chose not to include the $J_{\rm{UK}}$. Although we used \textit{Gaia} data to derive the criteria to identify foreground stars, the number of stars with \textit{Gaia} photometry in the majority of other nearby galaxies is limited. Thus, to ensure the applicability of our approach we discarded these bands also (which are partly covered by the Pan-STARRS bands). 

The IR catalogs were built upon source detection in both the [3.6] and [4.5] images (e.g., \citealt{Khan2015}), while the measurements in the longer bands were obtained by just performing photometry in those coordinates (regardless of the presence of a source or not). However, in most cases there is a growing (with wavelength) number of sources with only upper limits in the photometry. As these do not provide secure measurements for the training of the algorithm we could not use them. If we were to take into account sources with valid measurements up to the [24], the number of sources would end up dropping by more than 50\% for some classes (see in Table \ref{t:sample_phot_completness}, e.g., the corresponding fractions of WR and YSG with secure error measurements). As this is not really an option, we decided to remove all bands that contained a significant fraction of upper limits (i.e., above [5.8]). This, rather radical,  selection is also justified by the fact that the majority of the unclassified sources (in the catalogs to which were are going to apply our method) do not have measurements in those bands. It is also interesting to point out that the majority of the disregarded sources belong to the RSG class (the most populated), which means that we do not lose any important information (for the training process). 

From the optical set of bands, we excluded $g$ for two reasons. About 130 sources fainter than 20.5 mag tend to have systematic issues with their photometry, especially red stars for which $g-r$ turns to bluer values. Also, due to the lack of known extinction laws for most galaxies, and the lack of data for the many sources, we opted not to correct for it. As $g$ is the band most affected by extinction we opted to use only the redder bands to minimize its impact \citep{Schlafly2011, Davenport2014}. Therefore, we kept $r$, $i$, $z$, and  $y$ bands and we performed the same strict screening to remove sources with upper limits. In total, we excluded 44 sources, reflecting a small fraction of the sample ($\sim4.5\%$  
- treating both M31 and M33 sources as a single catalog). We show the final number of sources per class in Col. 5 in Table \ref{t:spectral_classes} summing to 932 objects in total. 
 
To remove any distance dependence in the photometry we opted to work with color terms and obtained the consecutive magnitude differences: $r-i$, $i-z$, $z-y$, $y - [3.6]$, $[3.6] - [4.5]$. We examined different combinations of these color indices, but the difference in the accuracy with respect to the best ones found is negligible. Those combinations contained color indices with wider wavelength range that are affected more from extinction that the consecutive colors. Moreover, they tend to be systematically more correlated, resulting in poorer generalization (i.e., when applied to the test galaxies; Sect. \ref{s:other_galaxies}). Some (less pronounced) correlation still exists in the consecutive color set also, because of the use of each band into two color combinations (except for $r$ and [4.5]), and due to the stellar continuum, since the flux at each band is not totally independent from the flux measured in other bands. We also noticed that more optical colors help to better sample the optical part of the spectral energy distribution and separate more efficiently some classes (BSG and YSG in particular). The consecutive color set seems as the most intuitive selection, including well-studied colors. Moreover, it represents how the slopes of the spectral energy distribution changes with wavelength.

We also experimented with other transformations of these data, such as fluxes, normalized fluxes, and standardizing data (scaling magnitudes around their mean over their standard deviation), but we did not see any significant improvement in the final classification results. Therefore, we opted for the simplest representation of the data, which is the aforementioned color set. 
 
In Fig. \ref{f:SEDs-all} we plot the color indices with respect to their corresponding wavelengths for both individual sources for each class and their average. In the last panel, we overplot all averaged lines to display the difference among the various classes. As this representation is equivalent to the consecutive slopes of the spectral energy distributions for each class we notice that the redder sources tend to have a more pronounced $y-[3.6]$ feature, a color index that characterizes the transition form the optical to the mid-IR photometry. The BeBR class presents the highest values due to the significant amount of dust (and therefore brighter IR magnitudes), followed by the GALs due to their Polycyclic Aromatic Hydrocarbons (PAH) emission, the (intrinsically redder sources) RSGs, and the WRs (due to their complex environments).
 

 \subsection{Implementation and optimization}

An important step of every classification algorithm is to tune its hyperparameters, that is, the parameters that control the training process. After having defined these the algorithm determines the values of the parameters used for each model (e.g., weights) based on the training sample. The implementation of all three methods (SVC, RF, and MLP) was done through the \texttt{scikit-learn} v.0.23.1\footnote{\url{https://scikit-learn.org/}} \citep{sklearn}\footnote{For the MLP/neural networks we experimented extensively with TensorFlow v1.12.0 \citep{tensorflow2015} and Keras v2.2.4 API \citep{keras2015}. This allowed us to easily build and test various architectures for our networks. We used both dense (fully connected) and convolutional (CNN) layers, in which case the input data are 1D vectors of the features we are using. Given our tests we opted to use a simple dense network, which can be easily implemented also within the \texttt{scikit-learn} that helps with the overall simplification of the pipeline.}.

For the optimal selection of the hyperparameters (and their corresponding errors), we performed a stratified K-fold cross-validation  (CV; \texttt{sklearn.model\_selection.StratifiedKFold()}). With this, the whole sample is split into K subsamples or folds (five in our case), preserving the fraction representation of all classes of the initial sample into each of the folds. At each iteration one fold is used as the validation sample and the rest as training. By permuting the validation fold, the classifier is trained over the whole sample. Since we performed a resampling approach to correct for the imbalance in our initial sample (see Sect. \ref{s:imbalance_treatment}), we note that this process was performed only in the training folds, while the evaluation of the model's accuracy was done on the (unmodified) validation fold. We stress that the validation fold remained "unmodified" (i.e., it was not resampled) in order to avoid data leakage and hence overfitting. The final accuracy score is the average value, and its uncertainty corresponds to the standard deviation across all folds. 

For the SVC process we used the \texttt{sklearn.svm.SVC()} function. We opted to train this model with the following selection of hyperparameters: \texttt{probability=True} to get probabilities (instead of a single classification result), \texttt{decision\_function\_shape = 'ovo'}, which is the default option for multi-class problems, \texttt{kernel = 'linear'}, which is faster than the alternative nonlinear kernels and proved to be more efficient\footnote{The ``linear'' kernel was more efficient in recovering the LBV class systematically in contrast to the default ``rbf'' option. }, and \texttt{class\_weight='balanced'}, which gives more weight to rarer classes (even after the resampling approach, as described in Sect. \ref{s:imbalance_treatment}). We also optimized the regularization \textit{C} parameter, which represents a penalty for misclassifications (i.e., the objects falling on the "wrong" side of the separating hyperplane). For larger values a smaller margin for the hyperplane is selected so that the misclassified sources decrease and the classifier performs optimally for the training objects. This may result in poorer performance when applied to unseen data. On the opposite, smaller values of \textit{C} leads to a larger margin (i.e., a loose separation of the classes) at the cost of more misclassified objects. To optimize \textit{C} we tested the result in the accuracy by changing the value of \textit{C} from 0.01 to 200 (with a step of 0.1 in log space). We present these results in Fig. \ref{f:optimizing-SVC-C}, where the red line corresponds to the averaged values and the gray area to the $1\sigma$ error. As the parameter reaches fast to a plateau, the choice of this particular value does not affect significantly the accuracy above $\sim25$, which is the adopted value. 

For the RF classifier we used  \texttt{sklearn.ensemble.RandomForestClassifier()}. To optimize it, we explored the following hyperparameters over a range of values: \texttt{n\_estimators}, which is the number of trees in the forest (10-1000, step 50), \texttt{max\_leaf\_nodes}, which limits the number of nodes in each tree (i.e., how large it can grow, from 2-100; step 2), and \texttt{max\_depth}, which is the maximum depth of the tree (1-100, step 2), while the rest of the hyperparameters were left to their default values. We present their corresponding validation curves as obtained from five-fold CV tests (with mean values as red lines and their $1\sigma$ uncertainty as gray areas) in Fig. \ref{f:optimizing-RF-curves}. Again, we see that above certain values the accuracy reaches to a plateau. Given the relative large uncertainties and the statistical nature of this test, the selection of the best values is not absolutely strict (they provide almost identical results). We opted to use the following values: \texttt{n\_estimators=400}, \texttt{max\_leaf\_nodes=50}, \texttt{max\_depth=30}. We also set \texttt{class\_weight}="balanced", similar to SVC, in addition to the resampling approach.

For the neural networks we used  \texttt{sklearn.neural\_network.MLPClassifier()}. In this case we performed a grid search approach (\texttt{sklearn.model\_selection.GridSearchCV()}). This method allows for an exhaustive and simultaneous search over the requested parameters (with a cost in computation time). We started first by investigating the architecture of the network (e.g., number of hidden layers, number of nodes per layer) along with the three available types of methods for weight optimization (\texttt{'lbfgs},' \texttt{‘sgd,’} and \texttt{‘adam’}). We tried up to five hidden layers with up to 128 nodes per layer, using \texttt{'relu'} as the activation function (a standard selection). We present the results of this grid search in Fig. \ref{f:optimizing-NN-structures} from which we obtained systematically better results for the \texttt{'adam'} solver \citep{Kingma2014}, with the (relatively) best configuration being a shallow network with two hidden layers with 128 nodes each. Given this combination we further optimized the regularization parameter (\texttt{alpha}), the number of samples used to estimate the gradient at each epoch (\texttt{batch\_size}) and the maximum number of epochs for training (\texttt{max\_iter}), with the rest of the parameters left to their default values (with \texttt{learning\_rate\_init}=0.001). Similarly to the previous hyperparameters selections, from their validation curves (Fig. \ref{f:optimizing-NN-curves}) we selected as best values: \texttt{alpha}=0.13, \texttt{batch\_size}=128, and \texttt{max\_iter}=560. The classifier uses the Cross-Entropy loss, which allows probability estimates.

\section{Results}
\label{s:results}

We first present our results from the individual applications of the different machine-learning algorithms to the M31 and M33 galaxies. Then, we describe how we combine the three algorithms to obtain a combined result. Finally, we apply the combined classifier to the test galaxies.  

\subsection{Individual application to M31 and M33}
\label{s:m31m33runs}

\subsubsection{Overall performance}

\begin{figure*}
\centering
    SVC \hspace{5cm} RF \hspace{5cm} MLP\\
    \includegraphics[width=0.3\linewidth]{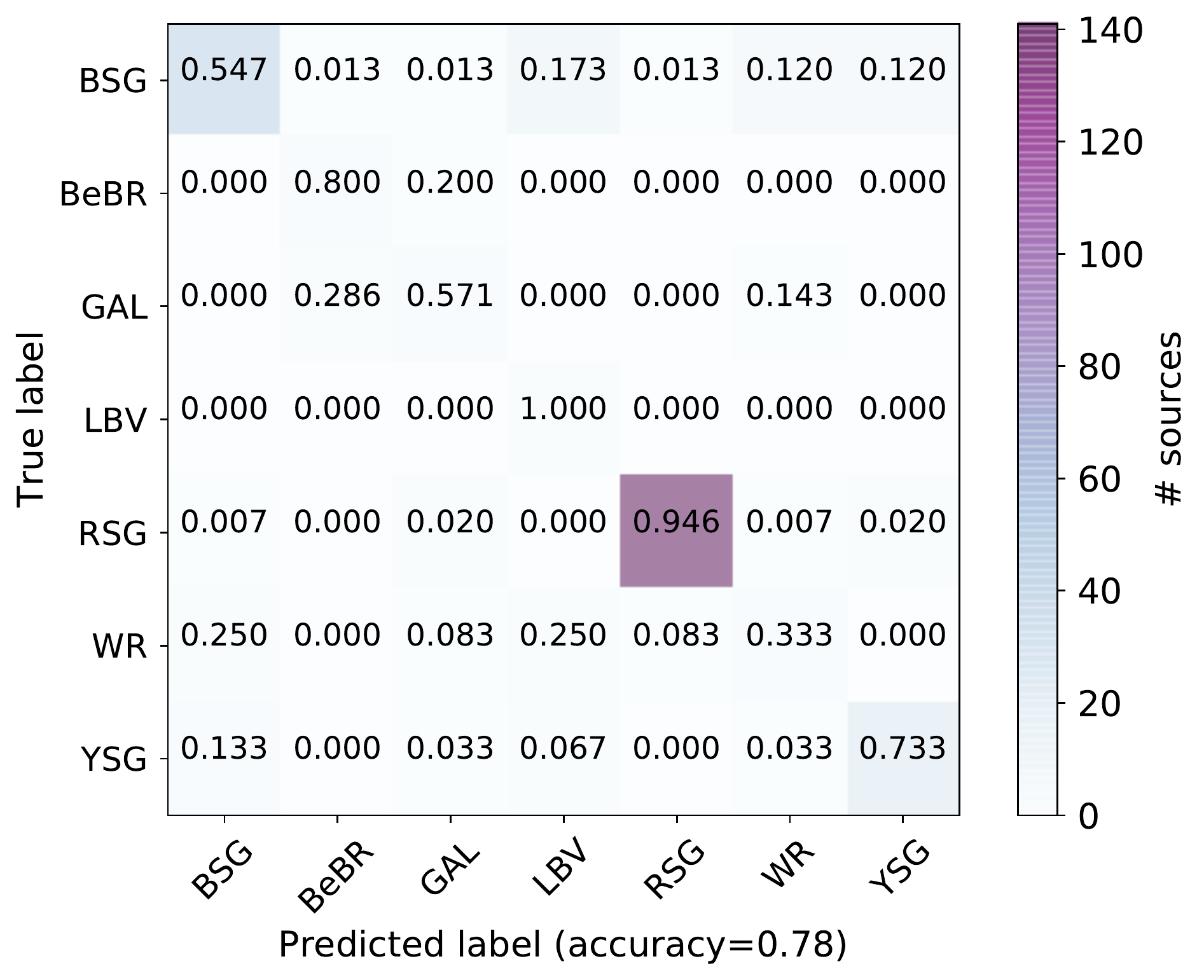}
    \includegraphics[width=0.3\linewidth]{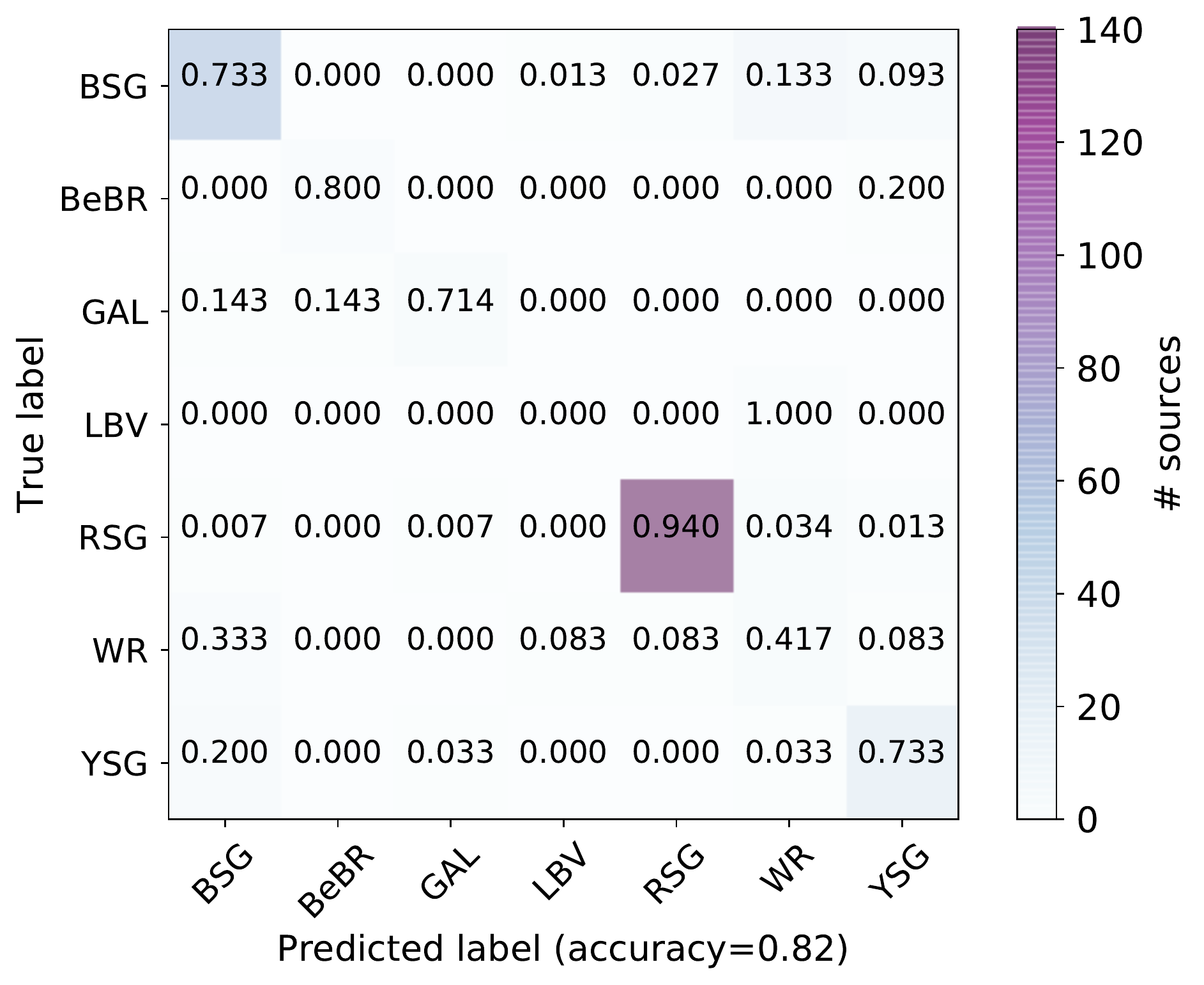} 
    \includegraphics[width=0.3\linewidth]{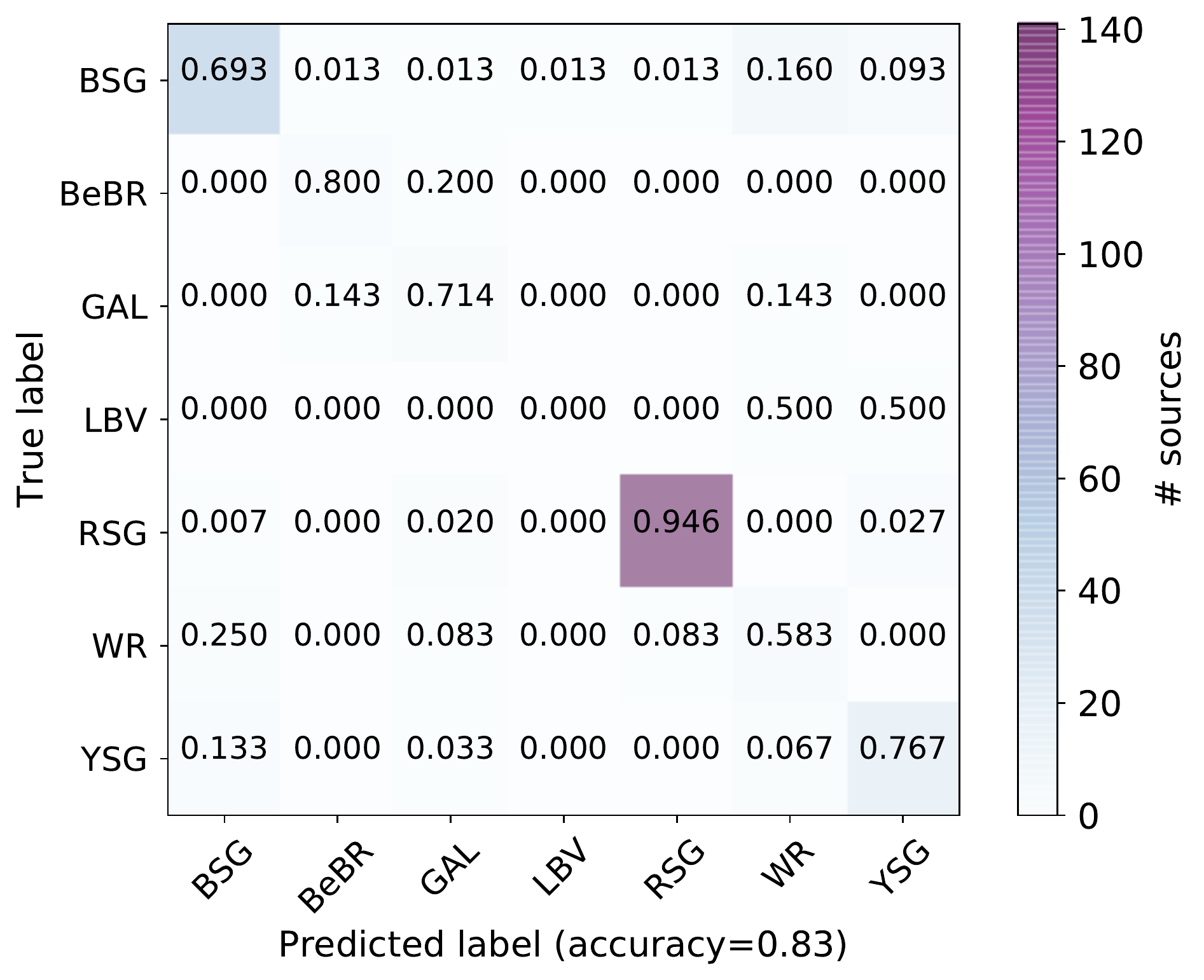} \\
    \includegraphics[width=0.3\textwidth]{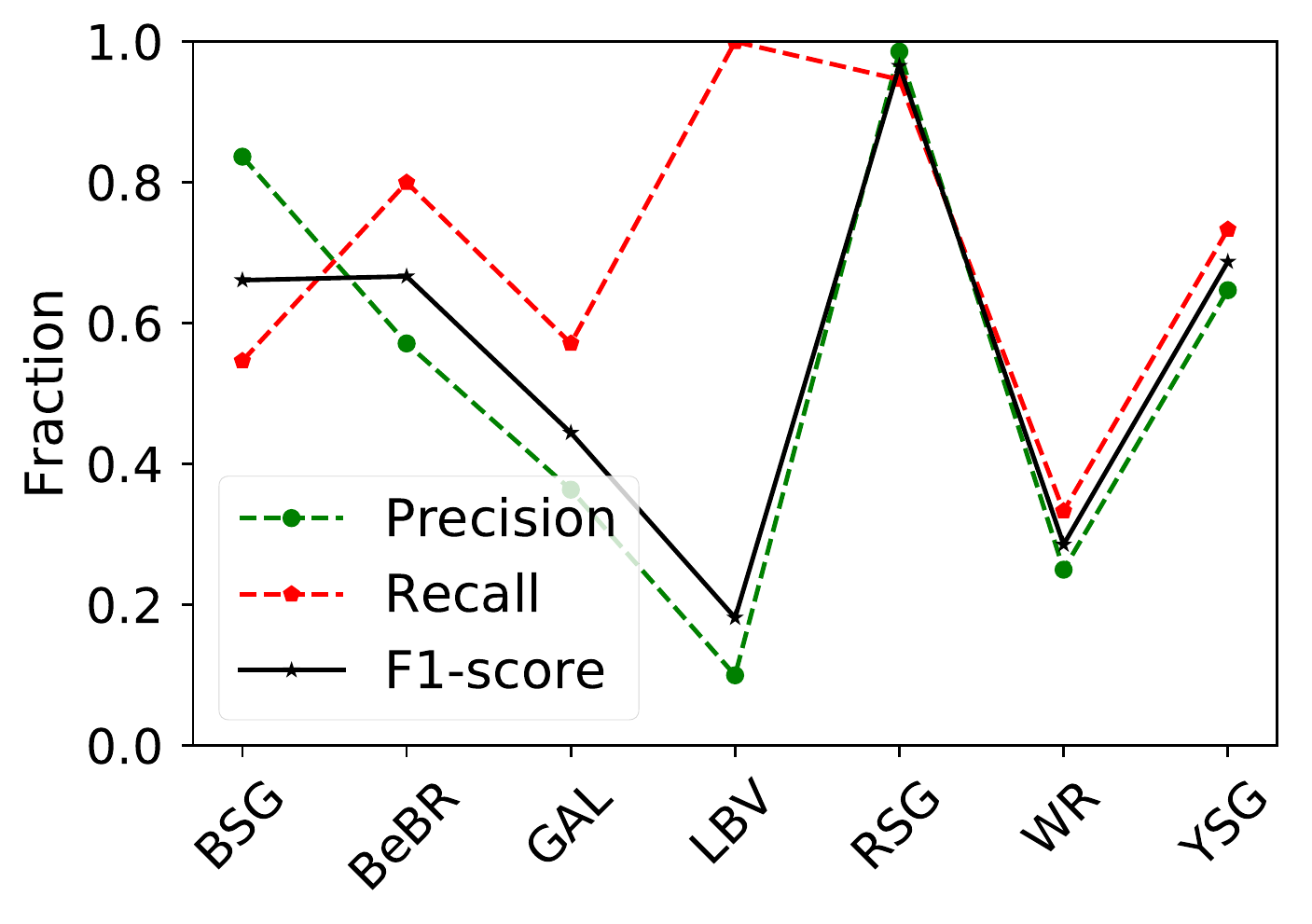}
     \includegraphics[width=0.3\textwidth]{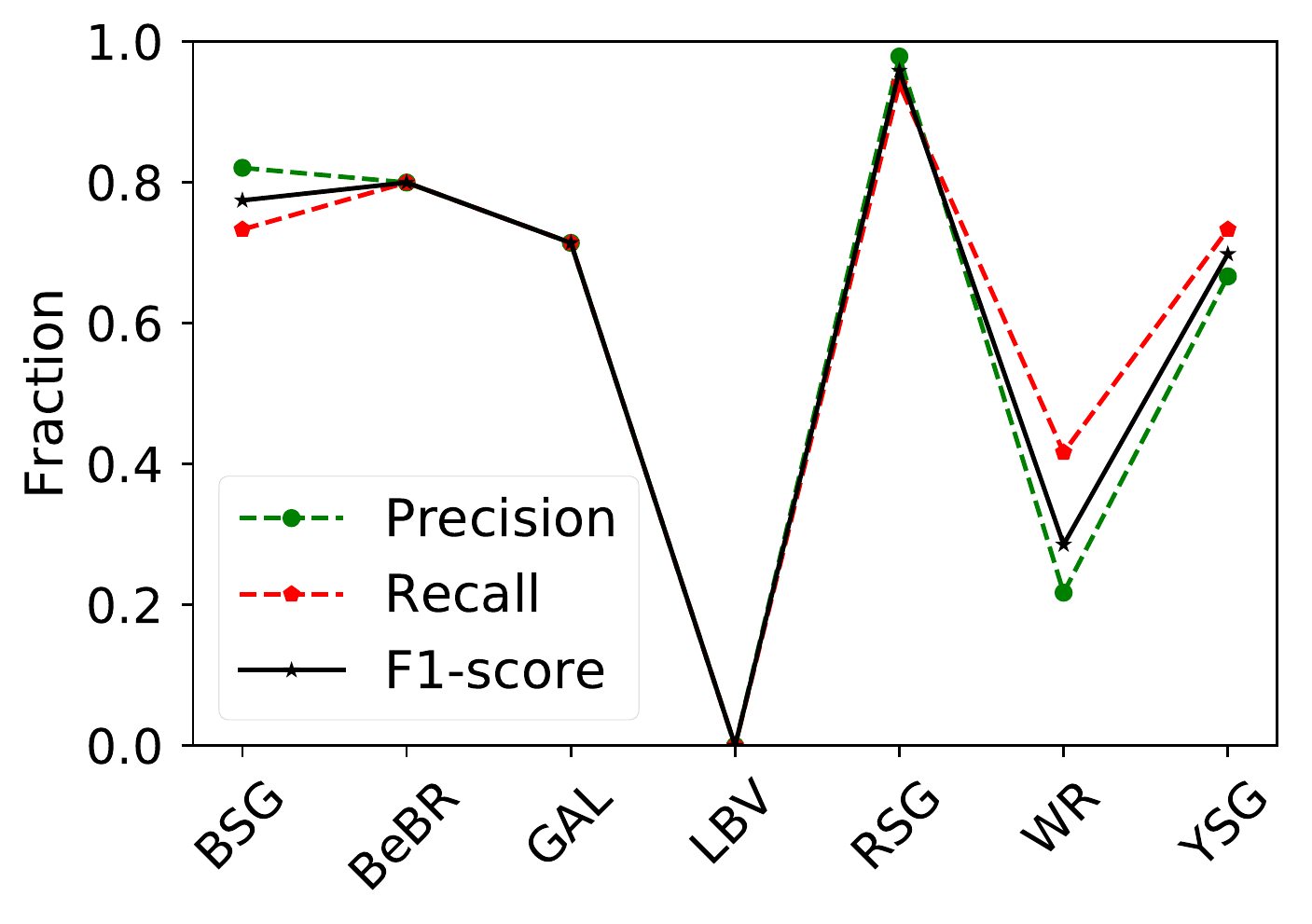}
     \includegraphics[width=0.3\textwidth]{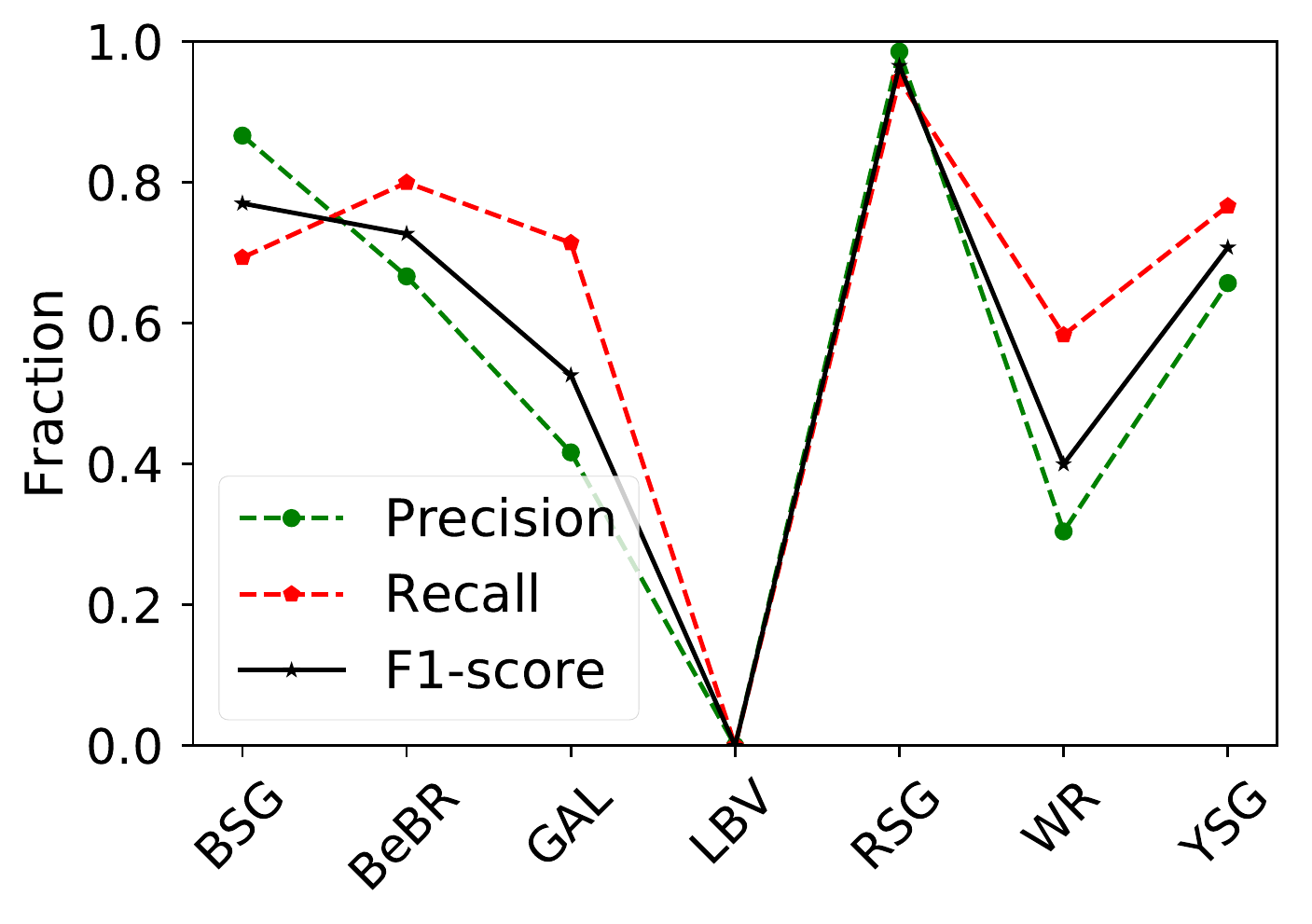}
    \caption{Confusion matrices (upper panels) for the SVC, RF, and MLP methods, respectively, along with the characteristic metrics (precision, recall, and F1 score; lower panels). These results originate from single runs, i.e., by using  70\% of the initial sample for the training sample, which is then resampled to produce a balanced sample before training each model and applying the model to the remaining 30\%\ of the sample (the validation). In general, the algorithms perform well except for the cases of LBVs and WRs (see Sect. \ref{s:m31m33runs} for more details).}
    \label{f:results_metrics}
\end{figure*}

\begin{figure*}
\centering
    SVC \hspace{5cm} RF \hspace{5cm} MLP\\
    \includegraphics[width=0.3\linewidth]{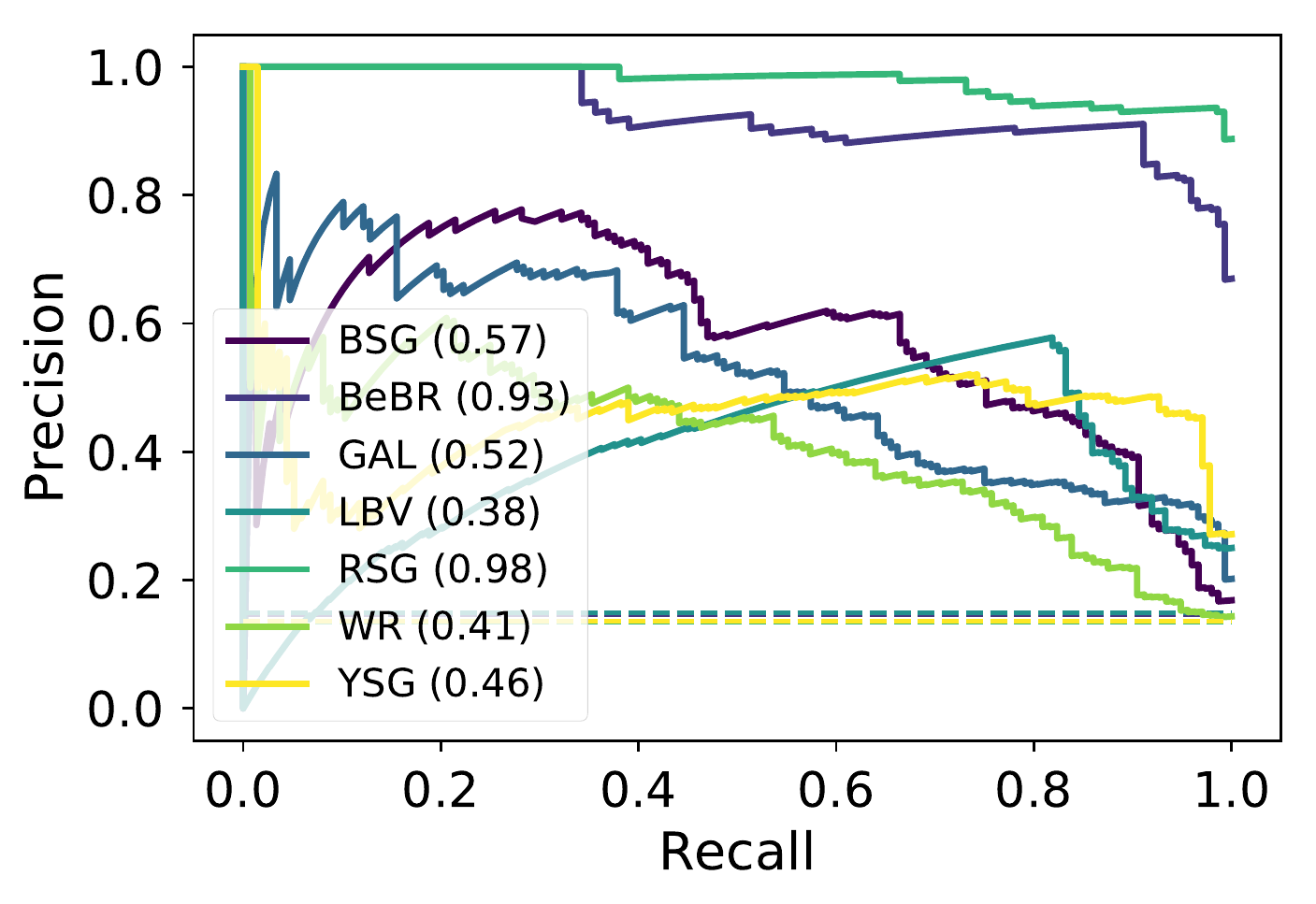}
    \includegraphics[width=0.3\linewidth]{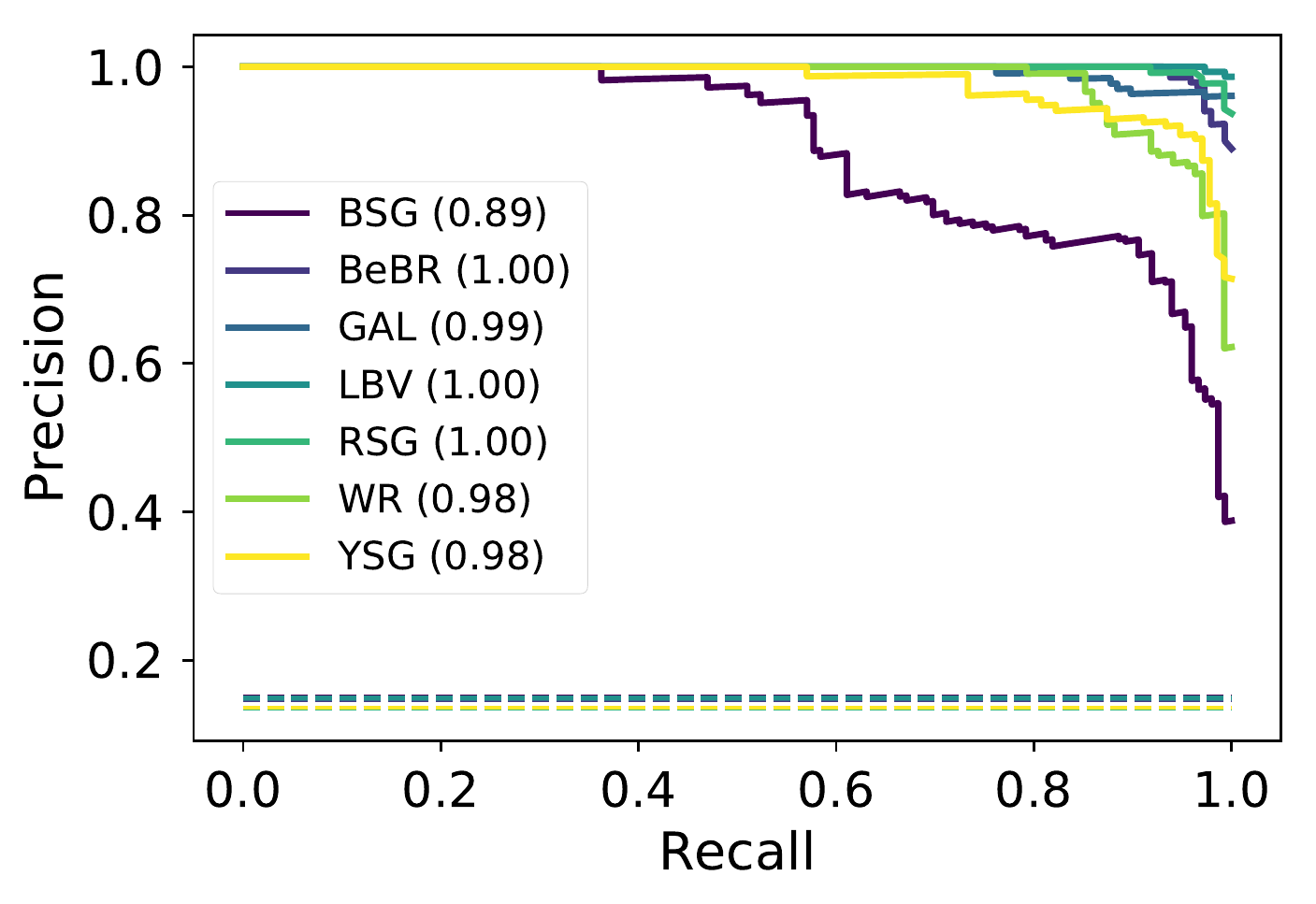} 
    \includegraphics[width=0.3\linewidth]{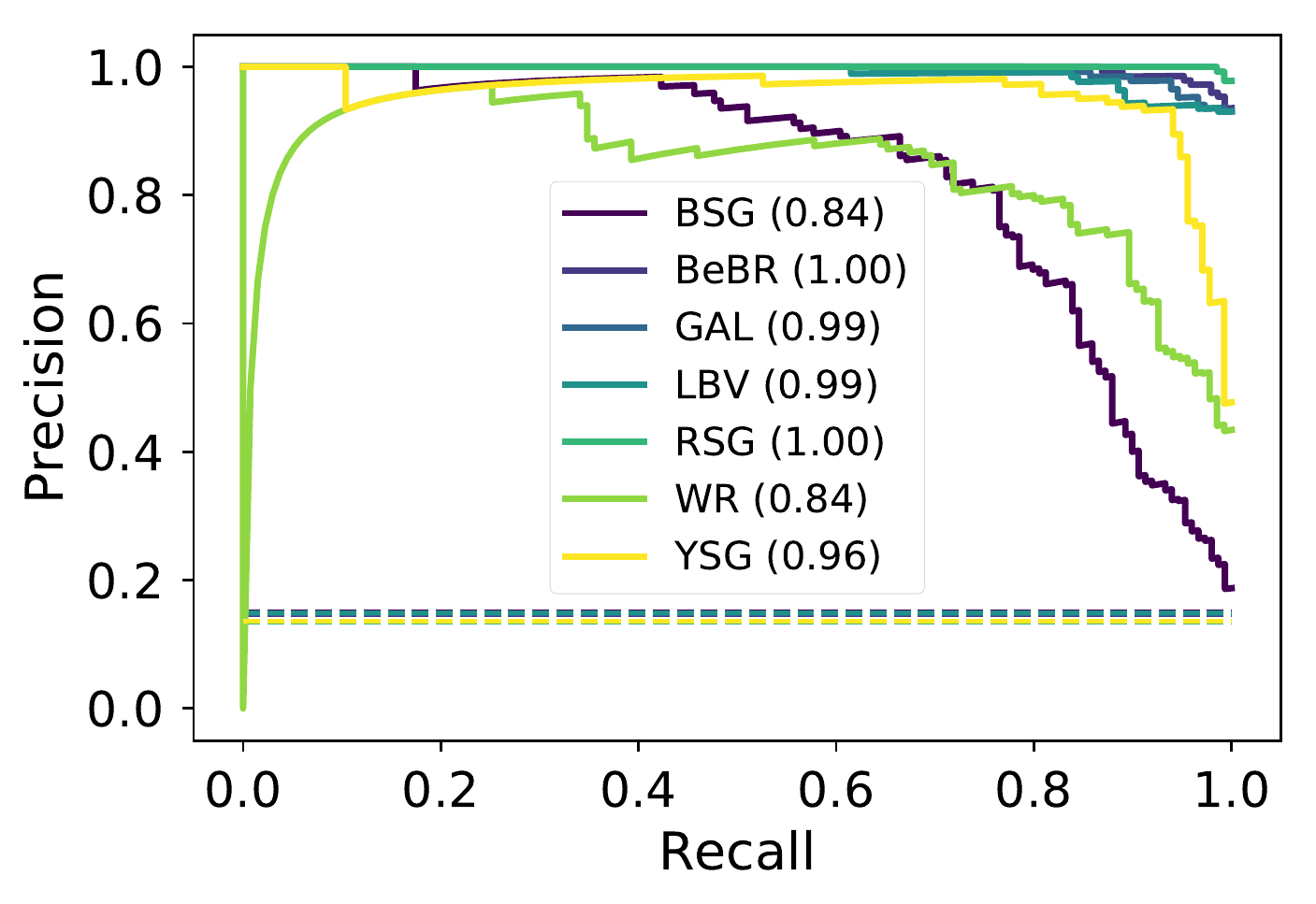}
    \caption{ Precision recall curves for the three methods, along with the values of the area under the curve for each class (in parentheses). In all cases, the results of the  comparison of each class against all others provide very good and consistent results, well above the random classifier (indicated for each class by the horizontal dashed lines; see Sect. \ref{s:m31m33runs}).} 
    \label{f:results_prs}
\end{figure*}

\begin{table*}
  \centering
  \caption{Performance for each method and per class, after a repeated K-fold cross validation (see Sect. \ref{s:m31m33runs} for details).}
  \label{t:recalls}
  \begin{tabular}{lcccc}
  \hline
  \hline
  Class & SVC & RF & MLP & combined  \\ 
  \hline
  overall & $0.78\pm0.03$ & $0.82\pm0.02$ & $0.82\pm0.02$ & $0.83\pm0.02$  \\
  \hline
  BSG & $0.58\pm0.08$ & $0.71\pm0.06$ & $0.71\pm0.07$ &  $0.71\pm0.06$\\
  BeBR & $0.80\pm0.23$ & $0.79\pm0.24$ & $0.73\pm0.25$ &  $0.81\pm0.17$\\
  GAL & $0.58\pm0.22$ & $0.63\pm0.21$ & $0.73\pm0.24$ & $0.71\pm0.17$ \\
  LBV & $0.28\pm0.43$ & $0\pm0$ & $0\pm0$ & $0\pm0$ \\  
  RSG & $0.93\pm0.03$ & $0.95\pm0.02$ & $0.94\pm0.02$ &  $0.94\pm0.02$\\
  WR & $0.43\pm0.15$ & $0.40\pm0.16$ & $0.46\pm0.19$ & $0.48\pm0.24$ \\
  YSG & $0.78\pm0.08$ & $0.75\pm0.10$ & $0.77\pm0.12$ & $0.80\pm0.08$ \\ 
  \hline
 \hline
  \end{tabular}
\end{table*}

Having selected the optimal hyperparameters for our three algorithms we investigated the individual results as obtained by directly applying them to our data set. For this we need to split the sample into a training set (70\%) and evaluate the results on a validation set (30\%), which is a standard option in the literature. The split was performed individually per class to ensure the same fractional representation of all classes in the validation sample. The resampling approach to balance our sample (as described in Sect. \ref{s:imbalance_treatment}) was applied only on the training set. The model was then trained on this balanced set and the predictions were made on the original validation set.  

Given a specific class, we refer to the objects that are correctly predicted to belong to this class as true positives (TPs), while true negatives (TNs) are those that are correctly predicted to not belong to the class. False positives (FPs) are the ones that are incorrectly predicted to belong, while false negatives (FNs) are the ones that are incorrectly predicted to not belong to the class. 

In Fig. \ref{f:results_metrics} we show example runs for the SVC, RF, and MLP methods. The first row corresponds to the confusion matrix, a table that displays the correct and incorrect classification results per class. Ideally this should be a diagonal table. The presence of sources in other elements provides information about the contamination of classes (or how much the method is miss-classifying the particular class). Another representation is given in the second row where we plot the scores of the (typically used) metrics for each class. Precision (defined as $\rm TP/(TP+FP)$) refers to the number of objects that are predicted correctly to belong to a particular class over the total number of identified  objects for this class (easily derived if we look at the numbers across the columns of the confusion matrix). Recall (defined as $\rm TP/(TP+FN)$) is the number of class objects over the total real population for this class (derived from the rows of the confusion matrix). Therefore, the precision indicates the ability of the method to detect an object of the particular class while recall its ability to recover the real population. The F1 score is the harmonic mean of the two previous metrics (defined as $\rm F1 score = 2 \times ( precision \times recall) / (precision + recall)$). In our case, we are mainly using the recall metric as we are interested to minimize the contamination and therefore to recover as many correct objects as possible. This is especially required for the classes with the smallest numbers, which reflect the rarity of their objects, such as the BeBR and LBV classes. We report our results using the weighted balance accuracy (henceforth, "accuracy"), which corresponds to the average of recall values across all classes, weighted by the number of objects per class. This is a reliable metric of the overall performance when training over a wide number of classes \citep{Grandini2020}.   

From Fig. \ref{f:results_metrics} we see that the accuracy achieved for SVC, RF, and MLP is $\sim78\%$, $\sim82\%$, and $\sim83\%$, respectively. These values are based on a single application of the algorithms, that is the evaluation of the models on the validation set (the 30\% of the whole sample). However, we left out an important fraction of information, which due to our small sample, is important. Even though we up-sampled to account for the scarcely populated classes, this happened (at each iteration) solely for those sources of the training sample, which implies that -- again -- only a part of the whole data set's feature space was actually explored. To compensate for that, the final model was actually obtained by training over the whole sample (after resampling). In this case there was no validation set to perform directly the evaluation. To address that we used a repeated K-fold CV to obtain the mean accuracy and the recall per class, which in turn provided the overall expected accuracy.  Using five iterations (and five folds per iteration) we obtained $78\pm3\%$,  $82\pm2\%$, and $82\pm2\%$ for SVC, RF, and MLP, respectively (the error is the standard deviation of the average values over all K-folds performed). In Table \ref{t:recalls} we show the accuracy (``overall''), and the recall as obtained per class. 

\cite{Dorn-Wallenstein2021}, using the SVC method and a larger set of features (12) including variability indices, achieved slightly better accuracy ($\sim90.5\%$) but for a coarser classification of their sources (i.e., for only four classes:  ``hot,'' ``emisison,`` ``cool,'' and ``contamination'' stars). When they used their finer class grid with 12 classes, their result was $\sim54\%$\footnote{The balanced accuracy reported by \cite{Dorn-Wallenstein2021} is the average recall across all classes, i.e., without weighting by the frequency for each class. This metric is insensitive to class distribution \citep{Grandini2020}. We converted the reported values to the weighted balanced accuracy to directly compare our results.}.

\subsubsection{Class recovery rates}

The results per class are similar for all three methods. They can recover the majority of the classes efficiently, with the most prominent class being the RSGs with $\sim95\%$ success (similar to \citealt{Dorn-Wallenstein2021}). Decent results are returned for the BSG, YSG, and GAL classes, within a range of $\sim60-80\%$.

The class for which we obtained the poorest results is the LBVs. The SVC is the most effective in recovering a fraction of the LBVs ($\sim30\%$, albeit with a large error of 43\%), while the other two methods failed. The LBV class is an evolutionary phase of main-sequence massive O-type stars before they lose their upper atmosphere (due to strong winds and intense mass-loss episodes) and end up as WRs. They tend to be variable both photometrically and spectroscopically, displaying spectral types from B to G. Hence,  physical confusion between WRs, LBVs, and BSGs is expected, as indicated by the lower recall values and the confusion matrices (see Fig. \ref{f:results_metrics}). Moreover, the rarity of these objects leads to a small-populated class for which their features are not well determined, and consequently, the classifier has significant issues distinguishing them from other classes. On the other hand, SVC examines the entire feature space, which is the reason for the (slightly) improved recall for LBVs in this case (\citealt{Dorn-Wallenstein2021} report full recovery but probably because of overfitting). Due to the small number and the rather inhomogeneous sample of WRs, all the classifiers have difficulties to correctly recover the majority of these sources. The best result is provided by MLP at $\sim46\%$, less than the $\sim75\%$ reported by \citealt{Dorn-Wallenstein2021}. Despite the small sample size of the BeBR class, it is actually recovered successfully ($>79\%$). As BeBRs (including candidate sources) form a more homogeneous sample than LBVs and WRs, their features are well characterized, which helps the algorithms separate them.

To better visualize the performance of these methods we constructed the Precision Recall curves, which are better suited in the case of imbalanced data \citep{Davis2006, Saito2015}. During this process, the classifier works in a one-versus-rest mode; that is to say, it only checks whether the objects belong to the examined class or not. In Fig. \ref{f:results_prs} we show the curves for each algorithm. The dashed (horizontal) lines correspond to the ratio of positive objects (per class) over the total number of objects in the training data. Any model found at this line or below has no ability to predict anything better than random (or worse). Therefore, the optimal curve directs toward the upper right corner of the plot (with precision=recall=1). In all cases the classifiers are better than random. RF displays systematically the best curves. In SVC, RSGs and BeBRs are almost excellent, and the rest of the classes display similar behavior. For MLP, all classes except BSGs and WRs are very close to the optimal position.

Another metric is obtained if we measure the fraction of the area under the curve. This returns a single value (within the 0-1 range) depicting the ability of the classifier to distinguish the corresponding class over all the rest. In Fig.  \ref{f:results_prs} we show these values within the legends. In general, we achieve high values, which means that our classifiers can efficiently distinguish the members of a class against all others. These consistent results add further support that the careful selection of our sample has worked and that the methods work efficiently (given the class limitations).

\subsection{Ensemble models}
\label{s:combining_models}

\subsubsection{Approaches}

\begin{figure*}
\centering
    \includegraphics[width=0.4\linewidth]{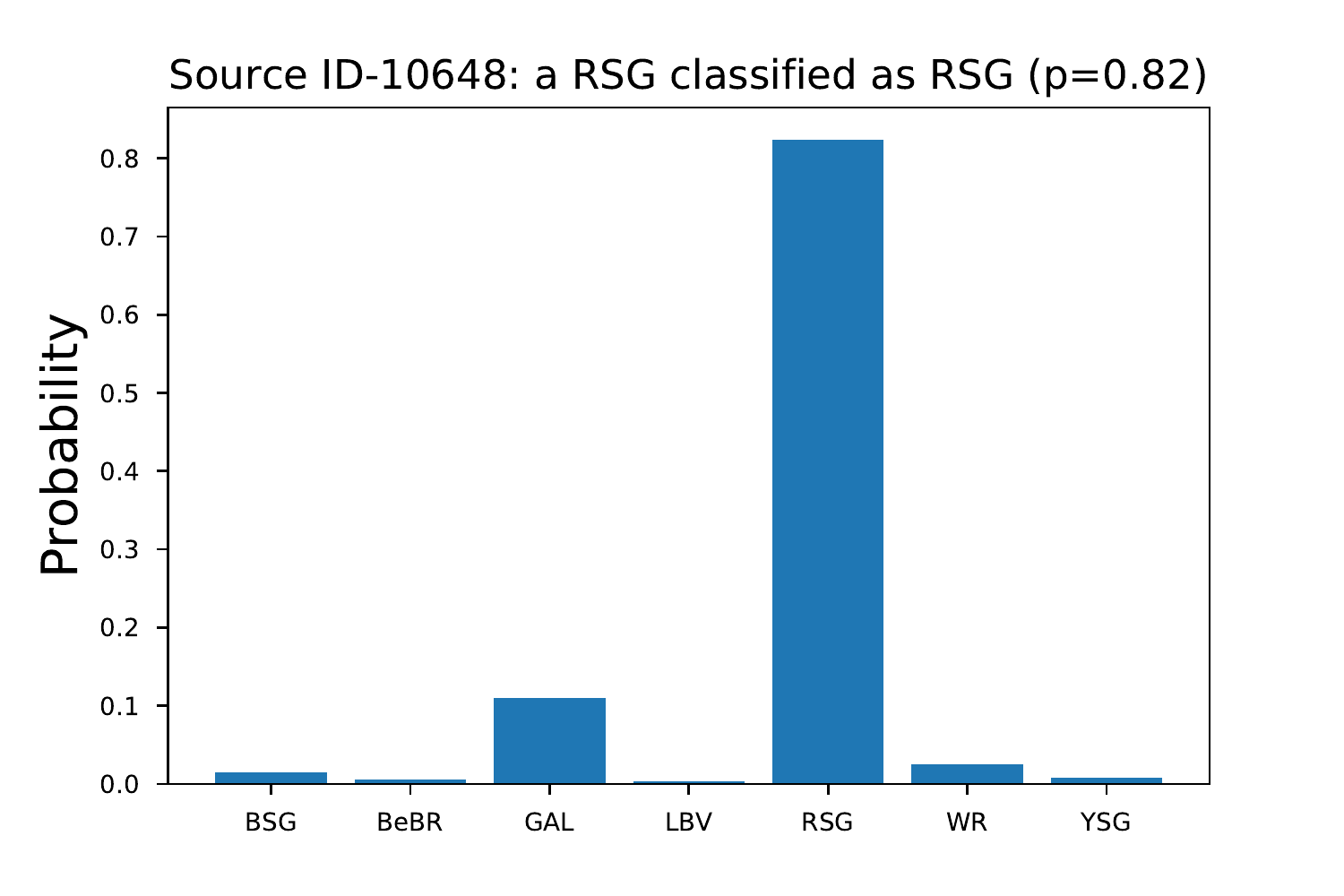}
    \includegraphics[width=0.4\linewidth]{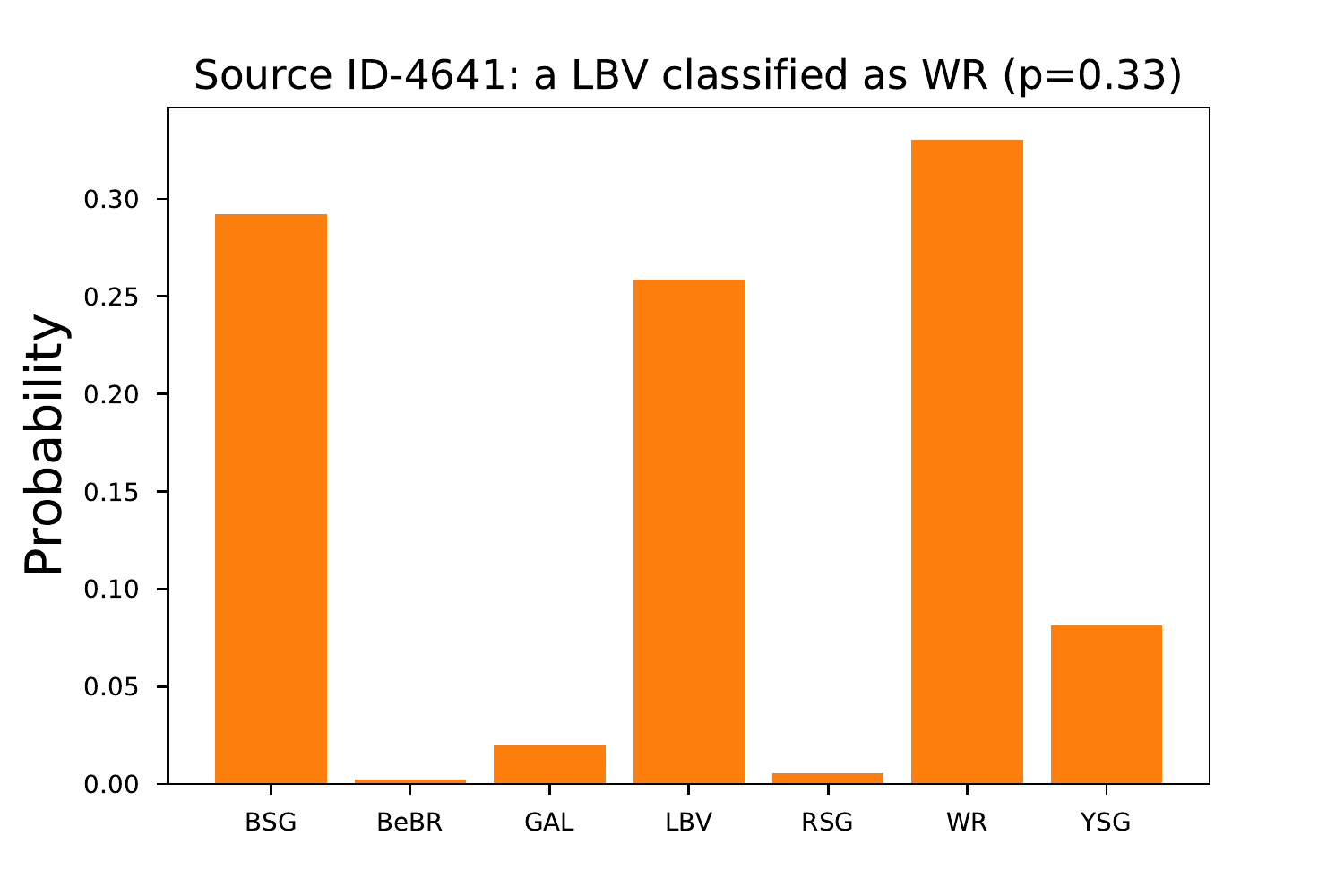}\\
    \includegraphics[width=0.4\linewidth]{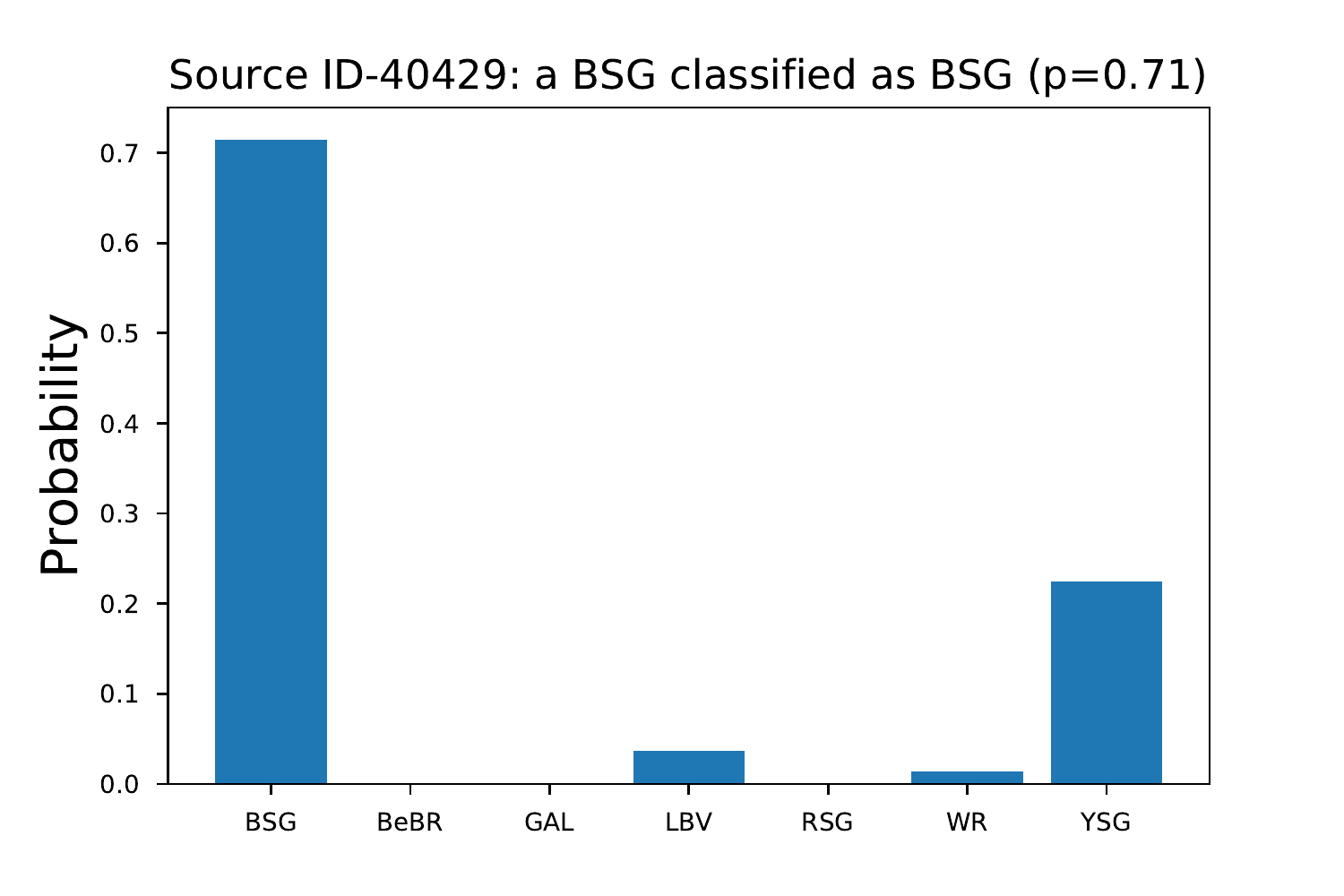} 
    \includegraphics[width=0.4\linewidth]{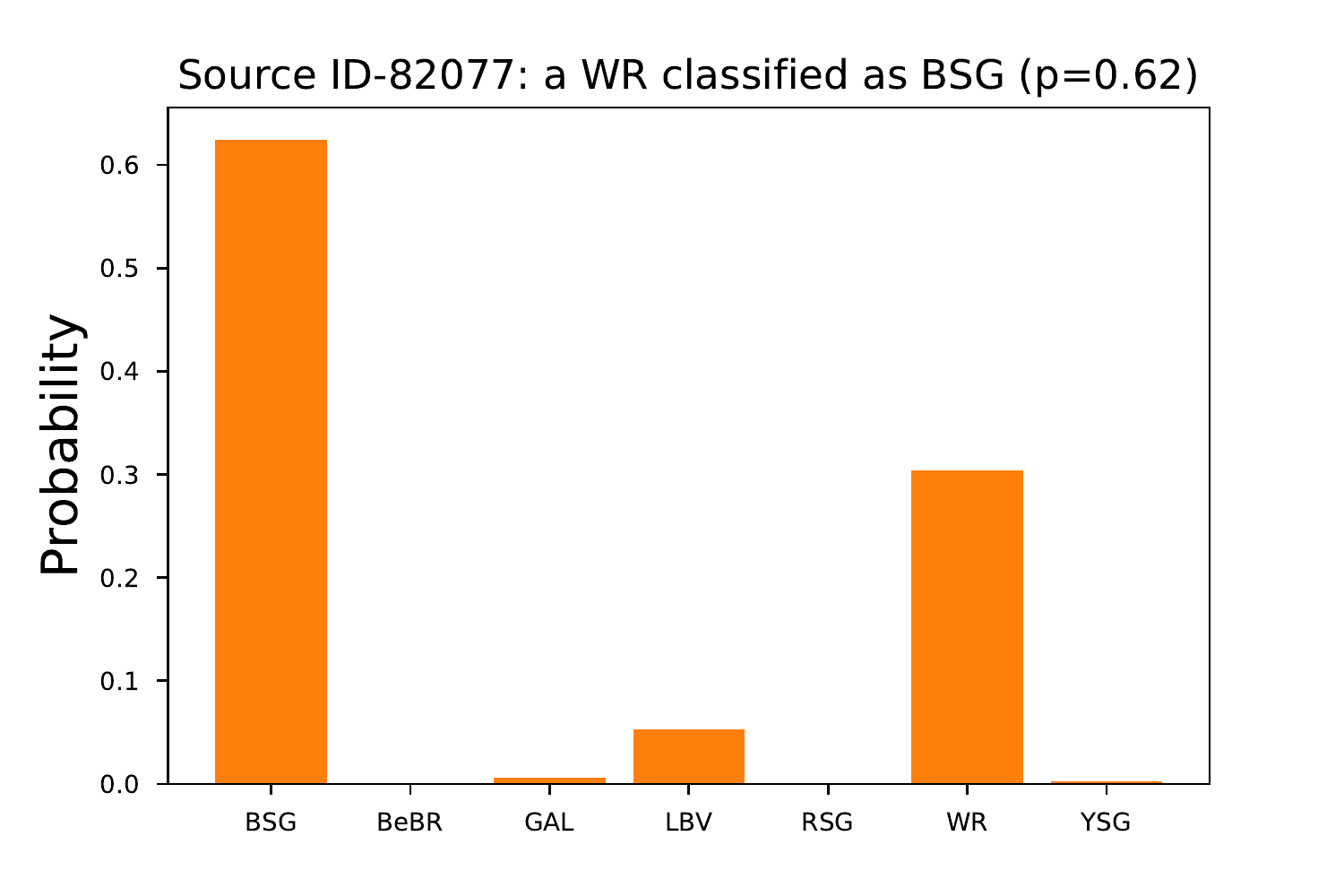}\\
   \caption{Examples of probability distributions for a number of objects with correct (left) and incorrect (right) final classifications.}
    \label{f:object_pdfs}
\end{figure*}

\begin{figure}
    \includegraphics[width=\columnwidth]{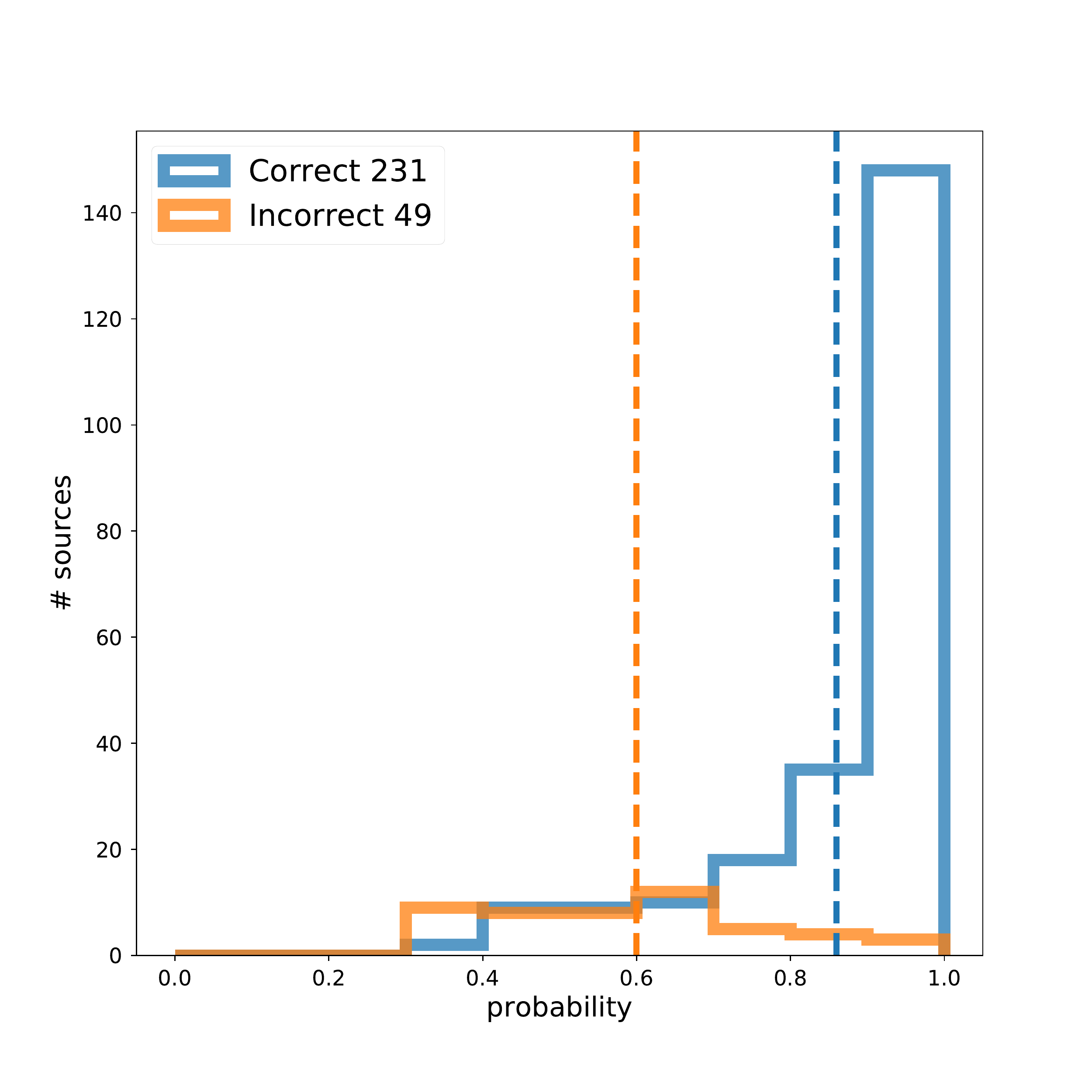}
    \caption{Probability distributions for sources classified correctly (blue) and incorrectly (orange), for the validation sample. We successfully recover the majority of the objects in the validation sample ($\sim83\%$). The dashed blue and orange lines correspond to the mean probability values for the correct (at 0.86) and incorrect (at 0.60) classifications, based on repeated five-fold CV tests (five iterations).} 
    \label{f:pdf_threshold}
\end{figure}

A common issue with machine-learning applications is choosing the best algorithm for the specific problem. This is actually impossible to achieve a priori. Even in the case of different algorithms that provide similar results it can be challenging to select one of them. However, there is no reason to exclude any. Ensemble methods refers to approaches that combine predictions from multiple algorithms, similar to combining the opinions of various "experts" in order to reach to a decision \citep[e.g.,][]{Re2012}. The motivation of ensemble methods is to reduce the variance and the bias of the models \citep{Mehta2019}. 

A general grouping consists of bagging, stacking, and boosting. Bagging (bootstrap aggregation) is based on training on different random subsamples whose predictions are combined either by majority vote (e.g., which is the most common class) or by averaging the probabilities (RF is the most characteristic example). Stacking (stacked generalization) refers to a model that trains on the predictions of other models. These base models are trained on the training data and their predictions (sometimes along with the original features) are provided to train the meta-model. In this case, it is better to use different methods that use different assumptions, so that to minimize the bias inherent by each method. Boosting refers to methods that focus on the improvement of the misclassifications from previous applications. After each iteration, the method will actually bias training toward the points that are harder to predict.

Given the similar results among the algorithms that we used, as well as the fact that they are trained differently and plausibly sensitive to different characteristic features per class, we were motivated to combine their results to maximize the predictive power and to avoid potential biases. We chose to use a simple approach, with a classifier that averages the output probabilities from all three classifiers.

There are two ways to combine the outputs of the models, either through "hard" or "soft" voting. In the former case the prediction is based on the largest sum of votes across all models, while in the latter the class corresponding to the largest summed probability is returned. 

To set an example with hard voting, if the results from the three models we used were BSG, BSG, and YSG, then the final class would be BSG. However, this voting scheme does not grasp all the information. Soft voting can be more efficient. Given the final summed probability distribution across all classes it is possible that the final classification may be different than the one that the hard voting would return. Additionally, it can solve cases when a single class cannot be determined, that is, when each of the three classifiers predicts a different class. With the final probability distribution we can also provide error estimates on the class predictions (and define confidence thresholds).

\subsubsection{Combined classifier}

The simplest approach is to combine the individual probabilities per class using equal weights per algorithm (since the accuracy of each is similar):  

\begin{equation}
    P_{\rm final} = (P_{\rm SVC} \times 1/3) + (P_{\rm RF} \times 1/3) + (P_{\rm MLP} \times 1/3).
\end{equation}In Fig. \ref{f:object_pdfs} we show some example distributions for a few sources with correct and incorrect final classifications. 

We performed a repeated (five iterations) five-fold CV test to estimate the overall accuracy of the combined approach at $0.83\pm0.02$ (see Table \ref{t:recalls}). The recall values are consistent with the results from the individual classifiers, with the highest success obtained for RSGs ($\sim94\%$) and BeBRs and YSGs ($\sim80\%$), while LBVs are not recovered. Despite this result, it is possible to get LBV classification at a probably lower significance (i.e., probability). However, even a small number of candidates for this class are important for follow-up observations, due to their rarity and their critical role in outburst activity \citep[e.g.,][]{Smith2014}.  

In Fig. \ref{f:pdf_threshold} we show the distributions of probabilities of the sources in a validation sample identified correctly (blue) and incorrectly (orange). The blue and orange dashed lines correspond to the mean probability values for the correct (at $0.86\pm0.01$) and incorrect (at  $0.60\pm0.03$) classifications. Although the distributions are based on a single evaluation of the classifier on the validation set, the values corresponding to these lines originate from a five-iteration repeated five-fold CV application. 



\subsection{Testing in other galaxies} 
\label{s:other_galaxies}

\begin{figure}
    \includegraphics[width=\columnwidth]{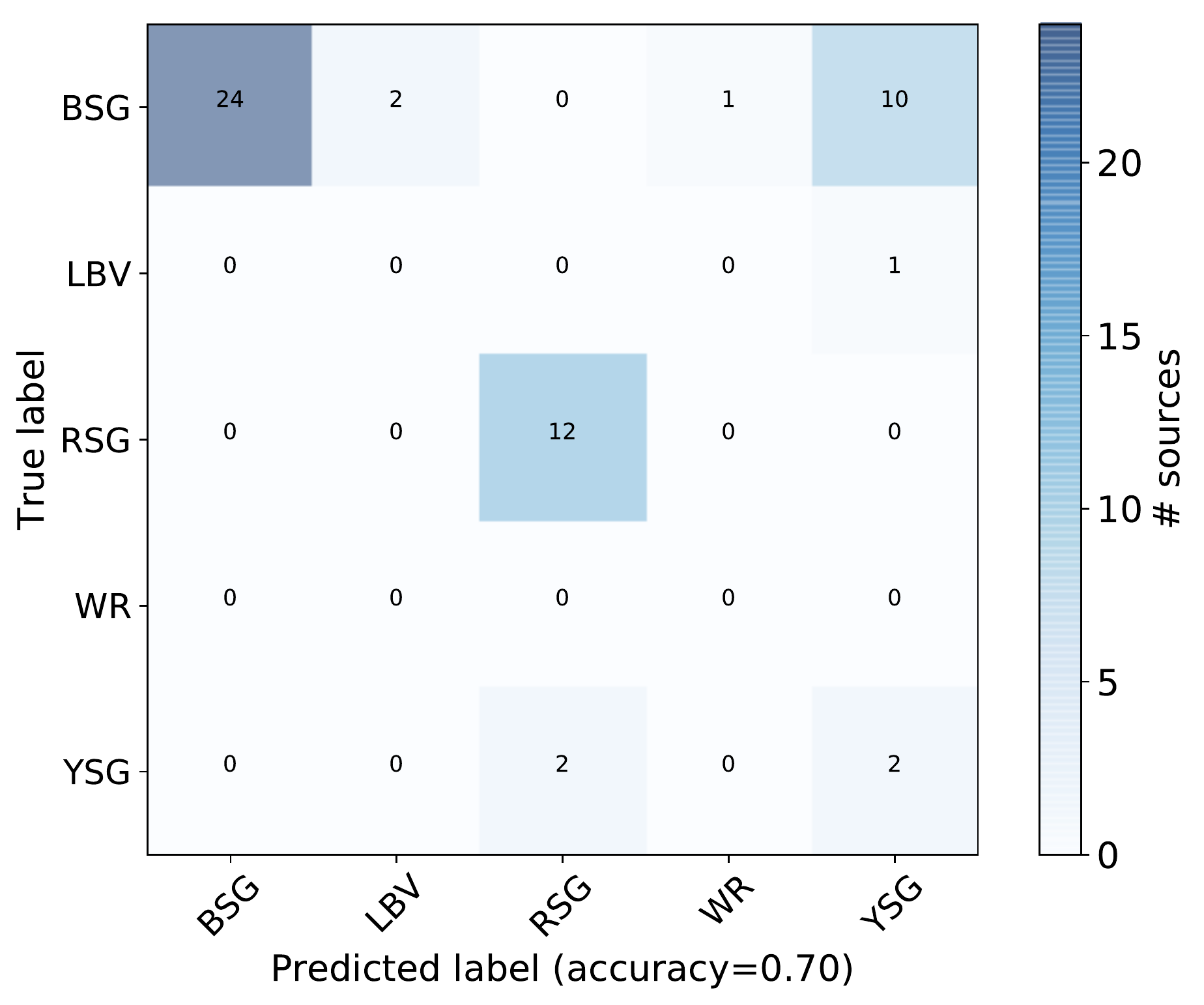}
    \caption{Confusion matrix for 54 sources without missing values in the three galaxies (IC 1613, WLM, and Sextans A). We achieve an overall accuracy of $\sim70\%$, and we notice that the largest confusion occurs between BSGs and YSGs. The overall difference in the accuracy compared to that obtained with the M31 and M33 sample is attributed to the photometric errors and the effect of metallicity and extinction in these galaxies.}
    \label{f:other_galaxies-cm}
\end{figure}

\begin{figure}[hbt!]
    \includegraphics[width=\columnwidth]{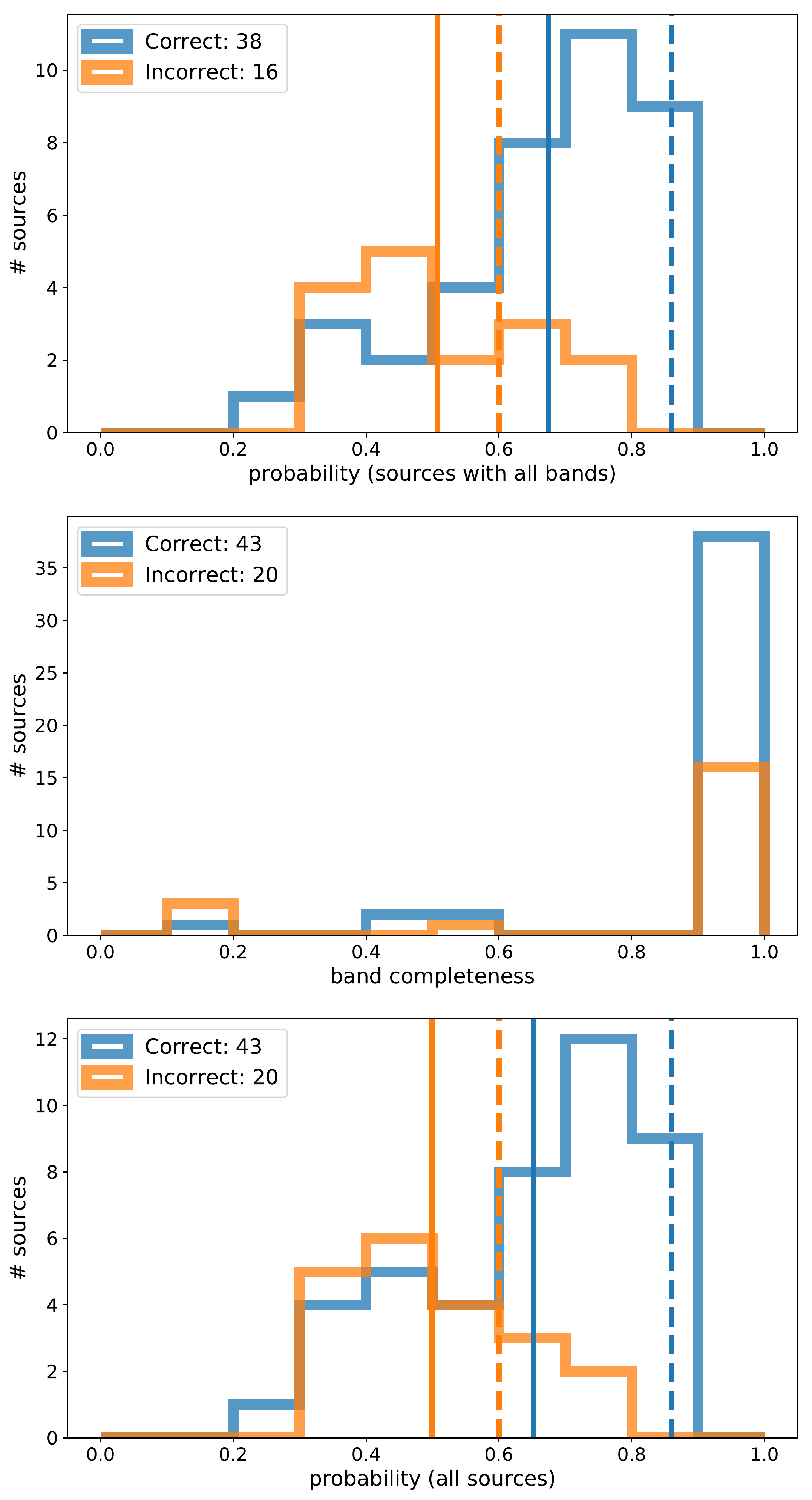}
    \caption{Probability and band completeness distributions for the sources of the three galaxies (IC 1613, WLM,  and Sextans A) with and without missing data. (Top) Probability distributions of the correct (blue) and incorrect (orange) final classifications for the total sample of stars with known spectral types and with measurements in all bands. We achieved a recovery rate of $\sim70\%$. The vertical dashed lines are the same as those in Fig. \ref{f:pdf_threshold}; the solid lines correspond to the peak of the probability distributions for the current sample. (Middle) Distribution of the band completeness, i.e., the fraction of features without missing values. (Bottom) Probability distributions for all sources, including those without measurements in multiple bands (vertical lines have the same meaning as in the top panel). The success rate of $\sim68\%$ is the same as in the top panel, indicating the effectiveness of the iterative imputer for missing data imputation.} 
    \label{f:other_galaxies}
\end{figure}

As an independent test we used the collection of sources with known spectral types in the IC 1613, WLM, and Sextans A galaxies (see Sect. \ref{s:data-spec_types}). In order to take into account all available information we resampled the whole M31 and M33 sample and we trained all three models.   

The application follows the exact same protocol as the training except from the resampling approach, which is used only for the training: (i) load the photometric data for the new sources, (ii) perform the necessary data processing to derive the features (color indices), (iii) load the fully trained models for the three classifiers\footnote{Using  Python’s built-in persistence model \texttt{pickle} for saving and loading.}, (iv) apply each of them to obtain the individual (per classifier) results, (v) calculate the total probability distribution from which we get the final classification result, and (vi) compare the predictions with the original classes. For the last step, we converted the original spectral types to the classes we formed while training. Out of the 72 sources we excluded nine with uncertain classifications: four carbon stars, two identified simply as "emission" stars, one with a "composite" spectrum, one classified as a GK star, and one M foreground star. 

In Fig. \ref{f:other_galaxies-cm} we show the confusion matrix for the sample of the test galaxies, where we have additionally (for this plot) excluded another nine sources with missing values (see next section). By doing this we can directly compare the results with what we obtained from the training galaxies M31 and M33. We successfully recovered $\sim70\%$, which is less than what we achieved for the training (M31 and M33) galaxies ($\sim83\%$). We note that due to the very small sample size (54 sources) even a couple of misclassifications can change the overall accuracy by a few percent. Nevertheless, a difference is still present. 

Evidently, the largest disagreement arises from the prediction of most BSGs as YSGs. These two classes do not have a strict boundary in the HRD, making their classification at larger distances even more challenging. Moreover, the sources in these galaxies are at the faint end of the magnitude distribution for the \textit{Spitzer} bands, which may influence the accuracy of their photometry. 

While M31 has a metallicity above solar and M33 a gradient from solar to subsolar \citep{Pena2019} the three test galaxies are of lower metallicity \citep{Boyer2015}. However, it is not certain how this influences the classification performance. Lower metallicity affects both extinction and evolution that could lead to shifts in the intrinsic color distributions. Currently, given the lack of photometric data and source numbers for lower metallicity galaxies, it is impossible to examine the effect of metallicity thoroughly.

In the upper panel of Fig. \ref{f:other_galaxies} we show the distribution of the probabilities of correct (blue) and incorrect (orange) classifications. The dashed lines represent the same limits as defined in Sect. \ref{s:combining_models} for the training sample (at 0.86 and 0.60, respectively), while the solid ones correspond to the mean values defined by the current sample, at 0.67 and 0.51 for correct and incorrect, respectively. These shifts of the peak probabilities, especially for the correct classifications, shows the increased difficulty of the classifier to achieve a confident prediction.

\subsection{Missing data imputation} 
\label{s:dat_imputation}

\begin{figure}[hbt!]
    \includegraphics[width=\columnwidth]{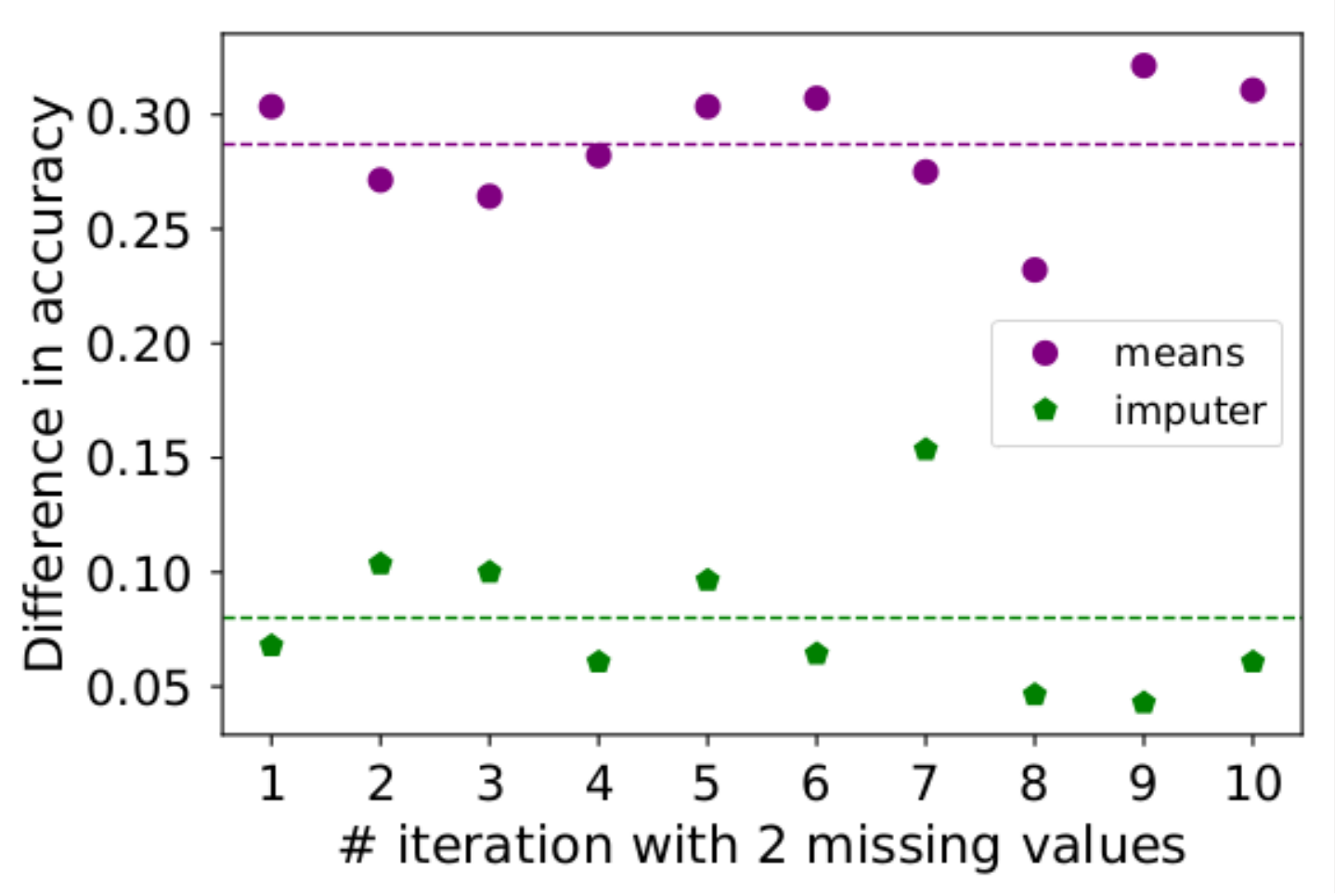}\\
    \includegraphics[width=\columnwidth]{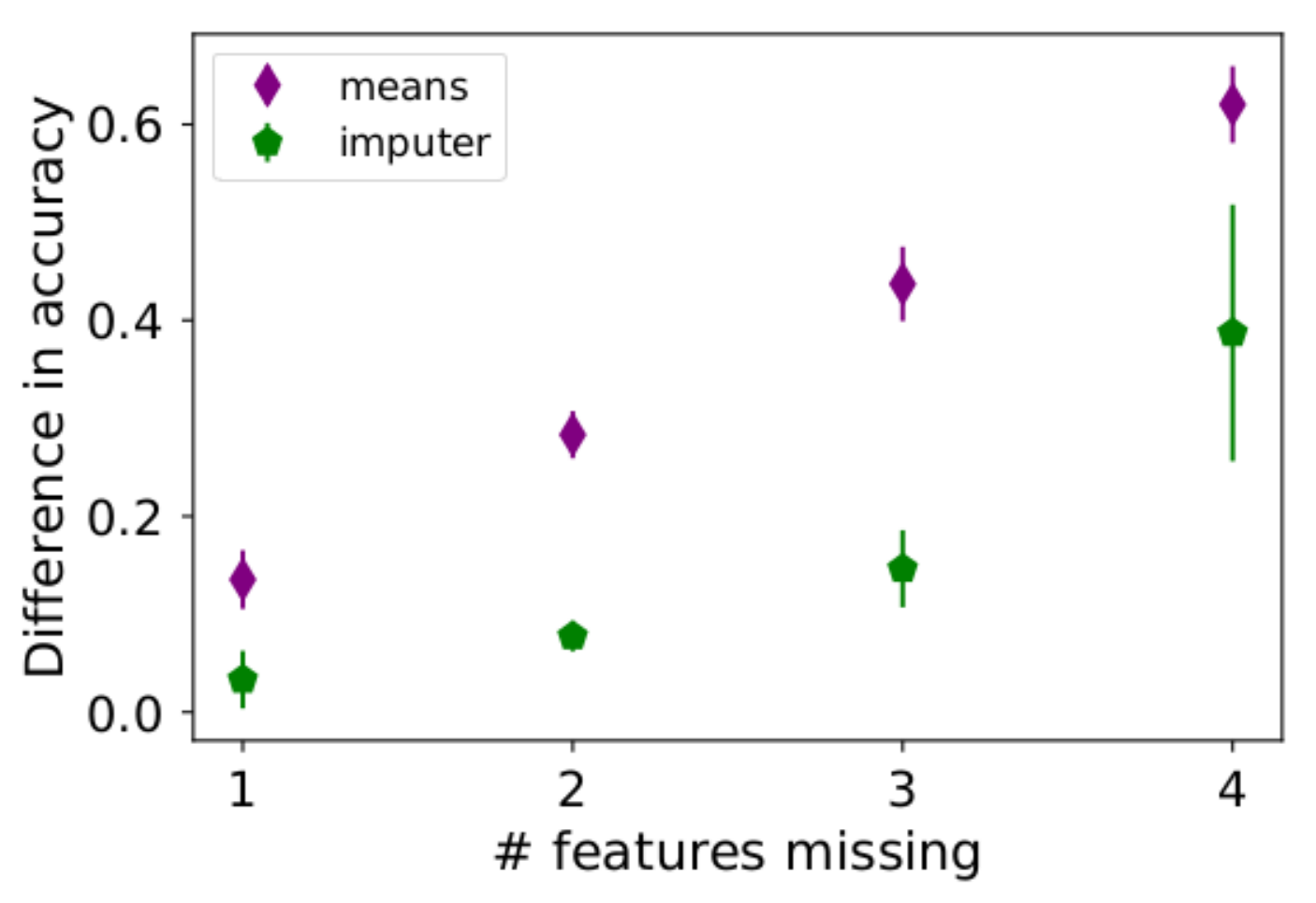}
    \caption{ Accuracy changes with missing features. (Top) Comparing the drop in accuracy from a typical ($30\%$ split) validation set without missing data to one where missing data have been generated by randomly selecting two features per object and replacing them with the corresponding mean values (purple circles) or the values imputed by the iterative imputer (green pentagons). The mean value obtained for the imputer is less than 0.1 and almost three times better than the mean drop for mean values. (Bottom) Iterative imputer, which is more capable of handling an increased number of missing features, with a limit at three (out of five available in total). The loss in accuracy is less than 20\%.}
    \label{f:missing_data}
\end{figure}

In the previous section we excluded nine sources that contain missing values, meaning they did not have measurements in one or more bands. This is important for two reasons. In order for the methods to work they need to be fed with a value for each feature. Simultaneously, the majority of the sources in the catalogs with unclassified sources (to which this classifier will be applied) do not possess measurements in all bands. 

To solve this, we performed a data imputation process in two ways. One, typical approach, is to replace missing values with a median/mean value. For this  we first derived the median value (to remove extremes) of each feature distribution per class and from all available sources in the training sample of M31 and M33. Then we took the mean of the feature's values over all classes. Another approach is to use  iterative imputation, in which each feature is modeled as a function of others originating from the multivariate imputation by chained equations (MICE; \citealt{Buuren2011}). This is a plausible assumption in our case, since we are dealing with spectral features that are indeed covariant to some degree (spectra do not fluctuate suddenly across neighboring bands unless a peculiar emission or absorption feature is present). It is hence plausible to impute a missing bandwidth value given the others. The imputation of each feature is done sequentially, which allows previous values to be considered as part of the model in predicting the subsequent features. The process is repeated (typically ten times), which allows the estimations of the missing values to be improved even more. We implemented this by using  \texttt{impute.IterativeImputer()} (with default parameters). 

To further investigate the influence of data with missing values, we ran a test by simulating sets with missing values from the original M31 and M33 sample. As usual, we  split the sample into training (70\%) and validation (30\%) samples. After resampling the training set, it was used to train the three classifiers, and an initial estimate of the accuracy was obtained with the validation sample. Then, according to how many features we can afford to "miss," we randomly selected the features of each object in the validation sample. We either replaced these features with the corresponding mean values or we applied the iterative imputer. Then the accuracy was obtained on this modified validation set. In the upper panel of Fig. \ref{f:missing_data} we show an example of the difference between the initial (unmodified) validation set and the ones with missing values, by randomly replacing two features (per object) and imputing data with the iterative imputer (green pentagons) and mean values (purple circles). The mean drop in accuracy (over ten iterations) is less than 0.1 for the imputer (green dashed line) but almost 0.3 for the means. In the bottom panel of Fig. \ref{f:missing_data} we show the drop in accuracy with increasing number of missing features. Obviously, the imputer is performing more efficiently and doing more than simply replacing missing features with mean values, and it can work with up to three missing features (out of five available in total).


We also quantified the fraction of missing values by defining a "band completeness" term, simply as $ 1 - N_{\rm bands\_without\_measurement } / N_{\rm total\_bands} $. In the middle panel of Fig. \ref{f:other_galaxies} we show the distribution of this completeness for correct and incorrect sources. Given that about half of the nine sources with missing values have band completeness $0.2$ (meaning only one feature is present) and the others are missing two to three, the success rate of five out of nine of these sources classified correctly ($\sim55\%$) matches what we would approximately expect from the bottom panel of Fig. \ref{f:missing_data}.

In the bottom panel of Fig. \ref{f:other_galaxies} we show now the probability distribution for all sources. The score is $68\%$, which is the same as the accuracy obtained for the sample without any missing values (at 70\%). The dashed and solid lines have the same meaning as previously, and there is no significant change (at 0.65 ad 0.59 for correct and incorrect classifications, respectively). In this particular data set the presence of a small number of sources (9 out of 63; $\sim14\%$) with missing values does not affect the performance of the classifier.

\section{Discussion}
\label{s:discussion}

In the following sections we discuss our results with respect to the sample sizes, label availability and feature sensitivity per class of our classifier. 
\subsection{Exploring sample volumes and class mixing}

\begin{figure*}
    \includegraphics[width=\textwidth]{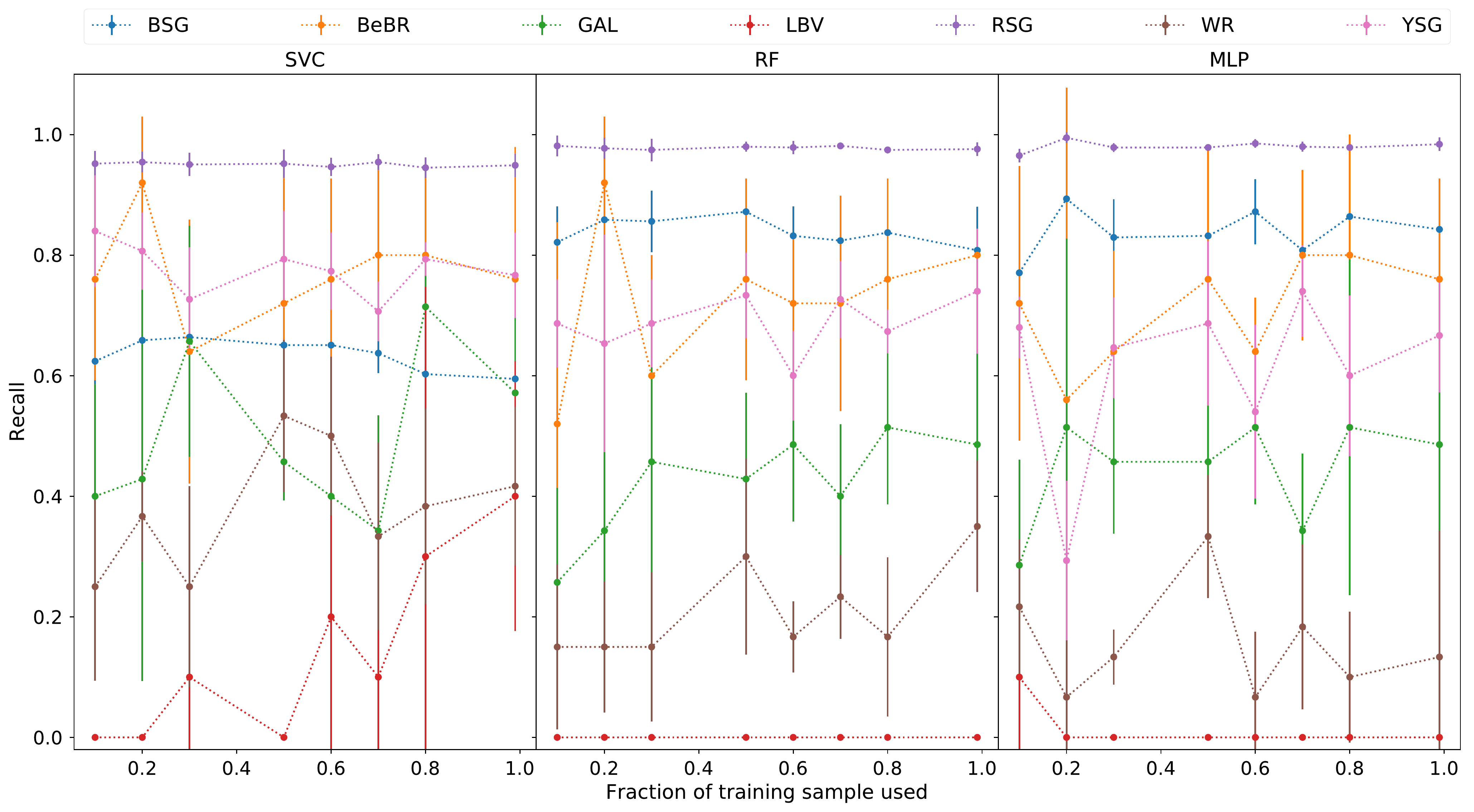}
    \caption{Recall vs. the fraction of the training sample used per class. We notice a significant improvement for BeBRs and YSGs with increased training samples. When the samples sizes are already adequate, the maximum possible value is achieved faster (e.g., for RSGs and BSGs). The GAL and WR classes show an increase, while the LBV sample is too small to produce meaningful results.}
    \label{f:sample_volume-recall}
\end{figure*}

One of the major concerns when building machine-learning applications is the representativeness of the samples used. To explore this we performed iterative runs for each algorithm by adjusting the size of the training sample used. 

At each iteration after the initial split into train (70\%) and validation (30\%) sets, we kept a fraction of the training. After randomly selecting the sources per class, we performed the resampling in order to create the synthetic data. However, we needed at least two sources per class for the SMOTE to run (for this process we adjusted \texttt{n\_neighbor=1}). Therefore, we started from $10\%$ up to the complete training sample. Especially for LBVs we added  an additional source by hand for the first two fractions (after 0.3 enough sources were selected automatically).

In Fig. \ref{f:sample_volume-recall} we plot the recall per class for each method (for completeness in Fig. \ref{f:sample_volume} we also present the precision and F1 score). We see an expected overall improvement with increasing sample size. This means that the larger the initial sample, the more representative it is of the parent population. The resampling method can interpolate, but does not extrapolate, which means that even though we are creating new sources they originate from the available information of the feature space. For example, \cite{Kyritsis2022} experimented with three different variants of RF to find that the results were dominated by the internal scatter of the features. Therefore, any limitations are actually transferred to the synthetic data. More information results in a better representation of their features by the classifier (leading to more accurate predictions).

\subsubsection{BSGs and RSGs}
 The BSGs and RSGs are the most populous classes, and they achieve a high accuracy much faster (except for BSGs in the SVC). The RSG class also performs well in the work of \cite{Dorn-Wallenstein2021}, at $\sim96\%$. In their refined label scheme they split (the equivalent to our) BSG sources into more classes, which results in a poorer performance. 

\subsubsection{BeBRs}
The careful reader will notice that the BeBR sample size is similar to that of the LBVs and smaller than the WR one. Despite that, we are able to get really good results due to the specific nature of these objects. The B[e] phenomenon (presence of forbidden lines in spectra) actually includes a wide range of evolutionary stages and masses, from pre-main-sequence stars to evolved ones, symbiotics and planetary nebulae \citep{Lamers1998}. The subgroup of evolved stars is perhaps the most homogeneous group, as they are very luminous $(\rm{log}(L/L_\odot) > 6.0)$ characterized by strong Balmer lines in emission (usually with P-Cygni profiles), narrow low-excitation lines (such as FeII, [FeII], and [OI]), and they display chemically processed material (such as TiO bands and $^{13}\rm CO$ enrichment) indicative of their evolved nature. Moreover, these spectral characteristics (arising from dense, dusty disks of unknown origin) are generally long-lived \citep{Maravelias2018} and these sources tend to be photometrically stable \citep{Lamers1998}. Those characteristics, along with strong IR excess due to their circumstellar dust (see \citealt{Kraus2019a}, but also \citealt{Bonanos2009, Bonanos2010}) make them a small but relatively robust class. Interestingly,  \cite{Dorn-Wallenstein2021} recover BeBR at the same accuracy with our approach.

\subsubsection{LBVs}
The LBV sample only shows a clear gain with an increased training sample for SVC, which is the most efficient method for recovering this class. When in quiescence, LBVs share similar observables (e.g., colors and spectral features) with BSGs, WRs, and BeBRs \citep[e.g.,][]{Weis2020, Smith2014}. Therefore, it is quite challenging to separate them, and the only way to certify the nature of these candidate LBVs is when they actually enter an active outburst phase. During this phase, the released material obstructs the central part of the star, changing its spectral appearance from a O/B type to an A/F type (which in turn would mix them with the YSG sources), while they can significantly brighten in the optical ($>2\,\rm{mag}$, but at constant bolometric luminosity; \citealt{Clark2005}). In order to form the most secure LBV sample, we excluded all candidate LBVs (some of which more resemble BeBRs; \citealt{Kraus2019a}), and we were left with a very small sample of six stars. The LBVs display an additional photometric variability of at least a few tenths of a magnitude \citep{Clark2005}. This information could be included as a  supplementary feature through a variability index (such as $\chi^2$, median absolute deviation, etc; \citealt{Sokolovsky2017}). However, this is not currently possible as the data sets we are using are very limited in the epoch coverage (for example, at the very best only a few points are available per band in the Pan-STARRS survey). Furthermore, the optical (Pan-STARRS) and IR (\textit{Spitzer}) data for the same source were obtained at different epochs, which may result the source's flux being sampled from different modes.  This effect, along with their small sample size (far from complete; \citealt{Weis2020}), may well explain the limited prediction capability of our method. On the other hand, \cite{Dorn-Wallenstein2021} took into account variability (using \textit{WISE} light curves) and they report a full recovery of LBVs, which might be due to overfitting. Because of the small size of their sample (two sources), they did not discuss it any further.  

\subsubsection{WRs}
In the single case scenario, LBVs are a phase in the transition of O-type stars before their  outer layers are stripped, because of the intense mass loss and/or massive eruptions \citep{Smith2014}. Binaries are another channel where efficient stripping can lead to WRs \citep{Shenar2020}. Depending on the metallicity and their rotation, WRs may also form directly from main-sequence stars \citep{Meynet2005}. As their evolution is highly uncertain, they can originate from both LBV or BSG stars. Stellar evolution is a continuous process that does not display strict boundaries between those groups in the HRD. Therefore, their features (color indices) can be mixed. They are bright sources and this has enabled the detection of almost their complete population (see \citealt{Neugent2019a} for a review) but the actual numbers are limited due to their rarity. Their small sample size -- which actually includes a number of different subtypes of WRs, such as the nitrogen or carbon rich ones, as well as some known binaries with O-type companions -- has an impact on our prediction capability, but it is better than for LBVs. We also note that their recall benefits from the increase in the training sample for SVC and RF, but not much for MLP.  \cite{Rosslowe2018} have shown that WRs and LBVs can be better distinguished  using near-IR (JHK bands), a region that is unfortunately excluded from our feature list because of the lack of extensive and consistent surveys for our galaxies (although 2MASS exists, it is not deep enough for our more distant galaxies). On the contrary, \cite{Dorn-Wallenstein2021} include these bands, which may explain their improved accuracy for WRs and (possibly) for LBVs.   


\subsubsection{YSGs}
The YSG class contains all sources that are found in between the BSG and the RSG classes. In general, this is a relatively short-lived phase as the star evolves off the main sequence or evolves back to hotter phases after the RSG phase (e.g., \citealt{Kourniotis2018,Gordon2019}; excluding the contamination by foreground sources, which we minimized by preprocessing with the \textit{Gaia} properties but definitely did not eliminate). However, it is hard (if not impossible) to get strict boundaries in the CMDs between the BSG and the YSG populations, as well as the YSG and the RSG ones. \cite{Yang2019} presents a ranking scheme that is based on the presence of each source in a number of multiple CMDs (c.f. fig. 16). With our current work we are able to remove this complexity as we take into account the information from multiple CMDs (through the color indices) at once. We are able to correctly predict the majority of this sample at $\sim73\%$, in contrast to the $\sim27\%$ from \cite{Dorn-Wallenstein2021}. The major factor in this case is the use of more optical colors, which helps in distinguishing YSGs from BSGs more effectively, while \cite{Dorn-Wallenstein2021} work mainly with IR colors.


\subsection{Label uncertainties}

Uncertainty in the labels (or classes) can come in two flavors, either because of classification errors (e.g., human bias, instrument limitations) or due to the natural mixing of these sources. After all, there are uncertainties in the evolution of massive stars after the main sequence, as we still lack robust knowledge with respect to the transition of these sources through the various phases. However, it is a typical prerequisite in supervised machine-learning applications that the labels are the absolute truth. This can lead to inaccurate predictions. \cite{Dorn-Wallenstein2021} comment specifically on this, as with their refined classes (containing 12 classes) they achieve an accuracy of $\sim53\%$ for the SVC, because their labels for Galactic sources are "derived inhomogeneously, and many are from spectroscopy that is now more than 50 years old." In our case, we have obtained a more homogeneous sample, since we are working with specific galaxies (distance uncertainties are minimized) and the results originate from consistent surveys and modern instruments and facilities. In other words, our labels are more secure and help us achieve a better result. A way to tackle this is by properly handling label uncertainties during the training process itself, which, however, is not a trivial task.

\subsection{Feature sensitivity}

\begin{figure}
    \includegraphics[width=\columnwidth]{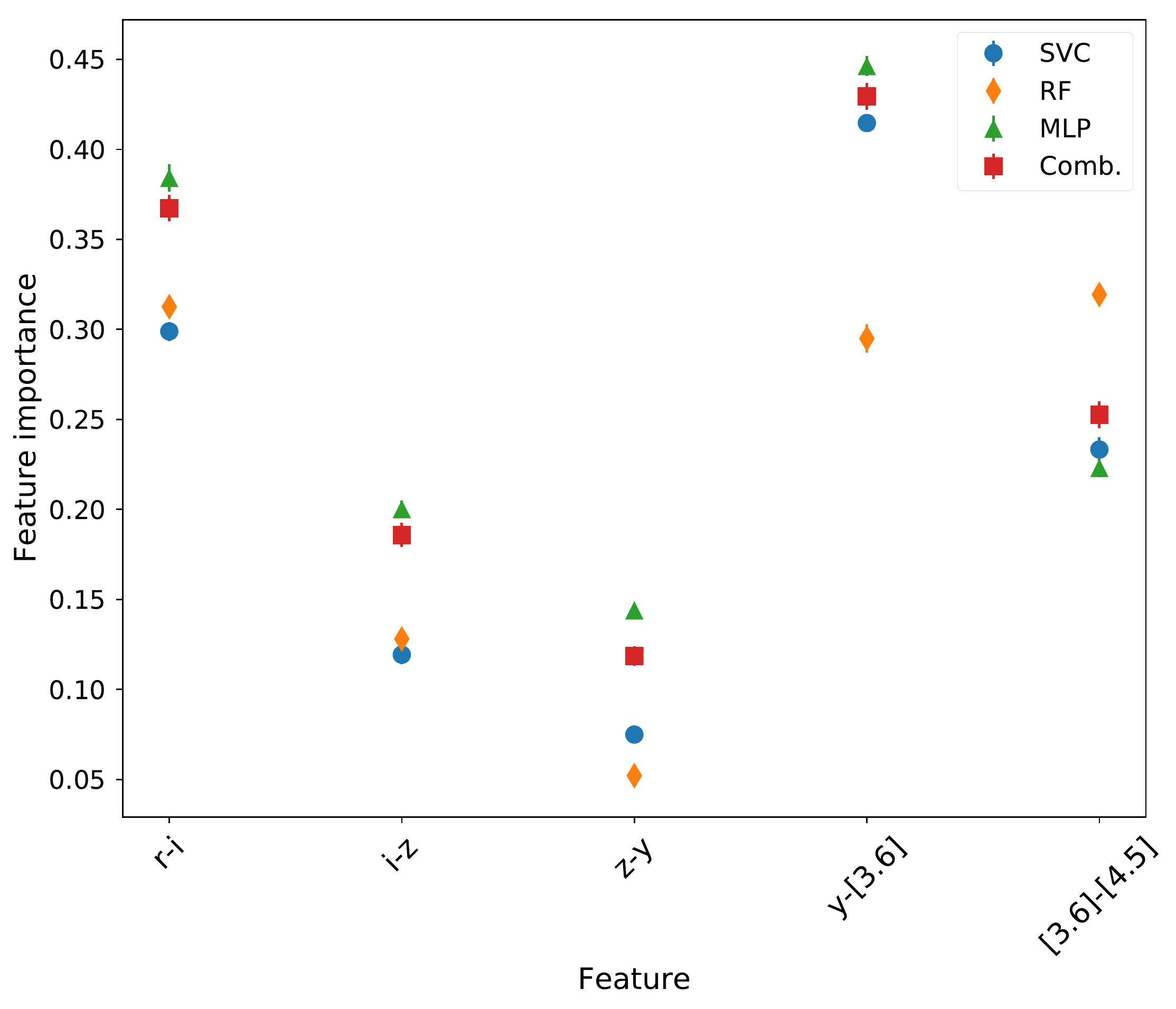}
    \caption{Permutation feature importance (i.e., the difference in accuracy between the original data set and the shuffled one) per feature for each classifier independently and the combined one. The features $r-i$, $y-[3.6]$, and $[3.6]-4.5$ consistently appear to be the most important.}
    \label{f:permutative_features}
\end{figure}

\begin{figure}
    \includegraphics[width=\columnwidth]{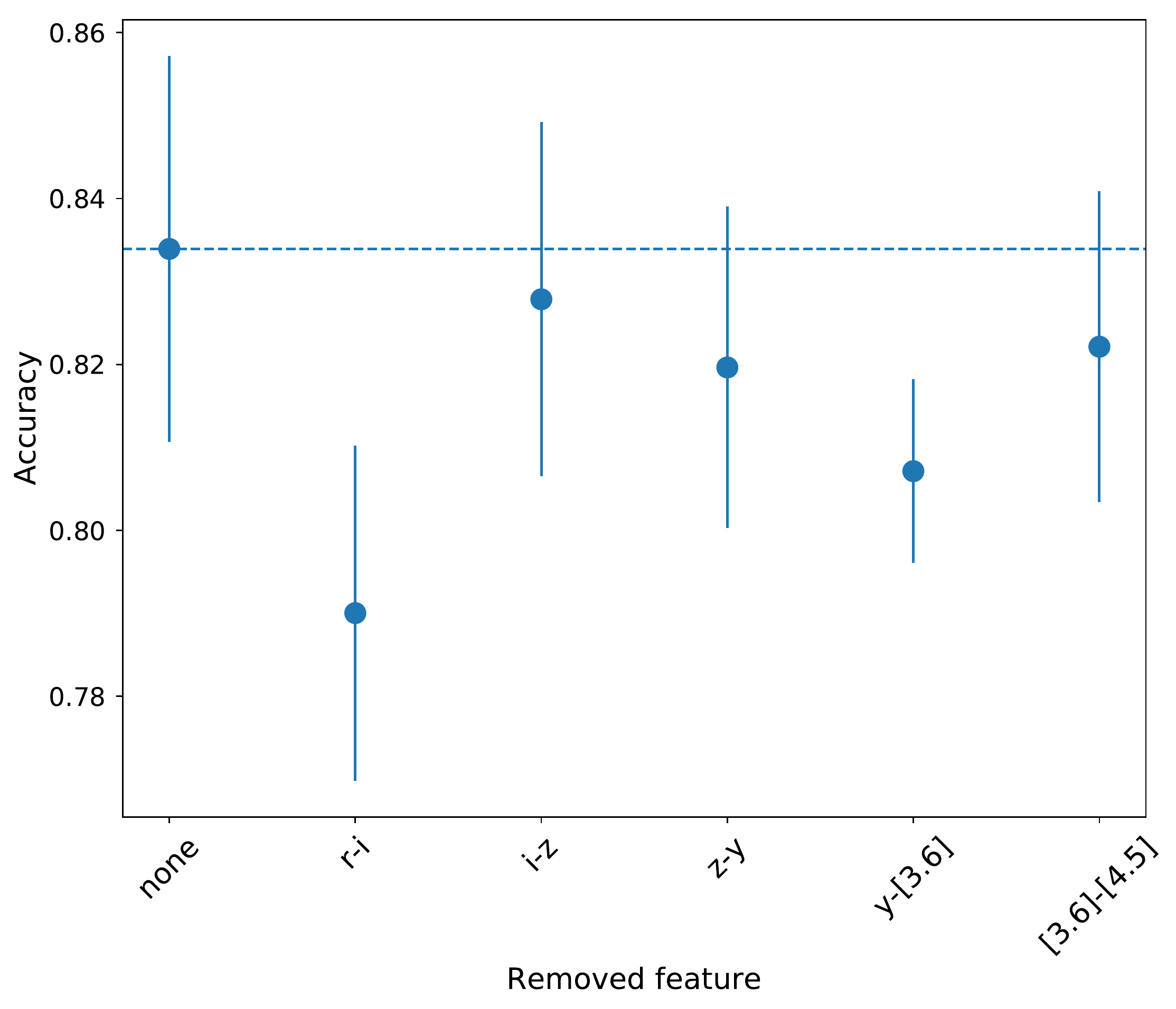}\\
    \includegraphics[width=\columnwidth]{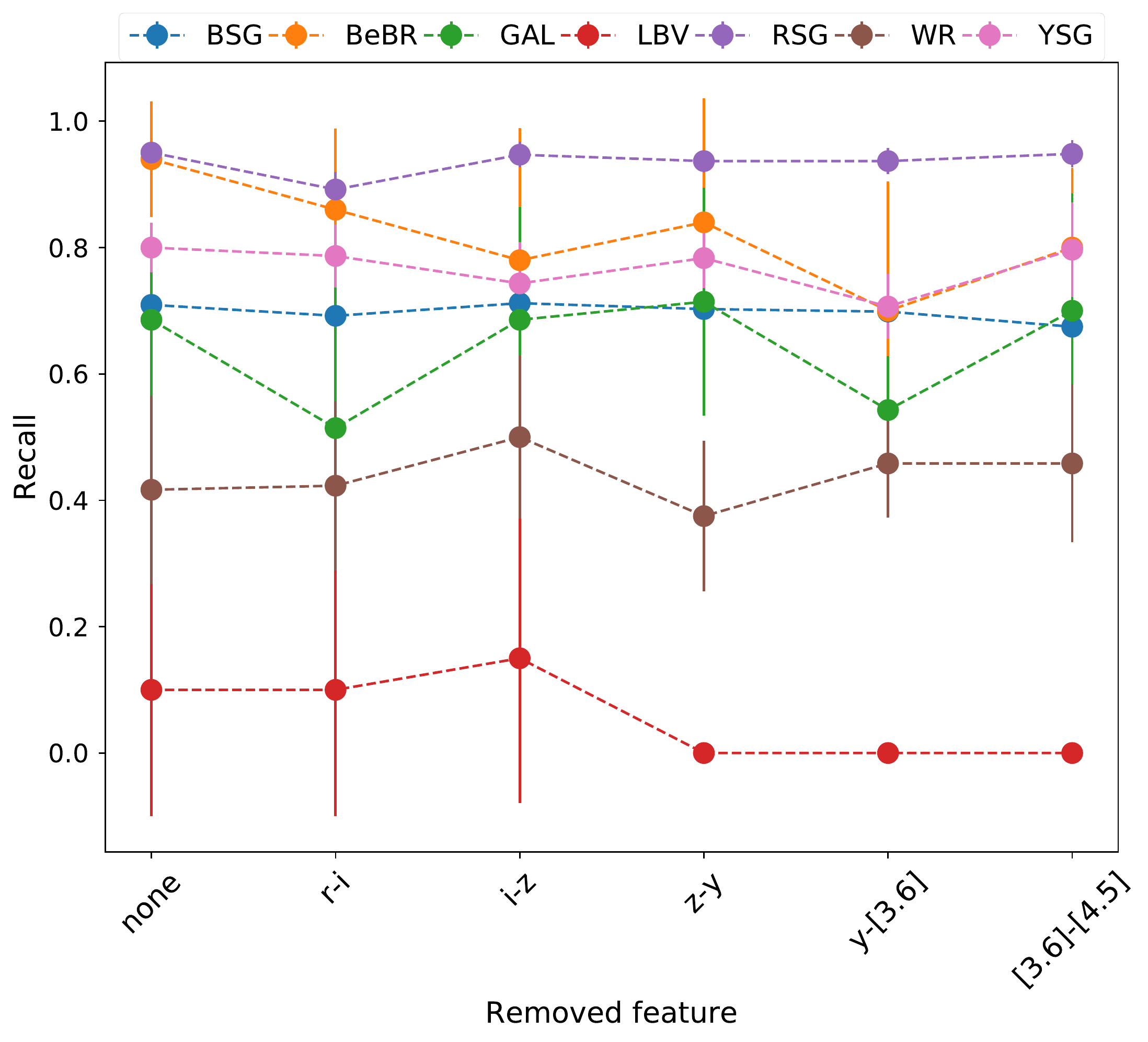}
    \caption{Feature importance per removed feature. The feature is removed and the whole combined model is retrained. The  first point corresponds to the full feature set. (Top) Considering the overall accuracy for the combined classifier, the most significant features are $r-i$ and $y-[3.6]$, consistent with the permutation  importance test. (Bottom) Recall per class. We see that different features become more significant for BeBRs, GALs, WRs, YSGs, and LBVs, while RSGs and BSGs do not show important changes (see text for more).}
    \label{f:featimp_remove}
\end{figure}

During the feature selection (Sect. \ref{s:feature_selection}) we disregarded bands that would significantly decrease our sample and/or they would introduce noise ($J_{\rm{UK}}$, \textit{Gaia}, \textit{Spitzer} [5.8], [8.0], and [24], Pan-STARRS $g$). The \textit{Spitzer} [3.6] and [4.5] bands are present for all of our sources (by construction of our catalogs), while the availability of optical ones (Pan-STARRS $r, i, z, y$) varies depending on the source. In order not to lose any more information, we included all optical bands (except for $g$) and performed missing data imputation whenever necessary. But then, the questions of how sensitive the classifier is to these features and which are more important per class, naturally follow.

We first investigate how the overall performance of the classifier depends on the features. For this we performed a permutation feature importance test (\texttt{sklearn.inspection.permutation\_importance()}). By shuffling the values of a specific feature we see how much it influences the final result. In this case the metric used is the difference between the accuracy of the original data set and of the shuffled one\footnote{The process is performed on the training sample, so we included all M31 and M33 sources in it and resampled accordingly.}. In the case of a nonsignificant feature this change will be small and the opposite holds for an important feature. In Fig. \ref{f:permutative_features} we show the results per classifier as well as their combined model (``all''). We notice that the most significant features are $r-i$, $y-[3.6]$, and $[3.6]-[4.5]$ while the least important are $i-z$ and $z-y$ and $[4.5]-[5.8]$. This is not a surprise actually since these features are the ones for which we have the largest separation among the averaged lines  of classes (see Fig. \ref{f:SEDs-all}). There are small differences between the individual algorithms but they are relatively consistent. They show similar sensitivity to the optical colors. The only exception is RF for $y-[3.6]$, which seems less sensitive than the others.

One key issue with this approach is that it is more accurate in the case of uncorrelated data. In our case there is some correlation both due to the fact that the consecutive color indices contain their neighboring bands and because the fluxes at each band are not totally independent from the others. An alternative and more robust way is by testing the accuracy of our model by dropping a feature each time. The general drawback in this approach is the computational time as it is needed to retrain (including resampling at each iteration) the model from the beginning, contrary to the previous test where only the values of a feature change and the model is applied. Fortunately, our training sample and modeling is neither prohibitively large nor complicated. Thus, using the combined classifier we iteratively removed one feature at the time. Then we calculated the metric of this iteration with respect to the initial feature set. In Fig. \ref{f:featimp_remove} (upper panel) we plot this accuracy, where $r-i$ and $y-[3.6]$ show (relatively) large deviation and seem to be the most important features. This is in agreement with what we found with the feature permutation approach. Interestingly, $r-i$ is the "bluest" feature and seems to be important for the overall classification (the optical part is excluded from the work of \citealt{Dorn-Wallenstein2021}).

When examining the results for the recall per class (Fig. \ref{f:featimp_remove}; lower panel) we see that for different classes are sensitive to different features. For BeBRs, $i-z$ and $y-[3.6]$ seem to be the most important, although smaller offsets are visible for the rest of the features also (the mean curve of BeBR peaks at this feature; see Fig. \ref{f:SEDs-all} This can be attributed to the overall redder colors because of the dusty environment around these objects. The GALs are sensitive to both $r-i$, the feature closer to the optical part, and $y-[3.6]$, partly due to the PAH component (GALs display the second strongest peak in Fig. \ref{f:SEDs-all}). 
Although not so significant $i-z$ seems to favor WR classification. The WR class is a collection of different flavors or classical (evolved) WRs, including binary systems. The YSGs are more sensitive to $y-[3.6]$ and a bit less in $i-z$, similar to BeBRs, as they also tend to have dusty environments. The BSGs and RSGs are the most populated classes, and they do not show any significant dependence. This might be because although distinct to the other classes they contain a wider range of objects that possibly mask significant differences between the bands (see \citealt{Bonanos2009, Bonanos2010}). For example, we included in the BSG sources with emission lines, such as Be stars that display redder colors. For LBVs, $i-z$ seems important but due to their small population the error is quite significant. Also, the redder features lie at zero, which may be due to the incapability of our model to predict these sources with higher confidence.  If we were to exclude any of these features, we would get poorer results for some of the classes. The inclusion of more colors would benefit the performance of our classifier as it would help with the sampling of the spectral energy distributions of the sources (going to the optical blue part will not help the redder sources but it would be  valuable for the hotter classes).

\section{Summary and conclusions}
\label{s:summary}

In this work we present the application of machine-learning algorithms to build an ensemble photometric classifier for the classification of massive stars in nearby galaxies. We compiled a \textit{Gaia} cleaned selected sample of 932 M31 and M33 sources, and we grouped their spectral types into seven classes: BSGs, YSGs, RSGs, B[e]SGs, LBVs, WRs, and background sources (outliers). To address the imbalance of the sample, we employed a synthetic data approach with which we managed to increase the underrepresented classes, although this is always limited by the feature space that the initial sources sample. We used as features the consecutive color indices from the \textit{Spitzer}  [3.6] and [4.5] and Pan-STARRS $r, i, z, $  and $y$  bands (not corrected for extinction). We implemented three well-known supervised machine-learning algorithms, SVC, RF, and MLP, to develop our classifier. The application of each of the algorithms results in fairly good overall results (recovery rates): BSGs, GALs, and YSGs from $\sim60\%$ to $\sim80\%$, BeBRs at $\sim73-80\%$, and WRs at $\sim45\%$, with the best results obtained for the RSGs ($\sim94\%$) and the worst for LBVs ($\sim28\%$ for SVC only). These results are on par with or improved compared to the results from \cite{Dorn-Wallenstein2021}, who worked with a much less homogeneous (with respect to the labels) but more populated Galactic sample. Given the similar performance of the three methods, and to maximize our prediction capability, we combined all outputs into a single probability distribution. This final meta-classifier achieved a similar overall (weighted balanced) accuracy ($\sim83\%$) and similarly good results per class. 

Examining the impact of the training volume size, we noticed that, as expected,  the sample size plays a critical role in the accurate prediction of a class. When many sources of a class are available (e.g., RSGs or BSGs), then the classifier works efficiently. In less populated classes (such as BeBRs and WRs), the inclusion of more objects increases the information provided to the classifier and improves the prediction ability. However, we are hampered by low-number statistics as these classes correspond to rare and/or short-lived phases. 

Additional information can be retrieved by using more features. We investigated the feature importance to find that, for the current data set, $r-i$ and $y-[3.6]$ are the most important, although different classes are sensitive to different features. Thus, the inclusion of more color indices (i.e., observations at different bands) could improve the separation of the classes. 

To test our classifier with an independent sample, we used data collected for IC 1613, WLM, and Sextans A sources, some of which ($\sim14\%$) had missing values. We performed data imputation by replacing the features' values using means and an iterative imputer. Although the missing values do not significantly affect the results for this particular data set, further tests showed that the iterative imputer can efficiently handle data sets with up to three missing features (out of the total five available). The final obtained accuracy is $\sim70\%$, lower than what we achieved for M31 and M33. The discrepancy can partly be attributed to photometric issues and the total effect of metallicity. The latter can modify the intrinsic colors of the sources and extinction due to the different galactic environments. Despite this, the result from this application is promising. In a follow-up paper we will present in detail the application of our classifier to previously unclassified sources for a large number of nearby galaxies. 

Currently, the metallicity dependence is impossible to address. For this we need larger samples of well-characterized sources in different metallicity environments. Although this is challenging because of the observing time required in large facilities, the ASSESS team is actively working toward this goal. A number of observing spectroscopic campaigns are completed and ongoing, which will provide the ultimate testbed of our classifier's actual performance along with opportunities for improvement.\\

\small
\noindent\textit{Acknowledgements} We thank the anonymous referee for their constructive comments and suggestions that helped us to improve this work. GM, AZB, FT, SdW, MY acknowledge funding support from the European Research Council (ERC) under the European Union’s Horizon 2020 research and innovation programme (Grant agreement No. 772086). GM would like to thank Alkis Simitsis, Thodoris Bitsakis, Elias Kyritsis, Andreas Zezas, Jeff Andrews, Konstantinos Paliouras for many fruitful discussions on machine learning and beyond. 

\textit{Facilities:} This work has made use of data from the European Space Agency (ESA) mission {\it Gaia} (\url{https://www.cosmos.esa.int/gaia}), processed by the {\it Gaia} Data Processing and Analysis Consortium (DPAC, \url{https://www.cosmos.esa.int/web/gaia/dpac/consortium}). Funding for the DPAC has been provided by national institutions, in particular the institutions participating in the {\it Gaia} Multilateral Agreement. The Pan-STARRS1 Surveys (PS1) and the PS1 public science archive have been made possible through contributions by the Institute for Astronomy, the University of Hawaii, the Pan-STARRS Project Office, the Max-Planck Society and its participating institutes, the Max Planck Institute for Astronomy, Heidelberg and the Max Planck Institute for Extraterrestrial Physics, Garching, The Johns Hopkins University, Durham University, the University of Edinburgh, the Queen's University Belfast, the Harvard-Smithsonian Center for Astrophysics, the Las Cumbres Observatory Global Telescope Network Incorporated, the National Central University of Taiwan, the Space Telescope Science Institute, the National Aeronautics and Space Administration under Grant No. NNX08AR22G issued through the Planetary Science Division of the NASA Science Mission Directorate, the National Science Foundation Grant No. AST-1238877, the University of Maryland, Eotvos Lorand University (ELTE), the Los Alamos National Laboratory, and the Gordon and Betty Moore Foundation. The UHS is a partnership between the UK STFC, The University of Hawaii, The University of Arizona, Lockheed Martin and NASA.

\textit{Software:} This research made use of Numpy \citep{numpy2020}, matplotlib \citep{matplotlib}, sklearn \citep{sklearn}, Jupyter Notebooks \citep{jupyter}, Mlxtend \citep{mlxtend}. This research made use of TOPCAT, an interactive graphical viewer and editor for tabular data \citep{topcat}. 

We wish to thank the "2019 Summer School for Astrostatistics in Crete"\footnote{\url{http://astro.physics.uoc.gr/Conferences/Astrostatistics_School_Crete_2019/}} for providing training on the statistical methods adopted in this work. We also thank Jeff Andrews for organizing the SMAC (Statistical methods for Astrophysics in Crete) seminar\footnote{\url{https://githubhelp.com/astroJeff/SMAC}}. We also acknowledge useful information provided by Jason Brownlee from his site Machine Learning Mastery\footnote{\url{https://machinelearningmastery.com}}

This research has made use of NASA's Astrophysics Data System. This research has made use of the SIMBAD database, operated at CDS, Strasbourg, France. This research has made use of the SVO Filter Profile Service (http://svo2.cab.inta-csic.es/theory/fps/) supported from the Spanish MINECO through grant AYA2017-84089.

%
%

\bibliographystyle{aa}
\bibliography{references} 

\begin{appendix}

\section{List of classified sources}

In this section we provide in detail the list of sources with spectral classifications presented in Table \ref{t:spectypes_refs}. For each source we provide a simple identification id with the galaxy and an increasing number (Col. 1), RA and Dec (Cols. 2 and 3) as obtained from their corresponding source, the spectral type (Col. 4) and the reference (Col. 5).

Only the first rows of the catalog is provided here for guidance, as it is published in its entirety in CDS.

\begin{table}
    \caption{Sources with known spectral types.} 
    \label{t:catalog_sptypes} 
    \begin{tabular}{lcccl} 
    \hline
    ID & RA & Dec & SpType & Ref \\
         & (J2000) & (J2000) & & \\
         &  (deg)  &  (deg)  & & \\
    \hline 
    \hline 
M31-1 &  9.26500  &  40.33747  &  RSG:  &  (1)   \\
M31-2 &  9.27583  &  40.02225  &  AI  &  (2)   \\
M31-3  &  9.30000  &  39.91256  &  YSG:  &  (2)   \\
M31-4  &  9.35208  &  40.30647  &  RSG:  &  (1)  \\
M31-5  &  9.35667  &  40.12550  &  YSG:  &  (2)  \\
M31-6  &  9.37083  &  40.33547  &  B1I  &  (2)  \\
M31-7  &  9.38875  &  40.01014  &  B2Ib-B5I  &  (2)  \\
M31-8  &  9.39250  &  40.01250  &  RSG  &  (2)  \\
M31-9  &  9.39333  &  40.02131  &  B0I  &  (2)  \\
M31-10  &  9.41625  &  39.97739  &  M1Ia  &  (2)  \\
M31-11  &  9.43875  &  39.97319  &  F5Ia  &  (2)  \\
M31-12  &  9.46625  &  39.98383  &  A2Ib  &  (2)  \\
M31-13  &  9.59500  &  40.54425  &  B9.5I+Dbl:  &  (2)  \\
M31-14  &  9.62167  &  40.51672  &  RSG:  &  (1)  \\
M31-15  &  9.63042  &  40.54169  &  RSG:  &  (1)  \\
M31-16  &  9.63458  &  40.51069  &  B7I  &  (2)  \\
M31-17  &  9.73610  &  40.57960  &  QSO  &  (3)  \\
M31-18  &  9.73833  &  40.52558  &  WN7:+Neb  &  (2)  \\
M31-19  &  9.73875  &  40.68153  &  M2I  &  (2)  \\
M31-20  &  9.75917  &  40.65200  &  M1I  &  (2)  \\
    \hline
    \hline
    \end{tabular}
\\
References: (1) \cite{Gordon2016}, (2) \cite{Massey2016}, (3) \cite{Massey2019}.\\
Note: The table is available in its entirety at the CDS. 
\end{table}

\section{\textit{Gaia} processing plots for M33}

\begin{figure*}
        \includegraphics[width=\textwidth]{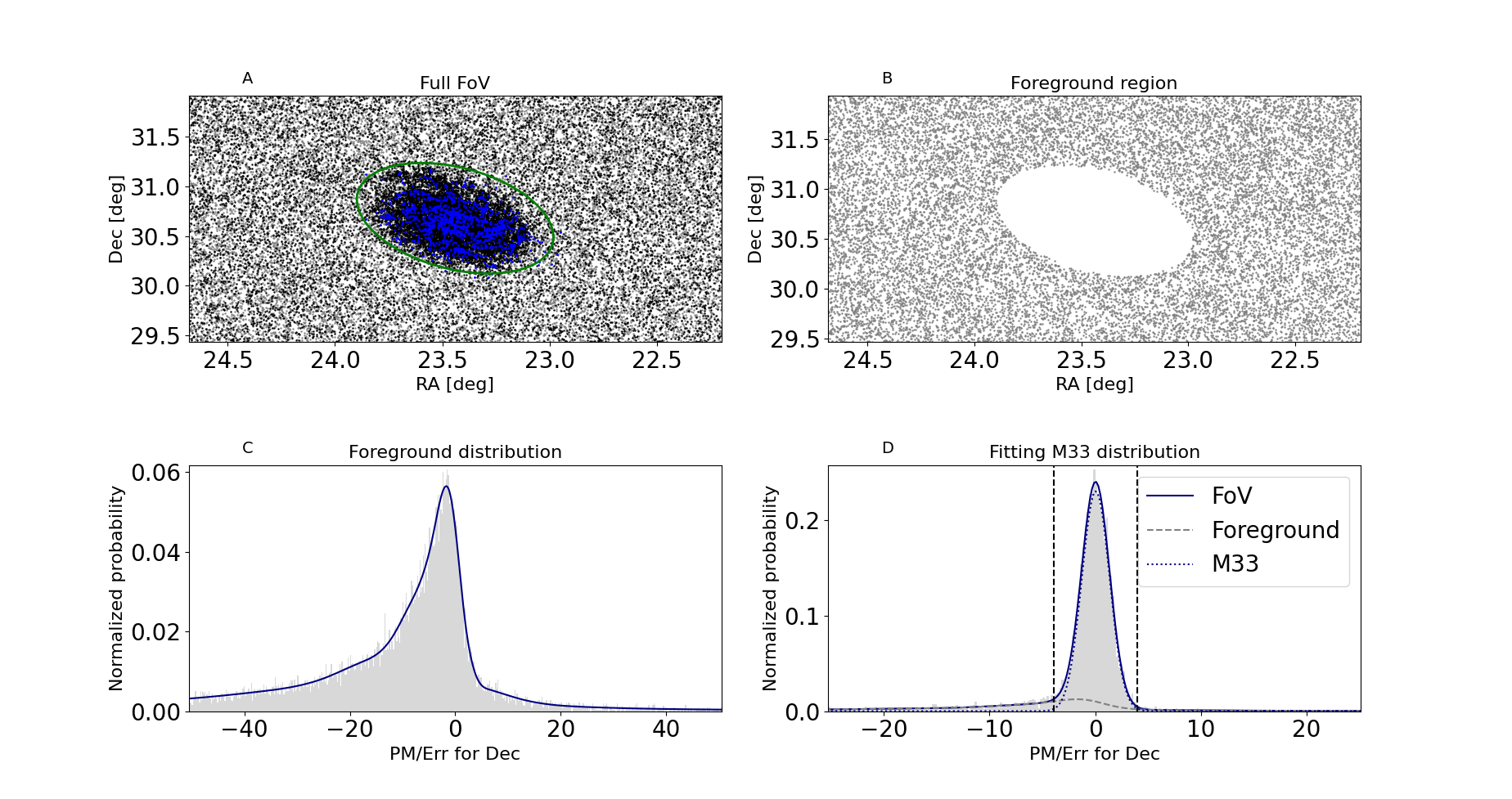}\\
        \includegraphics[width=\textwidth]{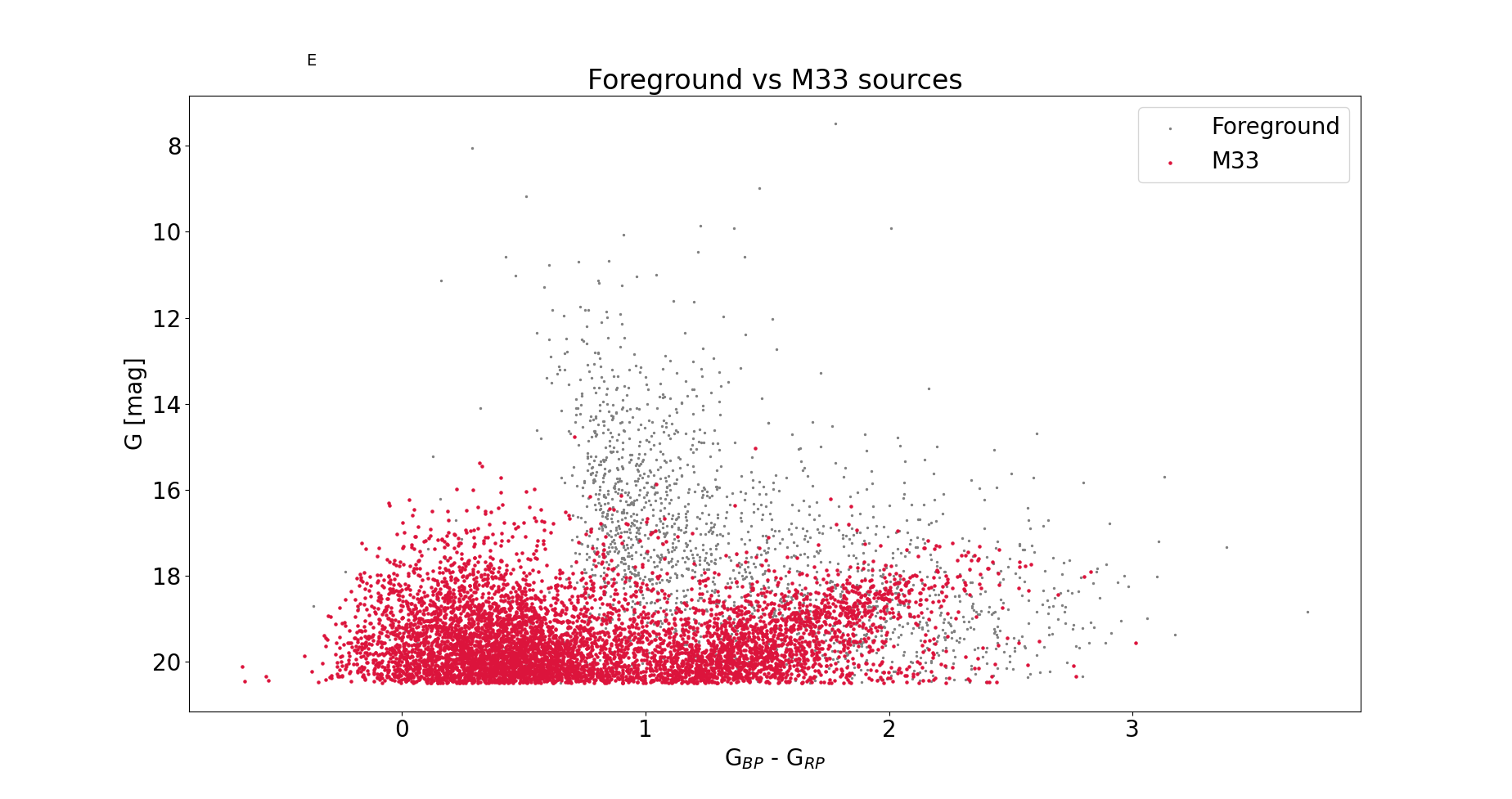}
        \caption{Using \textit{Gaia} to identify and remove foreground sources for M33. (A) Field-of-view of \textit{Gaia} sources (black dots) for M33, with the green ellipse marking the galaxy's boundary. The blue dots highlight the sources in M33 with known spectral classification. (B) Foreground region for M33. (C) Distribution of the proper motion over its error for Dec, for all \textit{Gaia} sources in the foreground region (fitted with a spline function). (D) Distribution of the proper motion over its error for Dec (solid line), for all sources along the line-of-sight of M33, including both foreground and galactic (M33) sources. The dashed line refers to the scaled spline (accounting for the number of foreground sources expected inside M33) and the dotted line to a Gaussian function. The vertical dashed lines correspond to the $3\sigma$ threshold of the Gaussian. Any source with values outside this region is flagged as a potential foreground source.  (E) \textit{Gaia} CMD of all sources identified as galactic (red points) and foreground (gray). The majority of the foreground sources lie on the yellow branch of the CMD, which is exactly the position at which the largest fraction of the contamination is expected. 
}
    \label{f:gaia_process-M33}
\end{figure*}

Similar to Fig. \ref{f:gaia_process} we present the corresponding plots for M33 in Fig. \ref{f:gaia_process-M33}. 

\clearpage

\section{Validation curves}

\begin{figure}
        \includegraphics[width=\columnwidth]{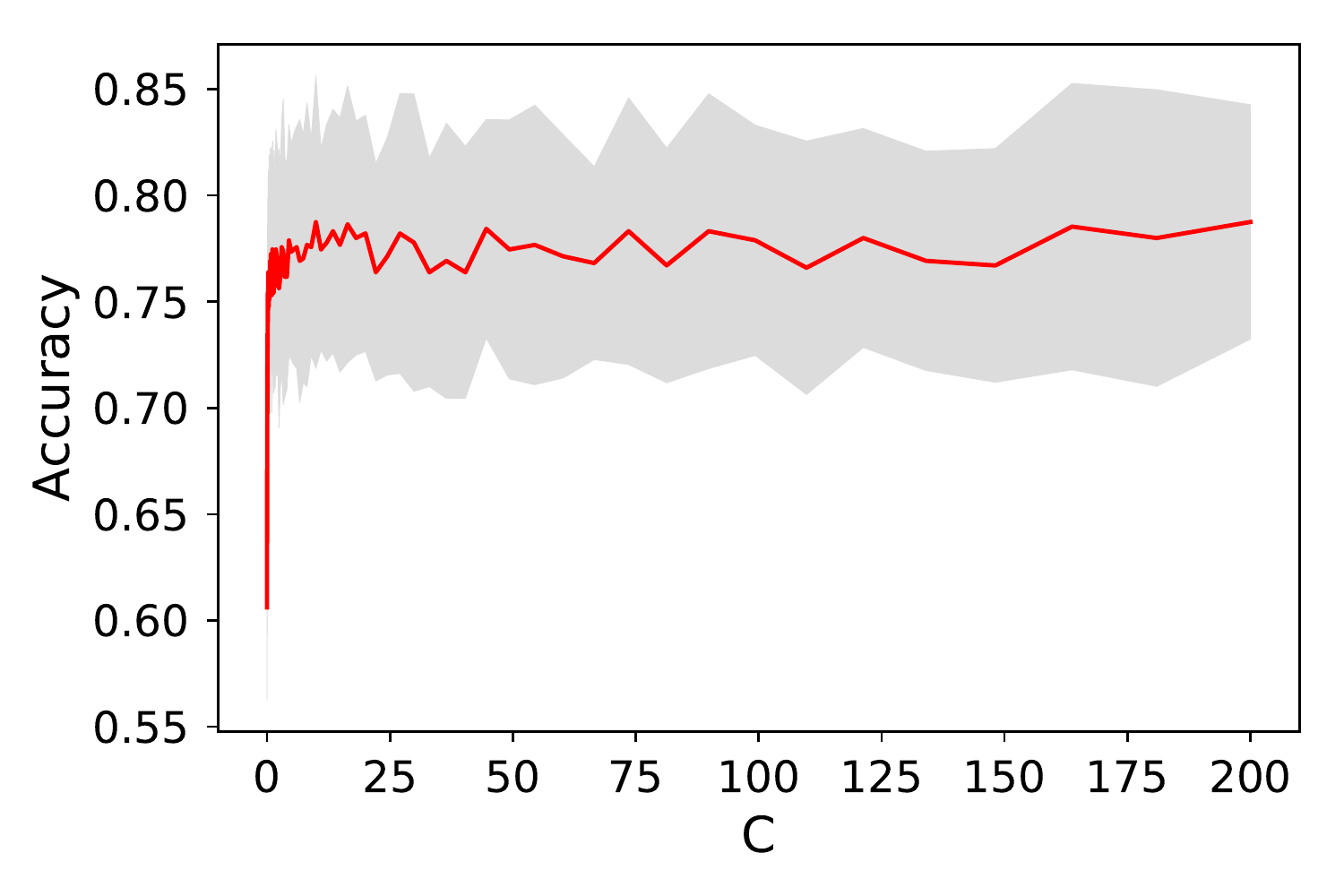} 
        \caption{Achieved accuracy with respect to the regularization parameter \textit{C} for SVC. The gray area corresponds to $1\sigma$. }
    \label{f:optimizing-SVC-C}
\end{figure}

\begin{figure}
        \includegraphics[width=\columnwidth]{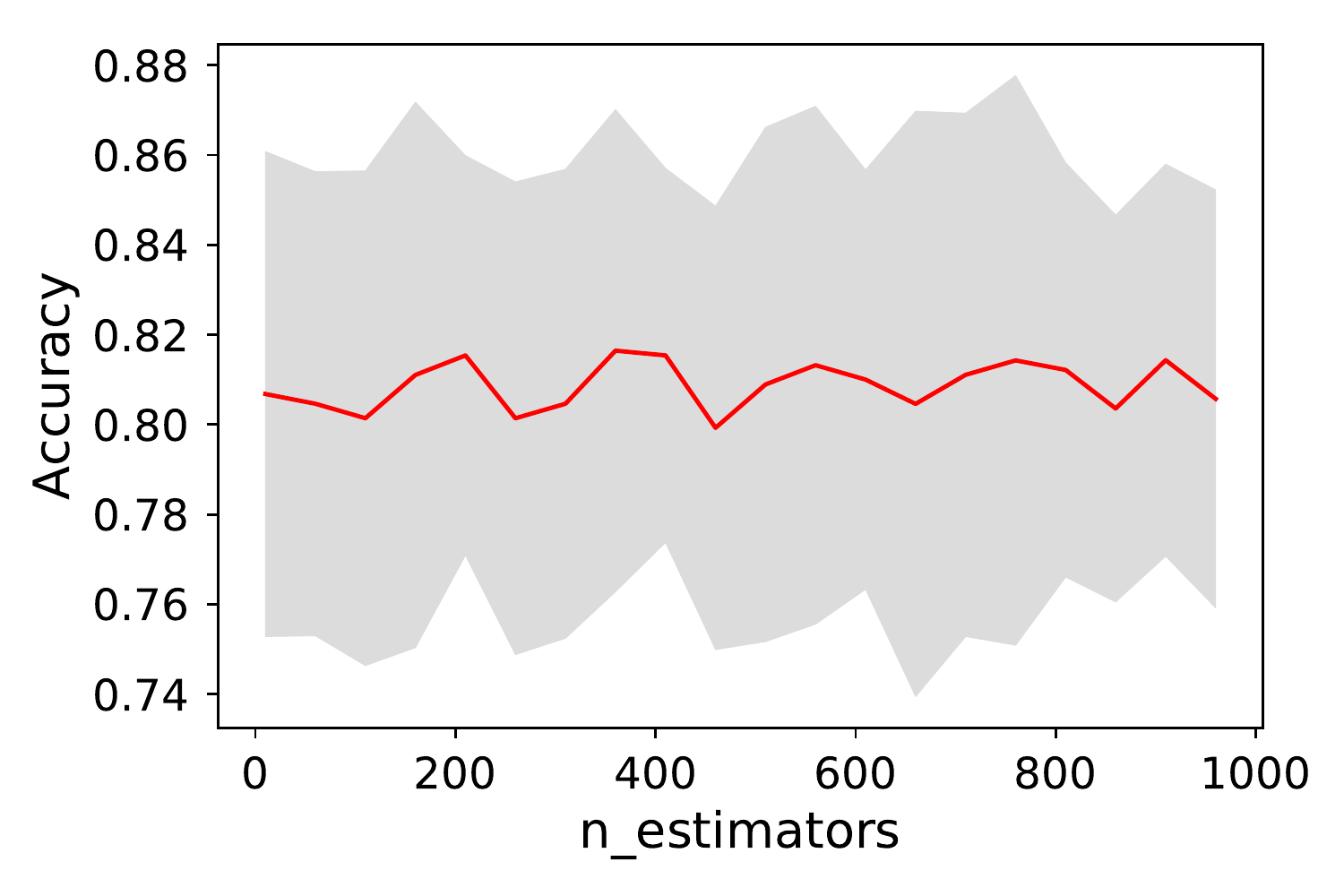}\\
        \includegraphics[width=\columnwidth]{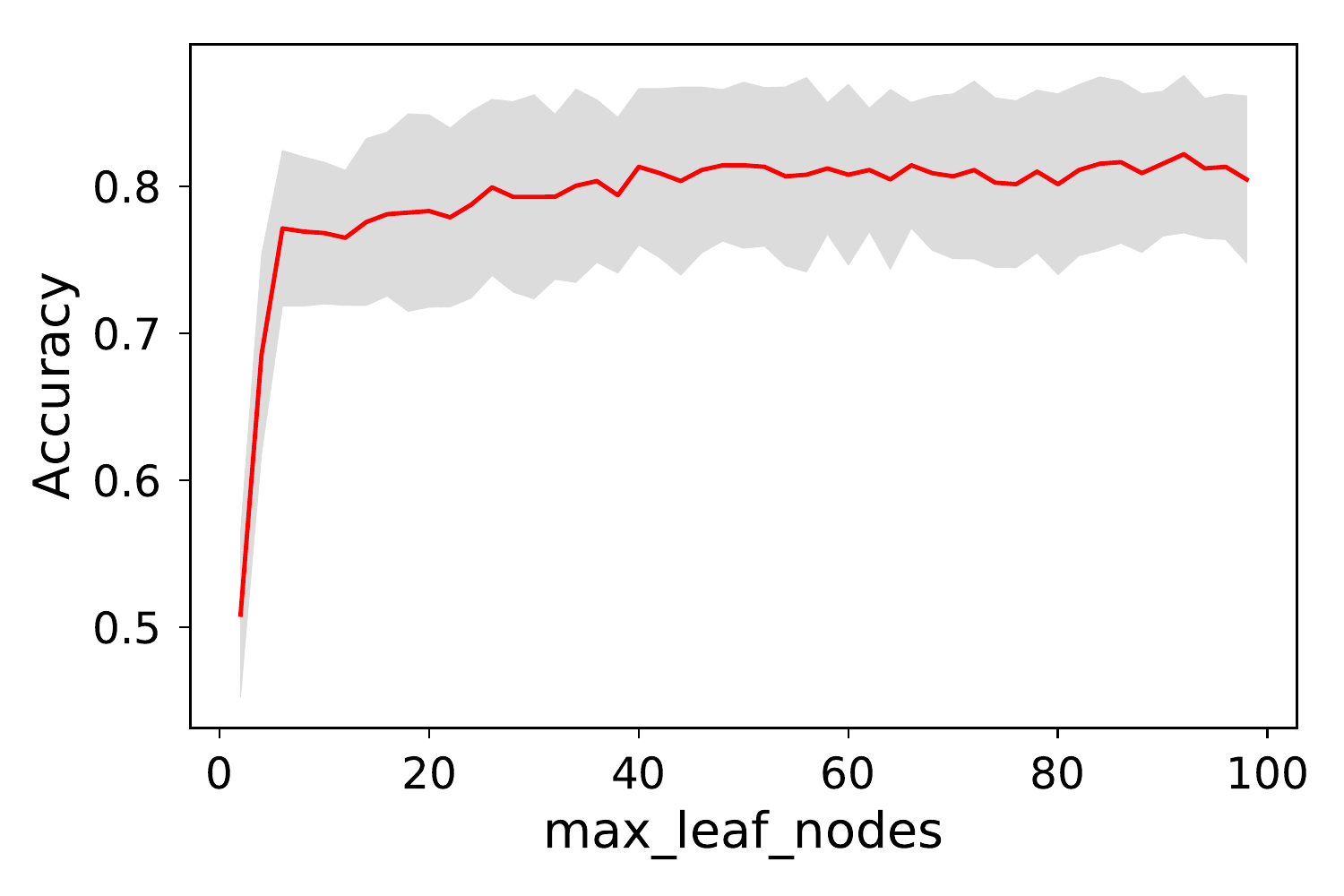}\\
        \includegraphics[width=\columnwidth]{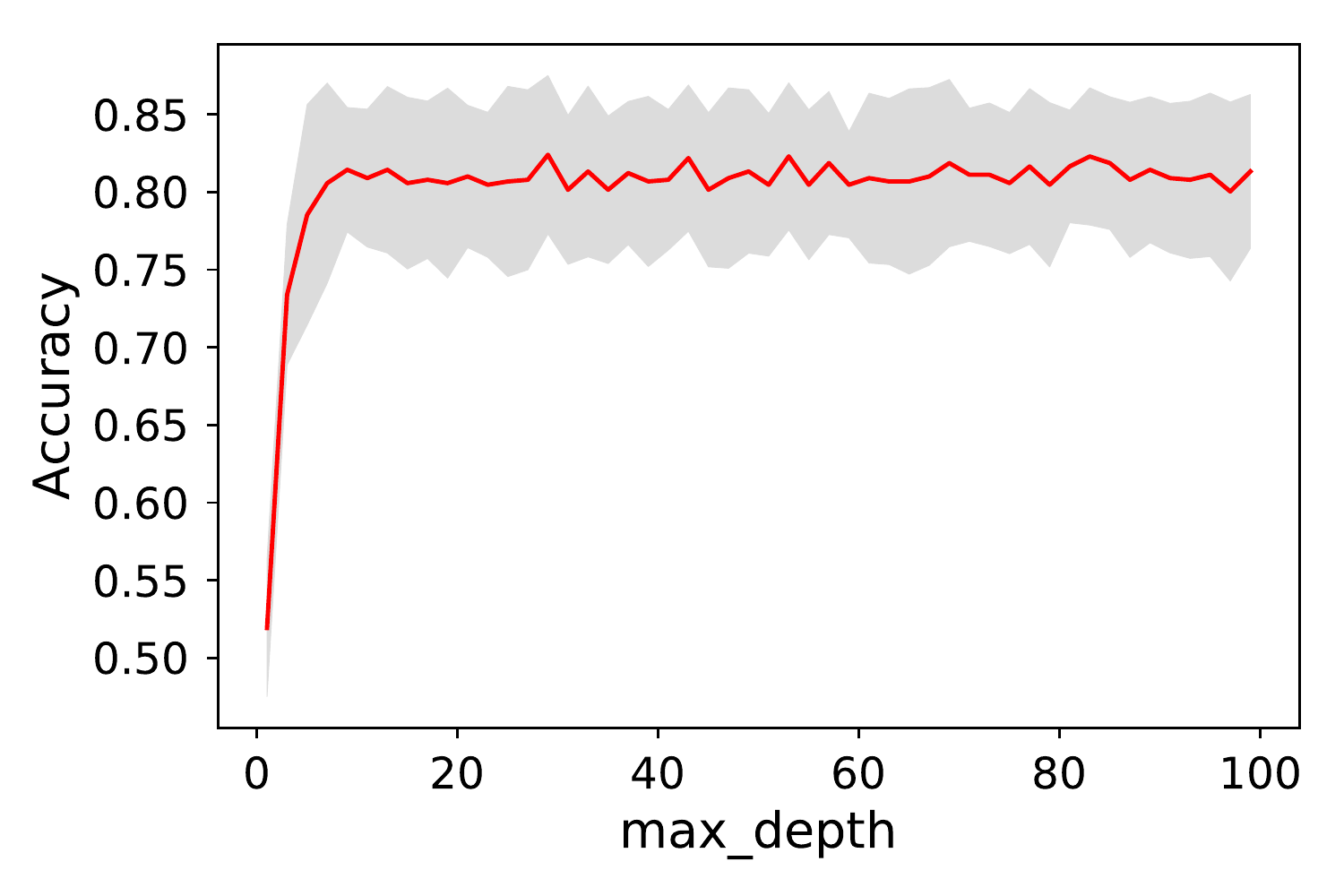}       
        \caption{Validation curves of selected hyperparameters for the RF model: \texttt{n\_estimators} is the number of trees in the forest (top), \texttt{max\_leaf\_nodes} is the number of nodes in each tree (middle), and \texttt{max\_depth} is the maximum depth of the tree (bottom). The other hyperparameters are left to their default values. The gray area corresponds to $1\sigma$.} 
    \label{f:optimizing-RF-curves}
\end{figure}

\begin{figure}
        \includegraphics[width=\columnwidth]{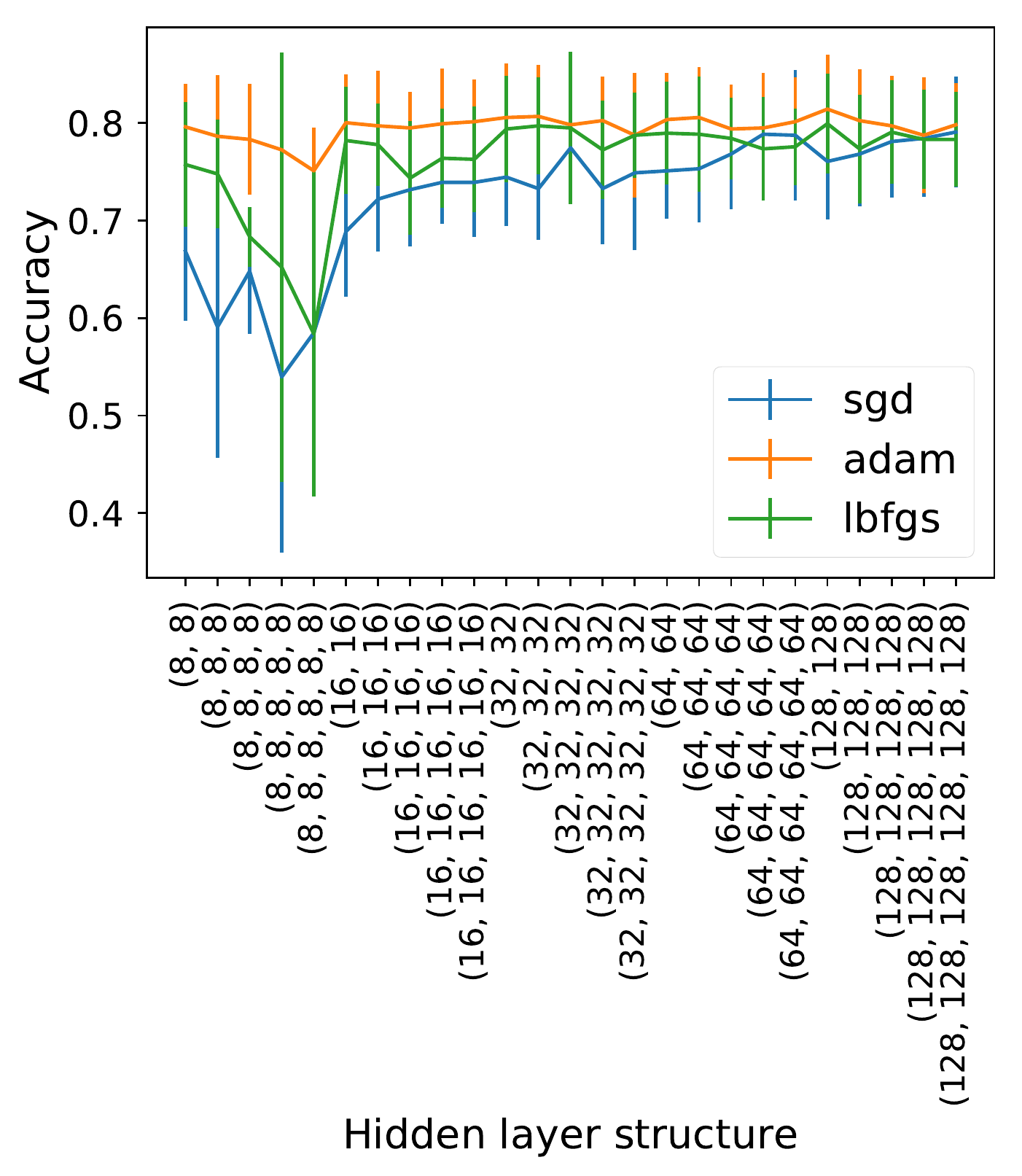} 
        \caption{Obtained accuracy for the different MLP architectures (number of hidden layers with the number of nodes) per solver. \texttt{'adam'} seems to systematically work better among the solvers, with the best accuracy achieved for a network with two layers with 128 nodes each. }
    \label{f:optimizing-NN-structures}
\end{figure}

\begin{figure}
        \includegraphics[width=\columnwidth]{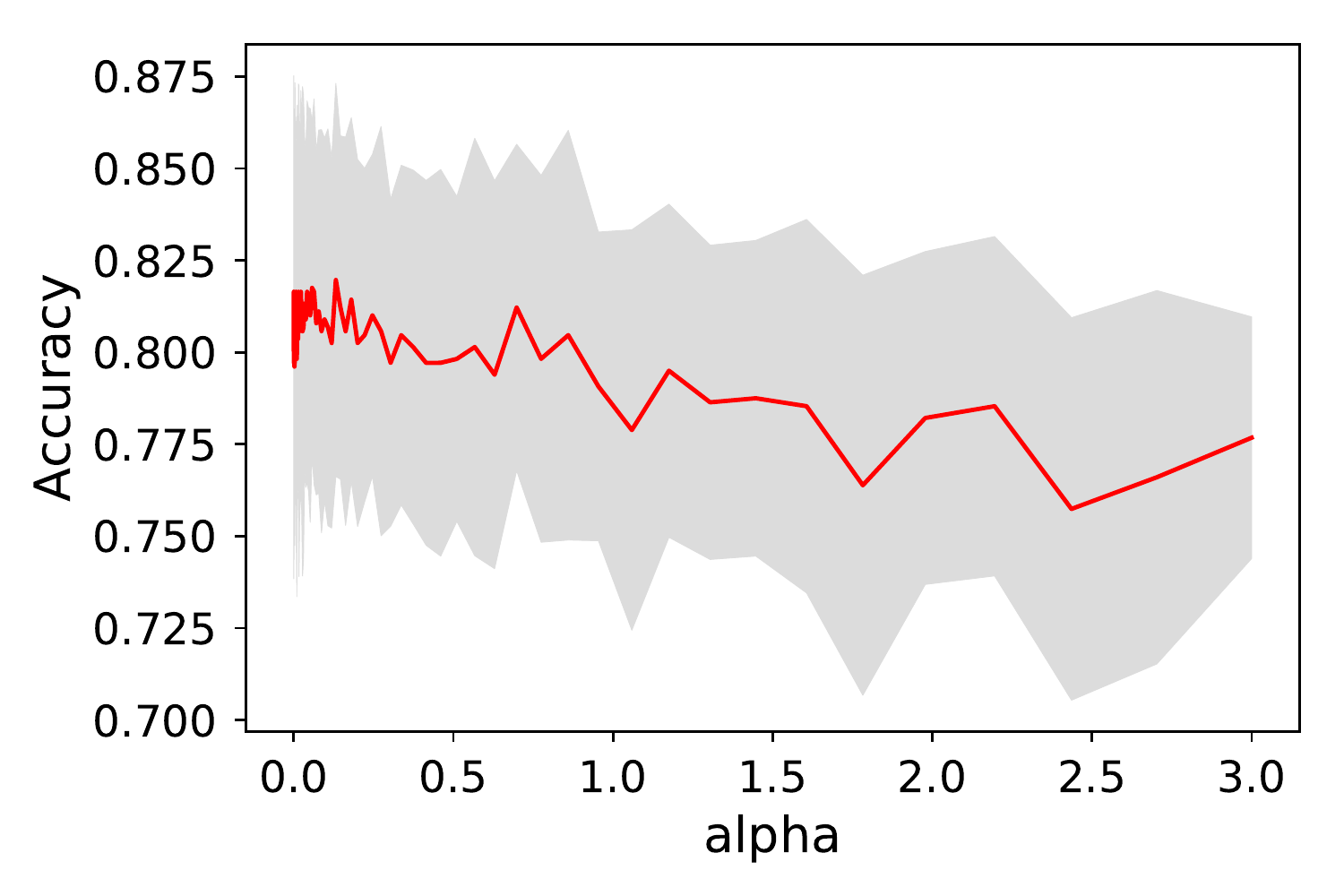}\\
        \includegraphics[width=\columnwidth]{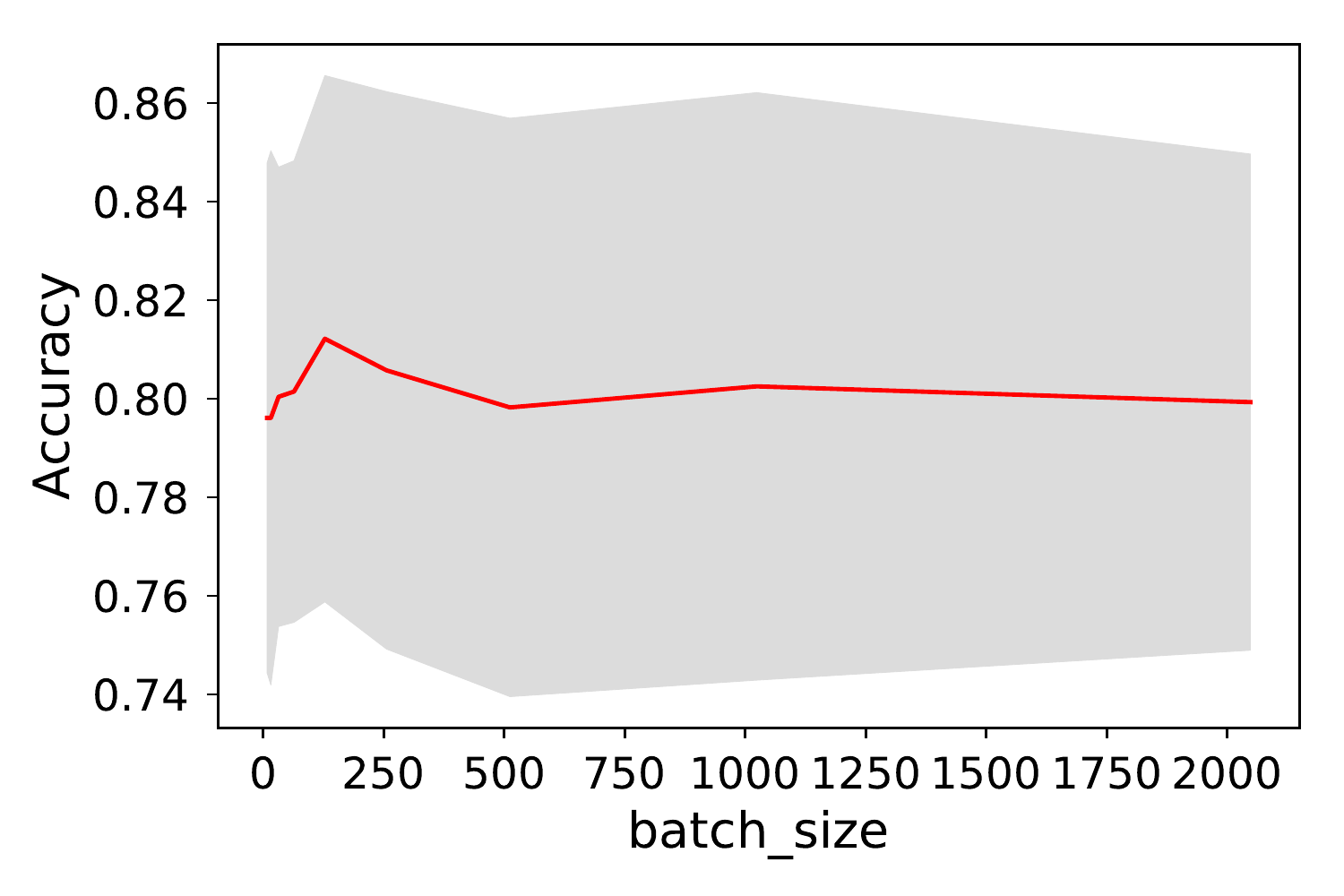}\\
        \includegraphics[width=\columnwidth]{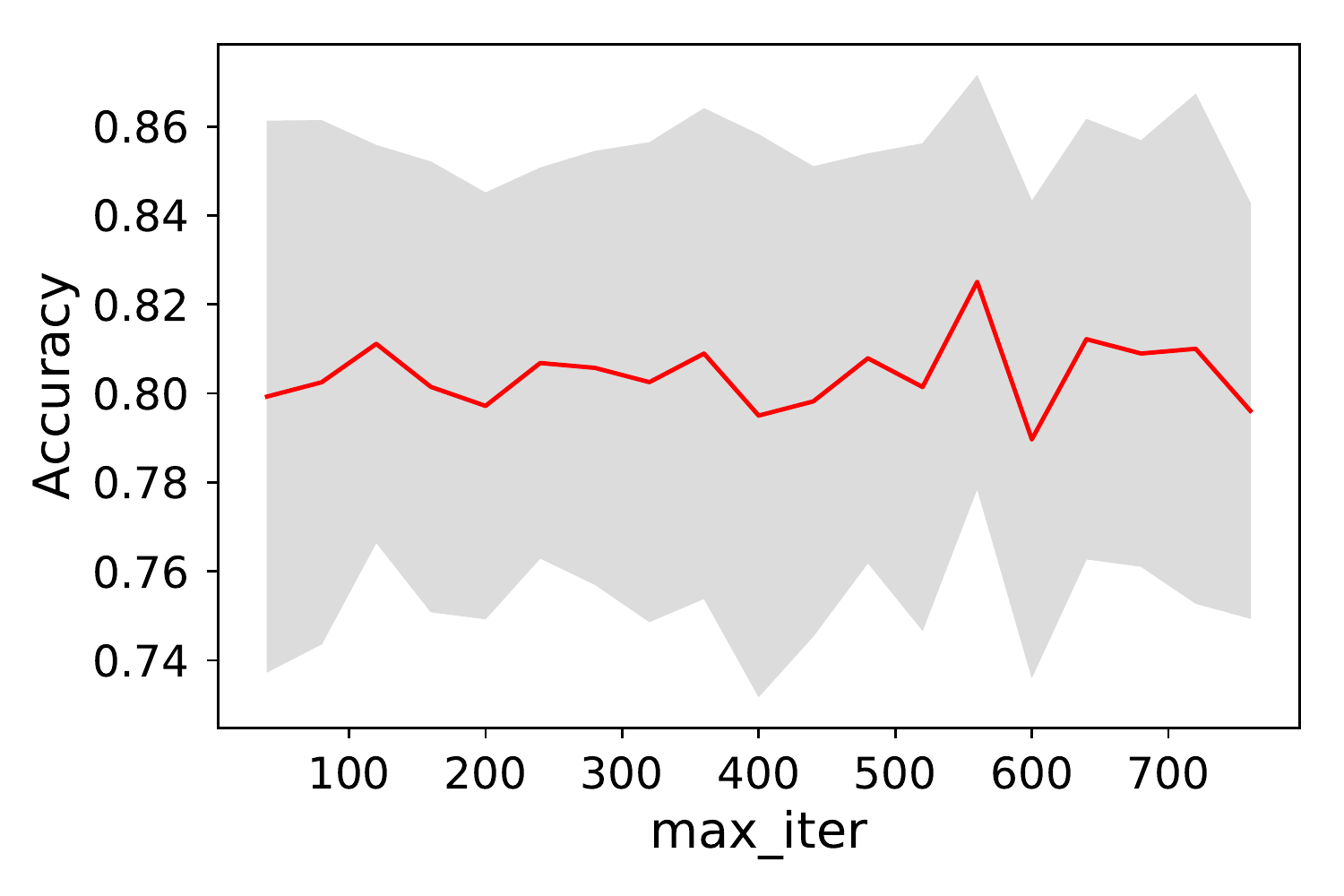}       
        \caption{Validation curves for the (64,64) architecture of the MLP, examining \texttt{alpha} (L2 regularization parameter) and \texttt{batch\_size} (number of samples to estimate the gradient) during the weight optimization of the MLP and \texttt{max\_iter} (maximum number of epochs during training), with optimal values defined as $\sim0.04$, 128, and 160, respectively. The other hyperparameters are left to their default values.} 
\label{f:optimizing-NN-curves}
\end{figure}

In this section we provide the plots for the hyperparameter optimization,  for the three algorithms we used. ig. \ref{f:optimizing-SVC-C} shows the validation curve for the \texttt{C} for SVC. Fig. \ref{f:optimizing-RF-curves} displays the basic parameters for the RF:  \texttt{n\_estimators}, \texttt{max\_lead\_nodes}, and  \texttt{max\_depth}. In Fig. \ref{f:optimizing-NN-structures} we  present the results from the optimal structure for a NN, while in Fig.  \ref{f:optimizing-NN-curves} we plot the validation curves for \texttt{alpha}, \texttt{batch\_size}, and \texttt{max\_iter}.      

\clearpage

\section{Metrics with sample volume}

\begin{figure*}
    \includegraphics[width=\textwidth]{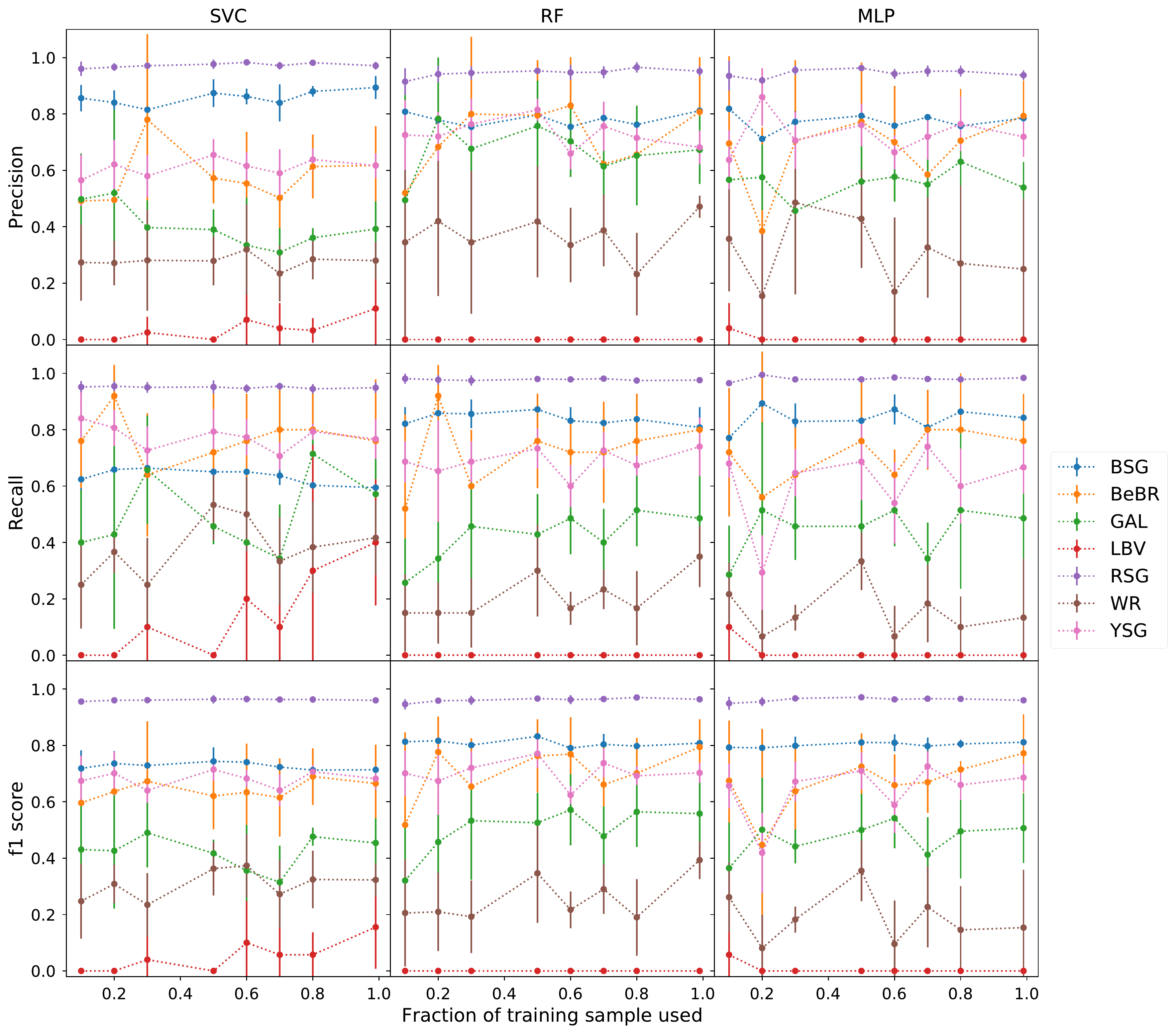}
    \caption{Precision, recall, and F1 score variation when adjusting the used training sample (per class). We notice that the metrics improve when the size of the used sample increases significantly, such as for BeBRs and YSGs. In cases where the samples sizes are already adequate (for BSGs and RSGs), the maximum possible value is achieved faster.}
    \label{f:sample_volume}
\end{figure*}

Complementary to Fig. \ref{f:sample_volume-recall} we plot all metrics (recall, precision, and F1 score) in Fig. \ref{f:sample_volume}. 

\end{appendix}

\end{document}